\documentclass[acmlarge]{acmart}

\AtBeginDocument{%
  \providecommand\BibTeX{{%
    \normalfont B\kern-0.5em{\scshape i\kern-0.25em b}\kern-0.8em\TeX}}}

\setcopyright{acmcopyright}
\copyrightyear{2021}
\acmYear{2021}
\acmDOI{00.0000/0000000.0000000}

\acmVolume{0}
\acmNumber{0}
\acmArticle{0}
\acmMonth{0}

\newcommand{\mypara}[1]{{\bf #1}\hspace{0in}}
\newcommand{\myname}[1]{{PhyMask}}
\usepackage{graphicx}
\usepackage{multirow}
\usepackage{caption}
\usepackage{subcaption}
\usepackage{array}
\usepackage[super]{nth}

\usepackage{amsmath}

\DeclareMathOperator*{\argmax}{arg\,max}
\usepackage{paralist}

\begin{document}

\title{PhyMask: Robust Sensing of Brain Activity and Physiological Signals During Sleep with an All-textile Eye Mask}


\author{Soha Rostaminia}
\orcid{0000-0003-0404-2729}
\email{srostaminia@cs.umass.edu}
\affiliation{%
  \institution{University of Massachusetts Amherst}
  \department{College of Information and Computer Sciences}
  \city{Amherst}
  \state{MA}
  \postcode{01003}
  \country{USA}
}
  
\author{Seyedeh Zohreh Homayounfar}
\email{shomayounfar@umass.edu}
\affiliation{%
  \institution{University of Massachusetts Amherst}
  \department{Department of Chemistry}
  \city{Amherst}
  \state{MA}
  \postcode{01003}
  \country{USA}
}

\author{Ali Kiaghadi}
\email{akiaghadi@umass.edu}
\affiliation{%
  \institution{University of Massachusetts Amherst}
  \department{Department of Electrical Engineering}
  \city{Amherst}
  \state{MA}
  \postcode{01003}
  \country{USA}
}
  
\author{Trisha Andrew}
\email{tandrew@umass.edu}
\affiliation{%
  \institution{University of Massachusetts Amherst}
  \department{Department of Chemistry}
  \department{Department of Chemical Engineering}
  \city{Amherst}
  \state{MA}
  \postcode{01003}
  \country{USA}
}

\author{Deepak Ganesan}
\email{dganesan@cs.umass.edu}
\affiliation{%
  \institution{University of Massachusetts Amherst}
  \department{College of Information and Computer Sciences}
  \city{Amherst}
  \state{MA}
  \postcode{01003}
  \country{USA}
}

\renewcommand{\shortauthors}{Rostaminia, et al.}

\begin{abstract}
Clinical-grade wearable sleep monitoring is a challenging problem since it requires concurrently monitoring brain activity, eye movement, muscle activity, cardio-respiratory features and gross body movements. This requires multiple sensors to be worn at different locations as well as uncomfortable adhesives and discrete electronic components to be placed on the head. As a result, existing wearables either compromise comfort or compromise accuracy in tracking sleep variables. We propose \myname{}, an all-textile sleep monitoring solution that is practical and comfortable for continuous use and that acquires all signals of interest to sleep solely using comfortable textile sensors placed on the head. We show that \myname{} can be used to accurately measure all the signals required for precise sleep stage tracking and to extract advanced sleep markers such as spindles and K-complexes robustly in the real-world setting. We validate \myname{} against polysomnography and show that it significantly outperforms two commercially-available sleep tracking wearables -- Fitbit and Oura Ring.
\end{abstract}



\maketitle

\section{Introduction}

There has been a significant commercial interest in measuring sleep given the wide-ranging effects of sleep disruptions, which includes diminished cognitive functioning, diabetes, high blood pressure, heart disease, obesity and depression~\cite{ferman1999rem, knutson2006role, javaheri2008sleep, kasasbeh2006inflammatory, taheri2006link, schwartz2005symptoms}. There is also growing interest in measuring sleep disorders which affects 50 to 70 million Americans of all ages and socioeconomic classes~\cite{altevogt2006sleep}. Therefore, it is crucial to scale accurate sleep monitoring such that it can be done less expensively at clinics and more comfortably at home.  
 
But achieving high-quality clinical-grade sleep monitoring with a comfortable wearable device is complicated by the number of sensor modalities and locations that need to be simultaneously monitored. Sleep is a complex process and monitoring it accurately necessitates many sensors placed at different places on the body. This includes sensors to measure brain activity, eye movement, muscle activity, as well as cardio-respiratory features and gross body movements. To monitor all of these variables, clinical-grade sleep monitoring systems (also referred to as polysomnography or PSG) place electrodes  on the head as well as several other locations on the body. However, this is impractical for daily use. 

\subsection{Limitations of sleep trackers} 

\begin{figure*}[t]
\centering
\includegraphics[width=1\linewidth, scale=1, bb=0 0 1000 500]{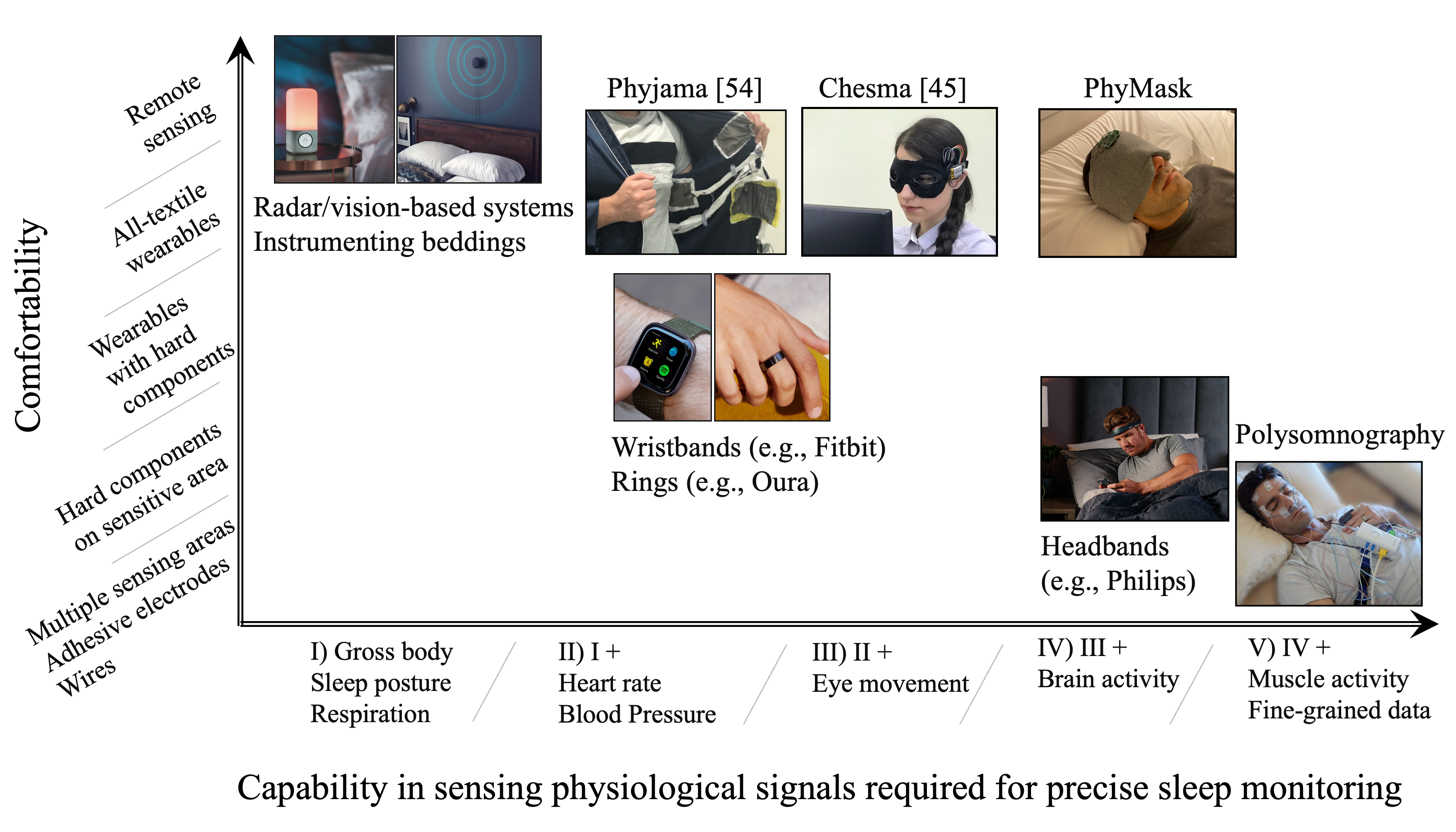}
\caption{Comparison of state-of-the-art sleep monitoring solutions in terms of comfort and capability in sensing various signals.}
\label{fig:related_work_graph}
\end{figure*}

Sleep sensing devices in the market therefore have to tradeoff between comfort and accuracy while monitoring sleep parameters (see Figure~\ref{fig:related_work_graph}). A key challenge is that measuring brain electrical activity (Electroencephalography, EEG) is both the most onerous aspect of sleep sensing (since it requires electrodes on the head) and the most valuable sensing modality (since sleep is best measured via brain activity). Thus, sleep sensing devices can be divided into two groups --- those that favor comfort over accuracy and rely on surrogate measures of sleep such as heart rate, breathing, and body movement signals rather than EEG, and those that favor accuracy over comfort and retain the use of electrodes to measure the electrical activity of the brain and eye movements induced by sleep but try to do so without compromising too much comfort.

\mypara{Measuring surrogate signals:} The first group, i.e. sleep monitors that use surrogate measures of sleep, includes contactless sleep monitors like instrumented bedding and bedside sleep monitors \cite{jia:2017, Rahman:2015, Liu:2015, li2016noncontact} as well as wearable devices such as the Fitbit and Apple Watch \cite{fitbit, applewatch}. 

While these are comfortable, they sacrifice fidelity and precision. Since the surrogate signals only capture a coarse temporal structure of sleep, they provide only coarse-level sleep metrics like sleep quality and, in some cases, macro-structural analysis of sleep such as sleep stages. In addition, metrics provided by these devices are accurate mostly for ``normal'' healthy individuals and not representative of individuals with sleep disorders. This is because these devices need to compensate for the imprecision of surrogate measures by relying on large population-level data analysis but these measures are erroneous for individuals whose sleep patterns do not follow population averages. For example, Rapid Eye Movement (REM) sleep stage is known as a state where the person experiences random/rapid movement of the eyes, accompanied with low muscle tone throughout the body, and the tendency to dream vividly. However, a person who is suffering from the REM sleep disorder, usually has violent arm and leg movements during REM sleep stage. Sleep monitors that use surrogate measures of sleep assume that the body normally freeze and does not move during REM since they solely measure at the cardio-respiratory features and gross body movement. As a result, these sleep trackers are inaccurate for individuals with substantial clinical need such as older adults with sleep disorders and medication-induced sleep disruptions \cite{kang2017validity}. They also fail to capture day/night (circadian) rhythm sleep patterns of individuals with sleep abnormalities (such as older adults with dementia) \cite{wennberg2017sleep}.

\mypara{Measuring EEG signal:} The second group, i.e. headworn devices with electrodes, includes headbands and masks that have become available for sleep tracking \cite{muse, Brainbit, dreem}. However the challenge in these devices is comfort --- headworn sleep trackers require rigid sensing elements that are directly pressed against the skin. For example, the Phillips Smart Sleep headband \cite{Phillips-Smart-Sleep} uses behind-the-ear sticker electrodes and the Muse \cite{muse} has an optical sensor on the forehead and EEG electrodes on a rigid frame. These rigid structures and the embedded hard-components on the head make such a device unnatural and uncomfortable. 

Another important limitation of these devices is that despite measuring EEG, these devices do not currently expose micro-structures of sleep such as spindles and K-complexes. This is a missed opportunity since micro-structures play an essential role in information processing and long-term memory consolidation \cite{jegou2019cortical} and potentially as biomarkers of Alzheimer's disease \cite{kam2019sleep} and seizures~\cite{el2008k}.

\subsection{Our contribution} 

In this work, we present a complete textile-based head-worn sleep sensing system that bridges the gap in existing sleep measurement devices in two ways. First, we show that a single head-worn device that leverages solely textile-based sensors can provide nearly all the parameters that are used in clinical-grade sleep monitoring without requiring rigid sensing elements and thereby without sacrificing comfort. Second, we show that such a device can provide accurate estimates of both macro and micro-structures of sleep including sleep stages, spindles and K-complexes.

Our work builds on our previous efforts on developing individual sensors that provide part of the solution. In the previous work~\cite{chesma}, we have introduced two sensors that we leverage in this work: a) a new thread-based, reusable wet electrode that achieves the high signal quality of commercial wet electrodes as well as the comfort and unobtrusiveness of dry electrodes (our preliminary results showed that these electrodes were viable for measuring eye movement patterns), and b) a fabric-based piezoionic pressure sensor~\cite{PressION} that is sensitive to ballistic signals from heartbeats.

In this work, we build on the above to provide a holistic multi-modal solution that can provide reliable and robust measures of sleep in a natural setting. Our work makes notable contributions towards a practical daily-use system that is comfortable yet accurate. From a device perspective, we show for the first time that a) a fabric-based electrode can be used to accurately measure EEG signals with high signal-to-noise, and that b) textile-based pressure sensors on the sleep mask can allow us to sense tiny head movements induced by heartbeats and respiration in various sleep postures. From an analytic perspective, we describe a signal processing and machine learning pipeline that allows us to extract both high-level physiological features such as heart rate and breathing rate, as well as micro-events of brain activity during sleep such as spindles and k-complexes. Put together, \myname{} is a comfortable textile-based sensing platform that can simultaneously sense brain activity (EEG), eye movement (electrooculography, EOG), physiological parameters (respiratory and cardiac rhythm), gross body movement, and sleep posture from the head and infer both macro- and micro-structural parameters of sleep.

\begin{figure*}[t]
\centering
\includegraphics[width=1\linewidth]{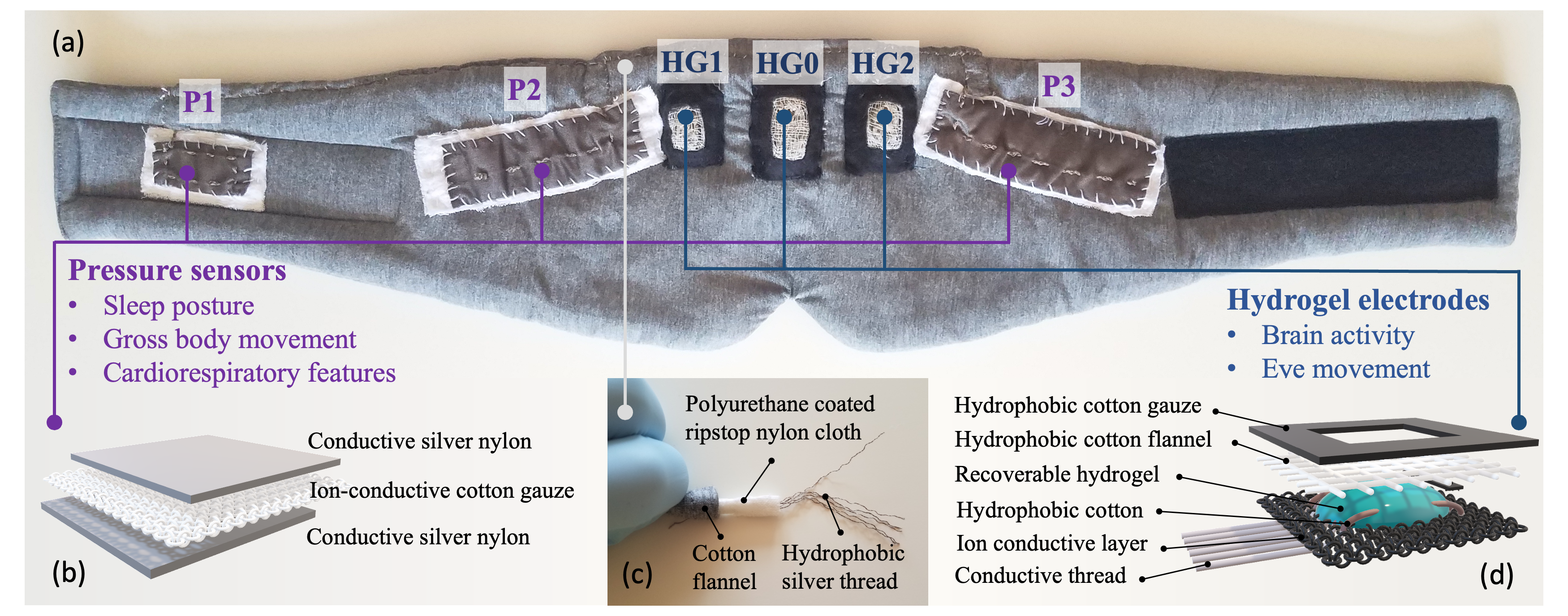}
\caption{PhyMask platform. (a) inside view of \myname{}, illustrating pressure sensors and hydrogel electrodes placement, (b) schematic of the layered structure of pressure sensor, (c) wiring constructed with isolated hydrophic conductive threads, and (d) the schematic of the layered structure of hydrogel electrode.}
\label{fig:PhyMask_platform}
\end{figure*}

We design and fabricate a fully functioning prototype of a comfortable all-textile sleep tracking device (shown in Figure~\ref{fig:PhyMask_platform}) that includes the fabric-based sensing elements, fabric-based wires, low-power circuit for signal amplification and data acquisition, a wireless radio for data transfer, and tailored algorithms for physiological signals and sleep micro-events tracking. 

We perform exhaustive data collection studies to validate our device and compare it against polysomnography ground truth and two popular sleep trackers, Fitbit wristband and Oura Ring, for continuous long-term sleep tracking.

Our results show that:
\begin{itemize}
    \item We have high accuracy for measuring heart rate (median error of  1.7 beats/min) and respiration (median error of 1 resp/min), and can easily capture gross motor activity including posture changes.
    \item The \myname{} can measure sleep stages almost as well as polysomnography ground truth with an F1 score of 0.91, and greatly outperforms Fitbit (F1 score of 0.64) and Oura Ring (F1 score of 0.7).
    \item The \myname{} can be used to detect advanced sleep markers such as spindles and K-complexes that are typically not provided by commercial wearable devices. We show that \myname{} detects spindles and K-complexes with accuracy of 85\% and 86\%, respectively. 
\end{itemize}

\section{Background and Related Work}
In this section, we first look at the key sleep markers that we need to measure in order to provide a reliable sleep tracking solution. We then look at the existing sleep tracking solutions with respect to two questions: a) which of these sleep markers can these solutions provide? and b) how comfortable and unobtrusive are these techniques? We explain the \myname{} solution and how it overcomes the accuracy and aesthetic shortcomings of current state-of-the-art sleep tracking devices. 

\begin{table}[ht]
\centering
\caption{The characteristics of important biosignals used in sleep analysis during each sleep stage~\cite{roebuck2013review_sleepfeatures}.}
\label{tab:sleepfeatures}
\renewcommand\arraystretch{1.5}
\begin{tabular}[t]{l*{5}{>{\centering\arraybackslash}p{0.16\linewidth}}}
\toprule
&Brain activity (EEG) & Eye movement (EOG) & Heart rate & Respiration & Gross motor activity\\
\midrule
N1 & Theta (4-8 Hz) & Slow cyclic eye rolling & $\downarrow$ & $\downarrow$ & Occasional muscle twitches\\
N2 & Theta (4-8 Hz) spindles \& K-complexes & $\downarrow$& $\downarrow$& $\downarrow$& $\downarrow$\\
N3& Delta (0.5-4 Hz) & Lowest& Lowest& Lowest& Relaxed\\
REM& Beta (13-32 Hz) & fast rapid eye movement& $\uparrow$& Fast \& irregular & Paralysis\\
\bottomrule
\end{tabular}
\end{table}
\subsection{Background: Sleep structure and biosignal characteristics}
\label{sec:BackgroundSleep}

\mypara{Sleep Macro- and Micro-structure:} Traditional sleep measures tend to focus on sleep macrostructure i.e. the longer-term measurements of sleep. A number of parameters are commonly used as outcome measures of sleep macrostructure including total sleep time (TST), sleep latency, sleep efficiency, wake time, and percentage of Stage 1 (N1), Stage 2 (N2), Stage 3 (N3) and REM sleep.

However, sleep is also rich in shorter timescale phasic EEG events such as K-complexes, spindles, delta wave bursts, and others \cite{terzano2000origin,malinowska2006micro}. While macrostructural variables have received more attention in past research, recent work has paid more attention to microstructural variables due to important clinical correlations to age-related sleep changes and other age-related disorders like Parkinsons and Alzheimers \cite{staner2003sleep}. Many age-related sleep changes are not captured by traditional sleep stage scoring --- for example, research has shown that aging-related changes were reflected particularly in fast spindle density, K-complex density, and delta power during N3 sleep—than on conventional sleep staging variables \cite{schwarz2017age}.

\mypara{Biosiognal characteristics during sleep:} A number of biosignals need to be measured to fully characterize micro and macro-structural parameters of sleep. Table~\ref{tab:sleepfeatures} summarizes the biosignals characteristics during each sleep stage i.e. Rapid Eye Movement (REM), non-REM N1, non-REM N2, and non-REM N3. Each sleep stage is characterized by brain waves of specific frequencies and/or amplitudes and is also associated with certain types of eye movements and muscle activities. Therefore, brain activity (EEG), eye movement (EOG) and muscle activity (EMG) measurements are required for accurate sleep staging. Brain activity, in particular, is considered the most important signal for high-quality sleep stage monitoring~\cite{roebuck2013review_sleepfeatures}. 

Laboratory-based polysomnography (PSG), which is considered the most clinically accurate method of monitoring sleep also includes monitoring of several other parameters which are relevant to holistic understanding of sleep. Cardio-respiratory parameters are particularly useful for studying and analysis of sleep disorders such as apnea. Thus, measurements of respiration and heart features are also included in laboratory-based polysomnography (PSG). In addition, gross body movements during sleep is also important when studying sleep disorders such as periodic leg movements.

\subsection{State-of-the-art in Sleep Tracking}

Sleep tracking devices typically capture a subset of the signals in Table~\ref{tab:sleepfeatures} due to limitations of comfort, sensor position, and other factors. We look at how well different types of sleep monitoring solutions can monitor these parameters.

\mypara{Non-wearable sleep trackers.} There has been a number of non-wearable sleep tracking approaches that attempt to monitor sleep indirectly by instrumenting the environment around the user. For example, Jia et al. leverage highly sensitive geophones and embed it into bedding to measure the seismic motions induced by individual heart beats and slow moving signals from respiration~\cite{jia:2017}. Commercial MEMS accelerometer-based and piezoelectric-based units are also available, that can measure body movement and heart rate based on ballistocardiography signals measured via the bed during sleep~\cite{Murata, beddit}. Several approaches also use vision-based and depth camera-based methods to find physiological variables such as respiration~\cite{bartula2013camera, martinez2012breath}, heart rate~\cite{li2016noncontact}, pulse oximetry~\cite{vogels2018fully}, head posture~\cite{deng2018design}, and body movement~\cite{yu2012multiparameter, nochino2019sleep} during sleep. Another body of work is on radar-based sensing of respiration and heart rhythm~\cite{fadel:2015, Niu:2018, Rahman:2015}. These methods use FMCW or UWB radars and measure changes in the displacement and the doppler shifts due to respiration and ballistics of the heart.

While non-contact sleep tracking solutions have the advantage of being unobtrusive and comfortable, they tend to be imprecise and noisy. The imprecision stems from the inability to observe the most valuable signal for sleep monitoring i.e. EEG, and relying on other parameters such as body movements (chest movement during respiration and gross body movements) to infer sleep markers. Noisy measurements results from the fact that the signal is confounded by proximate activity such as multiple individuals on the bed~\cite{Zhenhua:2017:sleepsharedbed, liu2015tracking} or other moving objects in the vicinity like a fan. 

\mypara{Wrist-worn wearable sleep trackers.} There are many wearable devices in the market for sleep sensing, most of which use hard electronic components such as photoplethysmography (PPG) and IMU to measure the pulse wave and body movement on the wrist or fingers (e.g. Fitbit~\cite{fitbit}, Polar Vivofit~\cite{vivofit}, Actiwatch~\cite{Actiwatch}, Whoop~\cite{WHOOP}, and Oura Ring~\cite{oura}). 

The main disadvantage of these solutions is the inability to monitor the EEG and EOG signals, which is considered critical for high-quality sleep analysis. Hence, they solely rely on cardiac, respiratory, and gross body movement information for sleep stage tracking which results in poor performance. Several studies in research settings have reported on the validity of the wearable sleep trackers compared to polysomnography (PSG). For example, Moreno-Pino et al. validate two models of Fitbit sleep trackers (Charge 2 and Alta HR) against PSG and find statistically significant differences between PSG and Fitbit measures for all sleep stages except for REM sleep~\cite{moreno2019validation}. Also, Miller et al. examine Whoop against PSG and report the sensitivity to light sleep, deep sleep, REM, and wake, 62\%, 68\%, 70\%, and 51\% respectively~\cite{whoop:evaluation:2020}.

\mypara{Head-worn wearable sleep trackers.} Another category of sleep tracking devices is headworn wearables. Headworn solutions are more promising as they allow us to obtain biopotential signals (EEG and EOG) from the head. Kim et al. leverage gel-based standard adhesive electrodes on a headband device and collect EEG, EOG, and EMG signals for sleep tracking purpose~\cite{kim2020wearable}. Phillips Smartsleep~\cite{Phillips-Smart-Sleep} also use sticky disposable electrodes placing behind the ear and obtain EEG signal during sleep. Despite the high signal quality of these electrodes, they are uncomfortable due to the adhesive and are not practical for long-term wear, since once the gel dehydrates, the electrode loses its functionality and should be replaced. Therefore, dry electrodes are often preferred in wearable devices. Kuo et al. use graphene-based electrodes in a sleep mask, for tracking eye movement signal during sleep~\cite{eogmask}. Furthermore, the commercially available computational headbands, Muse~\cite{muse}, Dreem \cite{dreem}, Brainbit \cite{Brainbit}, and Neuroon~\cite{neuroon}, use dry electrodes for obtaining EEG on the forehead. They also have a PPG sensor for cardiac and respiration rate tracking, and other sensors. 

The main drawback of these systems is the rigid structure of the device and the use of hard sensing components that touch the skin on the sensitive head areas. These can make the system highly uncomfortable in different sleep postures and not ideal for continuous long-term sleep tracking. Another debatable aspect of the current head-worn solutions is the quality of the acquired biopotential signals. Majority of these technologies use dry electrodes to avoid adhesives and increase the comfort factor, however, this results in considerable increase in motion artifact~\cite{wince}.

\mypara{PhyMask.} While most of the current solutions focus on collecting a few of the essential signals needed for precise sleep monitoring, \myname{} captures all these signals i.e. EEG, EOG, and heart and respiratory rate, while providing information on head posture and body movements.

This allows our device to extract information about sleep macrostructure as well as microstructure in contrast to existing devices which only provide macrostructural parameters. To the best of our knowledge, \myname{} is the first solution that can also accurately detect sleep micro-events, i.e. spindle and K-complex in addition to macrostructural parameters.

\subsection{State-of-the-art textile-based wearables}

Since \myname{} relies on textile-based sensing elements, we look at the state-of-the-art textile-based sensing solutions for monitoring these parameters.

\mypara{Physiological and physical signals sensing.}
Physiological and physical sensing through fabric-based elements has gained a lot of interest during the past decades. Some of the existing methods and requirements of smart textiles are surveyed in~\cite{andreoni2016defining}. Many of these devices are based on conductive fabric electrodes embedded into tight-fitting clothing. There has been some work on measuring impedance changes for physiological measurements --- for example, \cite{paradiso2005wearable} integrates piezoelectric elements in a garment and acquire electrocardiogram, respiration, and activity information. In our previous work, Phyjama~\cite{Kiaghadi:2019}, we also showed that we can measure heart rate and respiration as well as sleep posture through resistive pressure and triboelectric patches sewn into a loosely worn sleepwear. To the best of our knowledge \myname{} is the first system that captures the physiological and physical signals (cardiac, respiration, general body movement, and head posture) through fabric-based elements robustly from the \emph{head}.

\mypara{Biopotential signals sensing.} 
There has been some studies on design of textile-based dry electrodes. For example, silver-coated~\cite{liang2015development} or graphene-coated~\cite{golparvar2018electrooculography, sahito2015graphene} fabrics, or polymeric foams~\cite{lin2011novel} have been embedded as a dry electrode into a headworn to obtain EOG and EEG signals. Pani et al.~\cite{pani2015fully} use PEDOT:PSS-based electrodes to measure ECG signal. While dry electrodes are unobtrusive and comfortable to use, they are highly susceptible to noise. The fundamental challenge in reliably extracting biopotential signals through wearable electrodes is to design an electrode enjoying both the signal quality of traditional wet electrodes and the comfort of the dry ones. Shu et al.~\cite{shu2019multilayer} and Alba et al.~\cite{alba2010novel} propose semi-wet textile electrode for EEG measurement. However, to the best of our knowledge the existing solutions does not have the longevity, wash-stability, and recoverablity that are required for long-term use of the electrodes.

\mypara{PhyMask.} \myname{} presents a complete system that incorporates several of these textile-based sensor innovations in a single platform to monitor sleep. \myname{} uses the thread-based, reusable wet electrode to measure biopotential signals that has the signal quality of commercial wet electrodes and the comfort and unobtrusiveness of dry electrodes. We also use a pressure-sensitive, ionic fabric electrode to capture pulse wave-forms from the head. In order to further minimize using discrete electronic hard components at sensitive pressured locations, we inter-connect the all-textile patches by using silver-plated nylon threads as wires that are shielded in cotton.

\section{P\texorpdfstring{\MakeLowercase{hy}M\MakeLowercase{ask}}{PhyMask} Overview}

A key design objective of \myname{} is to sense the biosignals relevant to sleep while eliminate all hard sensing components and relying solely on textile-based and soft sensing elements that can be embedded in a sleep mask. While it is possible to obtain sleep markers using a combination of devices, say one on the head for EEG and EOG~\cite{muse, Phillips-Smart-Sleep}, and another on the body for cardio-respiratory rhythm and movement~\cite{oura, Kiaghadi:2019}, it is more ideal to obtain all metrics from a single device on the head. Since the head area is sensitive, we remove all hard sensing components from the head-worn device and rely solely on textile-based and soft sensing elements that can be embedded in a sleep mask. Such a design can eliminate rigid sensing elements from being in direct contact with the skin on the head area. We note that unlike sensing elements that need to be in contact with the skin, other rigid components like the microcontroller, radio and battery are easier to place in a more conducive location. We now describe the design of such a device and how we optimized it to extract the markers of interest.

\begin{figure*}[t]
\centering
\includegraphics[width=1\linewidth]{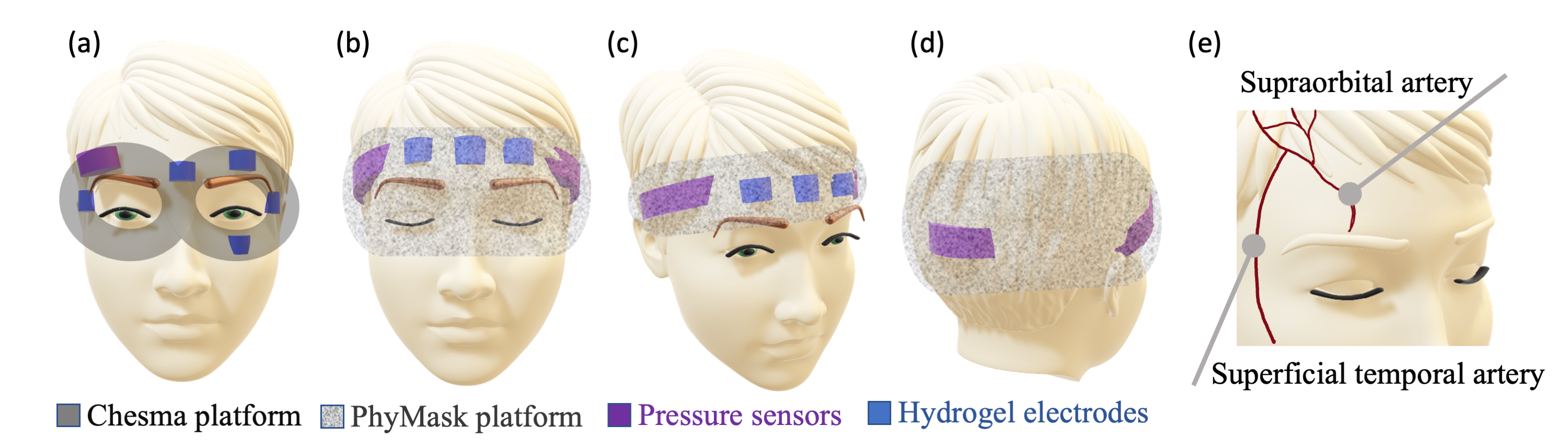}
\caption{Schematic drawings. (a) illustrates the sensors placements on our previous work, Chesma platform~\cite{chesma}. (b) Front, (c) side, and (d) rear views of \myname{}’s sensors placements are depicted. As it is shown in (c), \myname{} can be built in the form factor of both headband and eye mask. (e) illustrates schematic drawings of supraorbital and superficial temporal arteries.}
\label{fig:3d_sensors_placement}
\end{figure*}

\subsection{Robust sensing of biopotential signals with textile electrodes}
In order to extract biopotential signals such as EEG and EOG in a comfortable manner, we need a fabric-based electrode that is comfortable, robust, and provide reliable high signal-to-noise ratio (SNR). \myname{} leverages a recent innovation; a fabric-based, reusable wet electrode that has a layer of composite hydrogel, which when hydrated, mechanically behaves like the foams used in the standard electrodes~\cite{chesma}. This is sufficiently cushion-like to minimize motion artifacts in the absence of any harsh skin adhesives. The advantage of these electrodes is that they overcome the aesthetic drawback of requiring adhesives to have contact with the skin while simultaneously providing strong signal integrity. In addition, they address the lack of reusability of the commercial wet electrodes.

In order to reduce the computational cost of the system, we need to choose the minimum number of electrodes that can be embedded into a sleep mask and provide reliable EEG and EOG signals concurrently. The typical electrode placement for EOG and EEG can be too complex. For example, EOG alone requires five electrodes placed as illustrated in Figure~\ref{fig:3d_sensors_placement}a. 

To reduce the number of electrodes, we look at a more restricted placement. In order to be able to capture strong EEG signals, the electrodes should be placed closer to the brain area. Thus, we place one electrode on top of each left and right eyes (HG1 and HG2, respectively) and one in the middle (HG0), serving as both reference and common ground (shown in Figure~\ref{fig:3d_sensors_placement}b). HG1 and HG2 electrodes are respectively closest to the Fp1 and Fp2 sites (illustrated in Figure~\ref{fig:eeg_aasm}) according to the standard 10-20 EEG recording guideline~\cite{acharya2016american} and can conveniently capture horizontal EOG signal which is sufficient for the purpose of sleep stage tracking.

\begin{figure}
    \centering
    \begin{subfigure}[b]{0.49\textwidth}
        \centering
        \includegraphics[width=\textwidth]{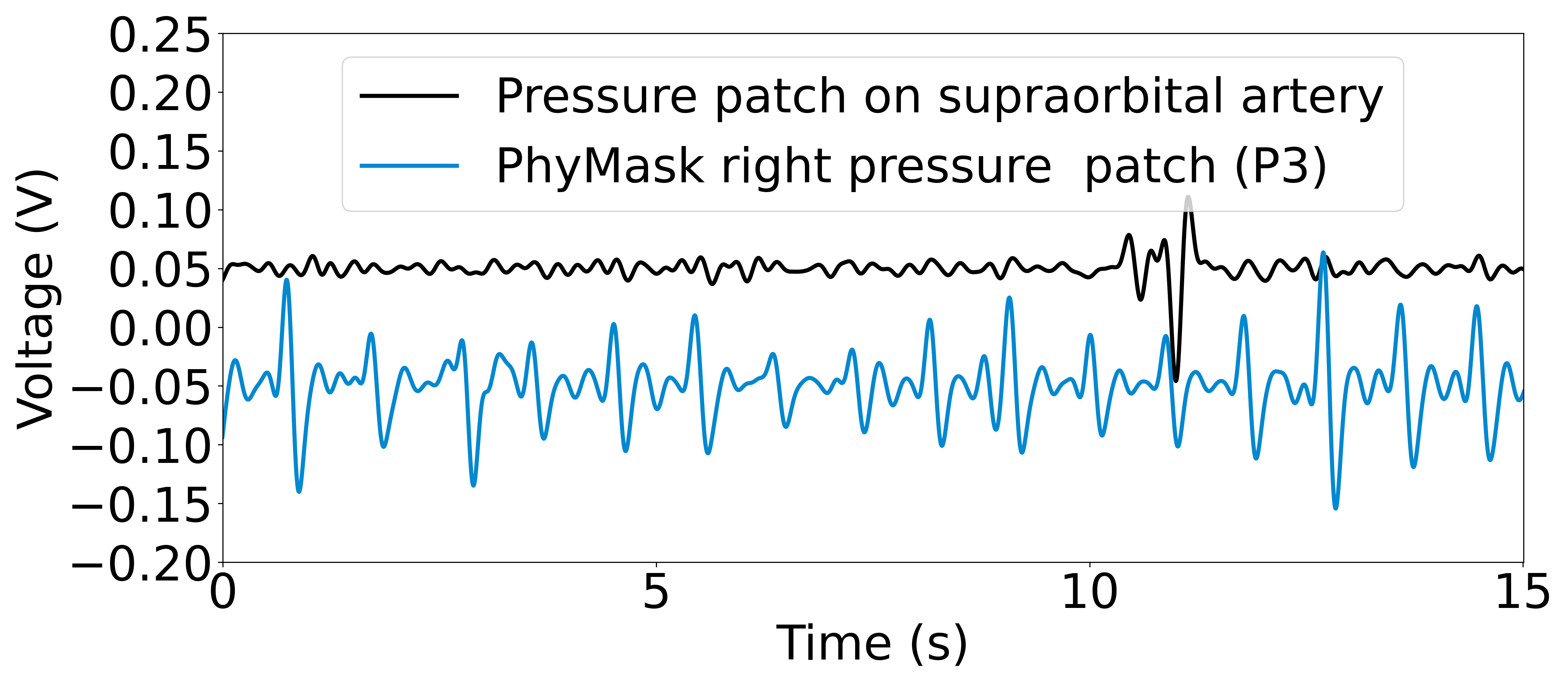}
        \caption{lying on right.}
        \label{fig:bcg_unusual_right}
    \end{subfigure}
    \hfill
    \begin{subfigure}[b]{0.49\textwidth}
        \centering
        \includegraphics[width=\textwidth]{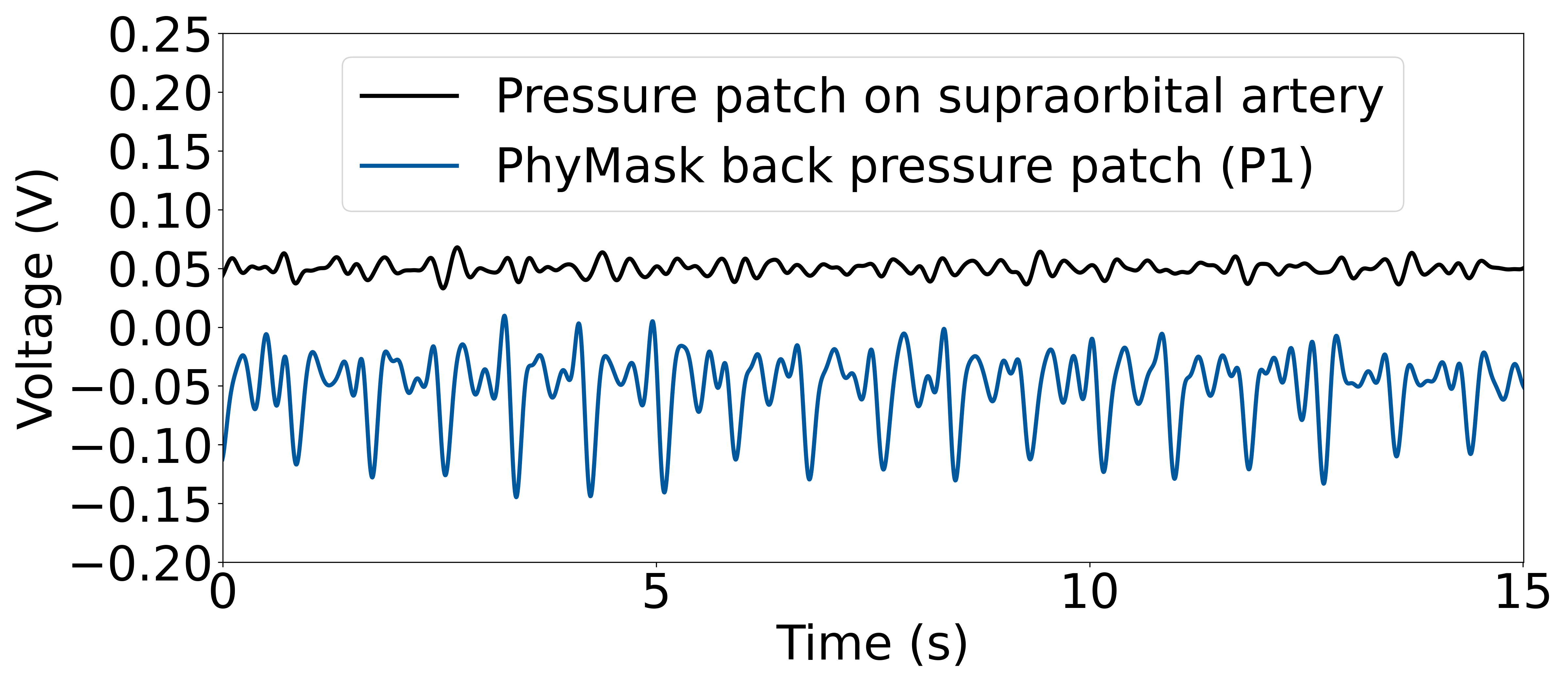}
        \caption{lying on back.}
        \label{fig:bcg_unusual_back}
    \end{subfigure}
    \caption{Comparing the measured pulse signal at a single location versus over a broader surface area in \myname{} platform, when the user is lying on their right and back, respectively. While the pressed pressure patch of \myname{} platform provides reliable pulse signal, the pressure patch placed on the supraorbital position fails in providing robust signal.}
    \label{fig:bcg_unusual}
\end{figure}

\subsection{Robust fabric-based physiological signals sensing on the head}
The next question is how to obtain robust physiological signals such as pulse and breathing, as well as gross motor activity on the head using comfortable textile-based sensing elements. 

Traditional methods for capturing cardiac signals on the head focus on placing sensors at the supraorbital artery position (shown in Figure~\ref{fig:3d_sensors_placement}e). The cardiac signal at this location can be measured with rigid optical sensors to obtain PPG (e.g. the MUSE headband). Furthermore, in the prior work on textile-based sensors, we have looked at using this artery to obtain pulse due to pressure changes (Chesma~\cite{chesma}). However, the issue is that textile pressure sensors do not work well when the signal is present in a small surface area like a single artery. This leads to a weak and noisy signal that is highly sensitive to small changes in sleep position.

We address this limitation using a new method for measuring cardiac signals in the head area while using textile-based sensors.  A key observation that we make is that the cyclical movement of blood from the heart to the head via the abdominal aorta and the carotid arteries causes the head to move imperceptibly in a periodic manner. Similar oscillations can also be observed during respiration since each inhale and exhale action results in body movements~\cite{Balakrishnan:2013}. Therefore, unlike the previous work which attempts to sense the artery pulse in a very localized region, we leverage fabric-based pressure sensors across a larger surface area to measure both the pressure exerted by blood pulsing through the facial artery as well as the small subtle head vibrations caused by blood pumping and respiration. 

In order to detect these vibrations under various sleep postures, we place the pressure patches on the back (P1), left (P2), and right (P3) of the sleep mask, where at least one of the patches is pressed on the pillow under the user’s head weight in different sleep postures (shown in Figure~\ref{fig:3d_sensors_placement}b-d). For precise placement of the pressure sensors and in order to achieve the maximum heart rate sensing ability, we look more closely at the physiology of the head. Figure~\ref{fig:3d_sensors_placement}e illustrates the anatomy of the arteries passing through the head. The superficial temporal artery is a major artery of the head. It arises from the external carotid artery when it splits into the superficial temporal artery and maxillary artery. Its pulse can be felt above the zygomatic arch, above and in front of the tragus of the ear. Therefore, P2 and P3 sensors are made bigger and placed slightly tilted ($\sim15^{\circ}$) on the temple area, allowing for the pulse waveform from the frontal branch of superficial temporal artery to be measured. This layout of the sensors also enables detecting the head posture on the pillow and can capture the pressure changes caused by gross body movement.

The difference between measuring pulse at a single location versus measuring head oscillations over a broader surface area is illustrated in Figures \ref{fig:bcg_unusual_right} and \ref{fig:bcg_unusual_back}, where the participant is lying on their right and back, respectively. We see that the pulse signal acquired with the sensors placed on the supraorbital artery is weak and imprecise, while \myname{}'s sensors that are optimized to extract head oscillations can pick up clear pulse signal.

\subsection{Fabrication of PhyMask}
Figure~\ref{fig:PhyMask_platform}a shows the integrated \myname{} platform, which incorporates three biopotential electrodes and three pressure sensors. We made two biopotential electrodes with functional area of 1 cm x 1 cm and one electrode with functional area of 1 cm x 2 cm, serving as reference. We made the reference electrode slightly bigger to ensure better skin-electrode contact and hence more stable signal during long wear. The layered structure of the electrode (shown in Figure~\ref{fig:PhyMask_platform}d) is comprised of an array of conductive silver-plated nylon threads, serving as the charge collector, on top of a hydrophobic backing fabric. This conductive array is then coated with silver chloride that provides the ion conductivity required for transducing signals from the ionic form in the body to the electrons in the wires. By taking advantage of the initiative chemical vapor deposition (iCVD) of poly(hydroxy ethyl acrylate) (pHEA) on a pharmaceutical grade of silver gel, we developed a reusable composite hydrogel on top of the electrode. Finally, an open-weave cotton gauze is placed on top and a cotton flannel framed the hydrogel to protect it from mechanical abrasions. The wide pores of the gauze fabric enable rehydration of the hydrogel and its direct contact with the skin while protecting it from being rubbed away. All the fabrics used in this electrode are coated with poly(perflurodecyl acrylate) using iCVD. The hydrophobic nature of this polymer prevents the fabrics from absorbing water/perspiration rather than the hydrogel.

As it can be seen in Figure~\ref{fig:PhyMask_platform}b, the pressure sensor is comprised of an ion conductive active layer sandwiched between two silver-coated conductive fabrics as electrodes. The active layer is made of a cotton gauze fabric coated with a poly(siloxane) polymer, poly(N-propylsilyl-N,N,N,-trimethylammonium chloride) through a solution-phase functionalization process. Through the application of a compression stress on the sensor, the mobile chloride counterions relocate on the surface of the fabric and the ions movement in addition to the reduction of the air gap between the layers lead to a decrease in the impedance of the pressure sensor. The functionalized ion-conductive layer was further encapsulated with a poly(perfluoroalkylsiloxane) coating through iCVD. The hydrophobic nature of this coating protects the ion-conducting fabric against common aging processes, such as erosion during laundering or air oxidation, ensuring that the ionic conductivity of the sensor will not be washed away or diluted if the pressure patch comes into contact with sweat~\cite{chesma}. We made a large patch of functionalized ion-conductive layer, and then cut it into four 12 cm x 4 cm and two 6 cm x 4 cm sheets, each of which was sewn around the perimeter onto a sheet of silver fabric. Sewing together each pair of these joined gauze-silver sheets yielded three resistive sensors with a 4-layer structure.

All of these fabric electrodes and sensors are tightly sewn onto a commercially-available sleep mask -- purchased from Amazon 4.5/5 with more than 4000 reviews, it was chosen because of its lightweight and adaptability to a wide variety of head shapes and sizes. For maximum comfort and to minimize the number of hard electronic components, we avoided using wires in our design. Instead, we used silver-plated nylon threads. In order to shield our wiring system from electromagnetic noises and make it laundering-stable, we cladded the silver threads with a three-layer fabric-based shield, comprising of polyurethane coated ripstop nylon cloth as the first layer, followed by a nanoscale (40–50 nm) coating of a hydrophobic polymer, PFDA as second, and encased within cotton piping as the last layer (shown in Figure~\ref{fig:PhyMask_platform}c). Our wiring system serves as lightweight and flexible interconnects between the six sewn-on electrodes and the circuit board (PCB) microcontroller (MCU).

\section{P\texorpdfstring{\MakeLowercase{hy}M\MakeLowercase{ask}}{PhyMask} System Design}

The \myname{} system pipeline has two main components: (1) the software running on a host device to process the raw data and extract the sleep and physiological features of interest, and (2) the hardware for acquiring sensors signals and real-time Bluetooth communication. In this section, we explain these components in detail.

\subsection{\myname{} Processing}
We start by describing our methodology for sleep micro-events (i.e. spindle and K-complex) detection from biopotential signals, as well as sleep posture, respiratory and cardiac rhythm extraction from the pressure signals.

\begin{figure}
     \centering
     \begin{subfigure}[b]{0.245\textwidth}
         \centering
         \includegraphics[width=\textwidth]{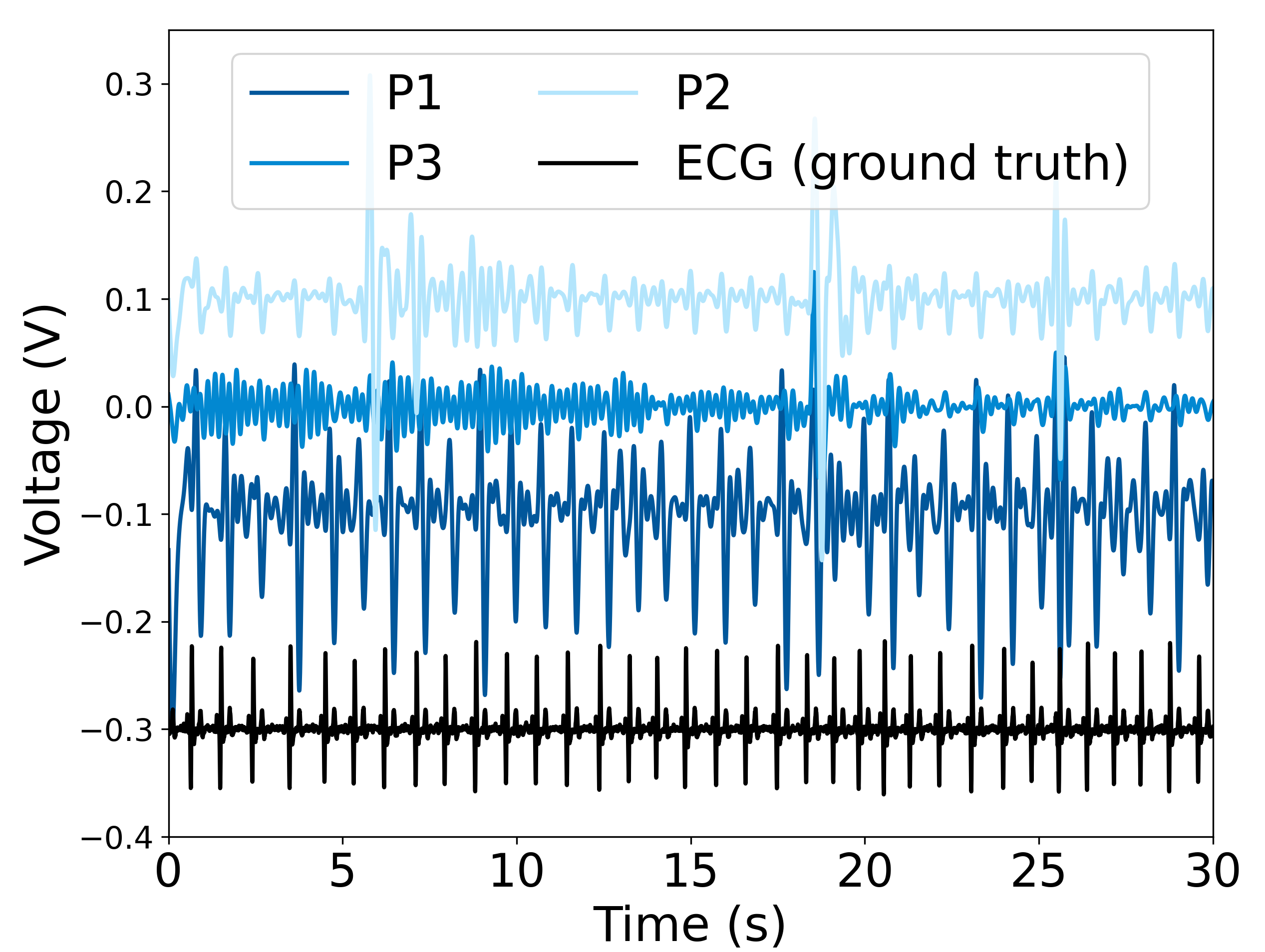}
         \caption{BCG}
         \label{fig:signal_examples_bcg}
     \end{subfigure}
     \hfill
     \begin{subfigure}[b]{0.245\textwidth}
         \centering
         \includegraphics[width=\textwidth]{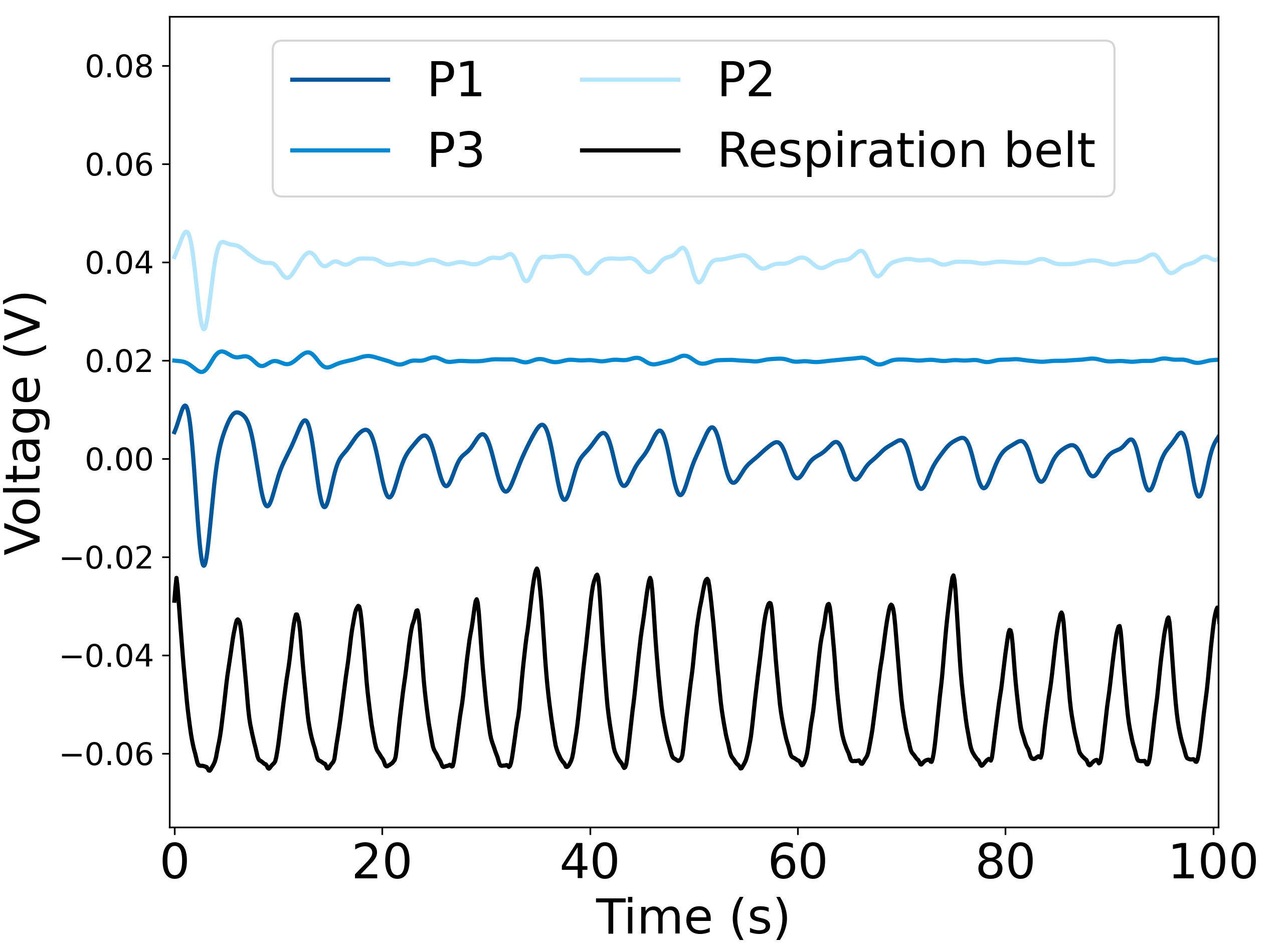}
         \caption{Respiration}
         \label{fig:signal_examples_resp}
     \end{subfigure}
     \hfill
     \begin{subfigure}[b]{0.245\textwidth}
         \centering
         \includegraphics[width=0.95\textwidth]{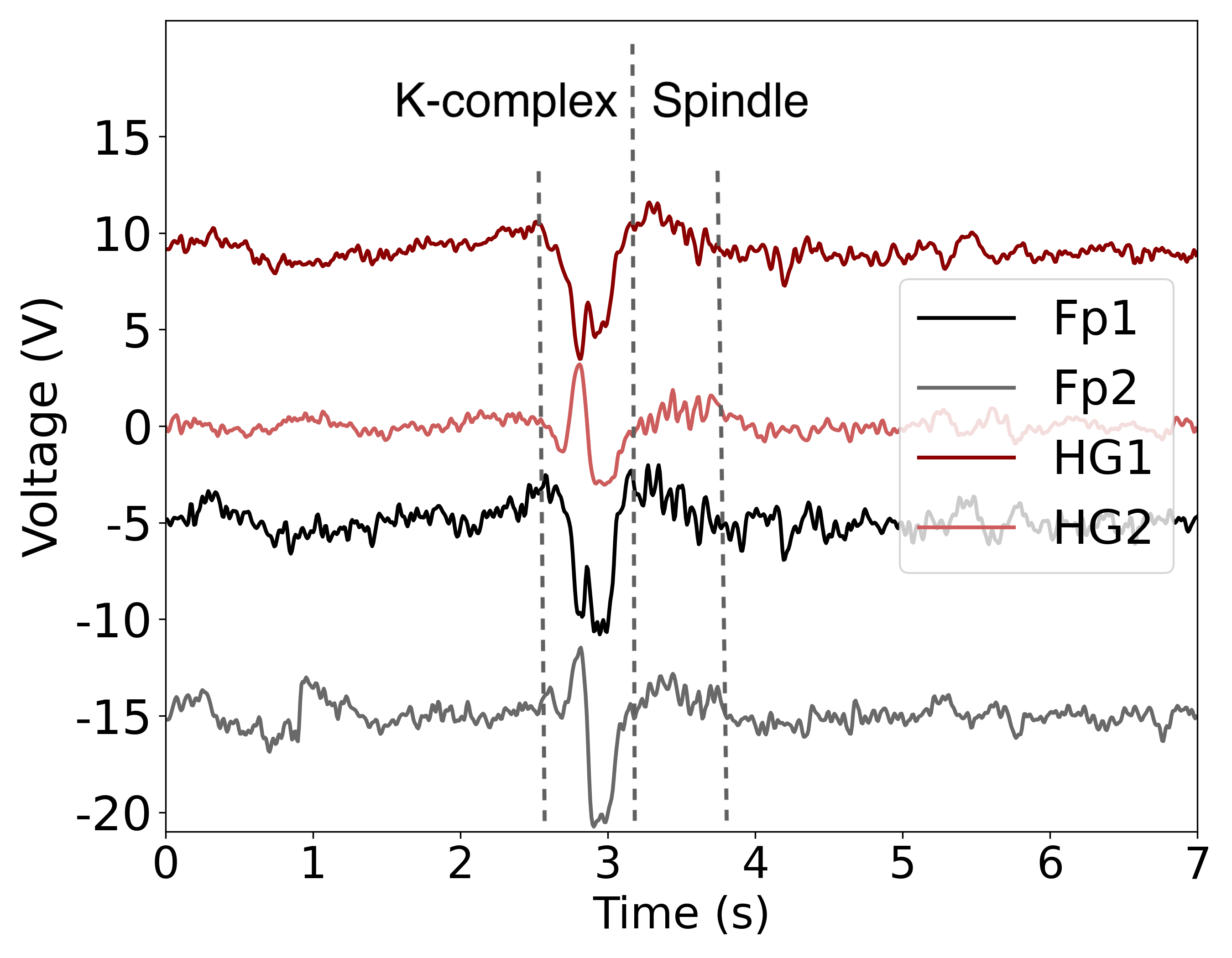}
         \caption{EEG}
         \label{fig:signal_examples_eeg}
     \end{subfigure}
     \hfill
     \begin{subfigure}[b]{0.245\textwidth}
         \centering
         \includegraphics[width=0.95\textwidth]{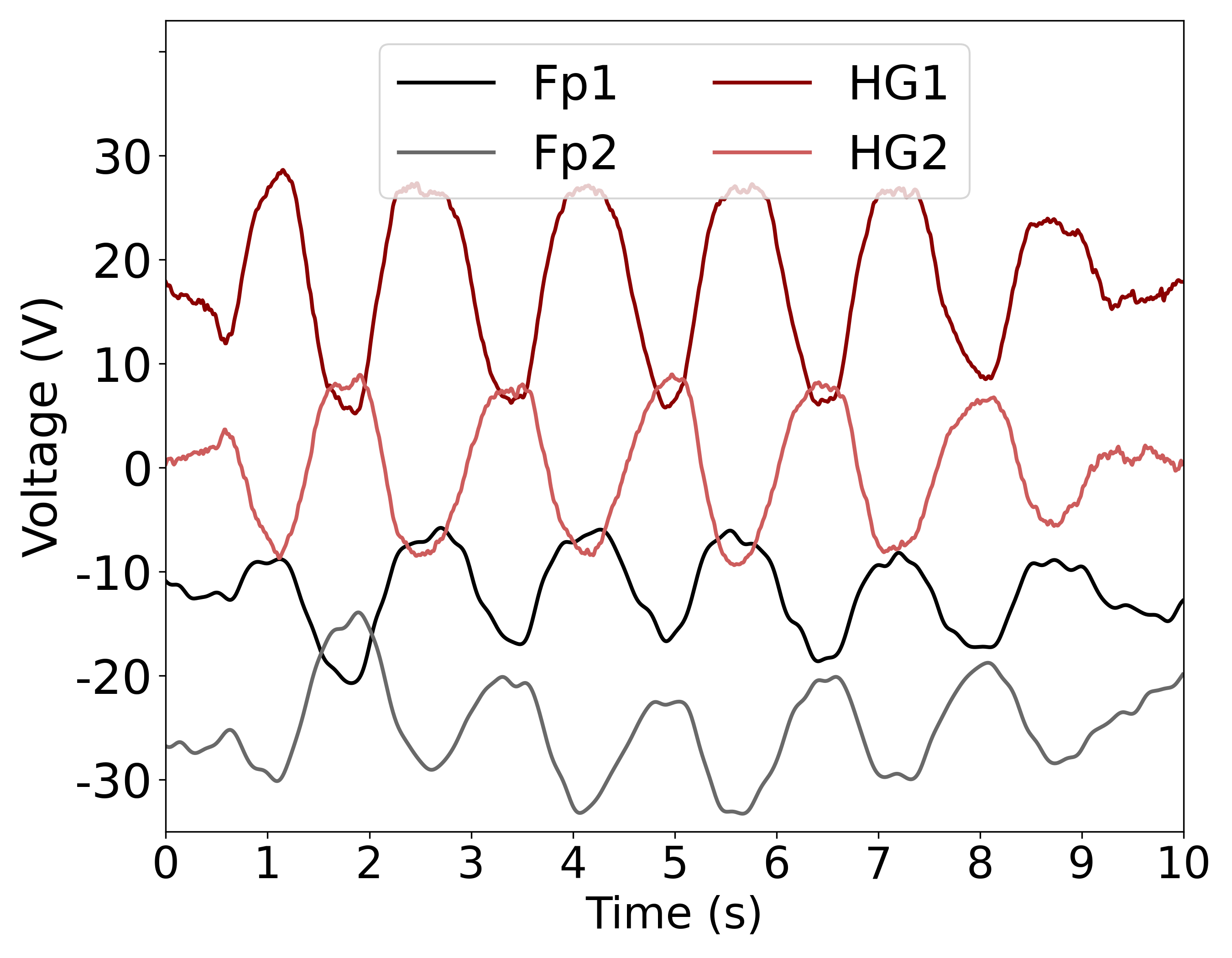}
         \caption{EOG}
         \label{fig:signal_examples_eog}
     \end{subfigure}
        \caption{\myname{} signals. (a) The BCG signal captured by the pressure patches along with the ECG ground truth. (b) The breathing signal along with the ground truth collected with respiration belt. In both (a) and (b) scenarios the user is lying on their back, hence the back patch captures stronger cardiac and respiratory signals. (c) EEG signals captured from right and left hydrogel electrodes along with the ground truth signals captured from standard wet electrodes. A K-complex followed by a spindle is presented at time 3s. (d) The EOG signal along with the ground truth signals. In this example, the user is looking at right and left repeatedly with eyes closed.}
        \label{fig:signal_examples}
\end{figure}

\subsubsection{Sleep Spindle and K-complex Detection}\label{sec:eeg_analysis}
As discussed earlier, one of the main advantages of acquiring the EEG signal during sleep is that we can detect and analyze the sleep micro-events such as spindle and K-complex. This is important since their characteristics, e.g. frequency of their occurrence, can unveil a great amount of information regarding the cognitive state of the mind~\cite{hennies2016sleep}. In this section, we describe the \myname{} approach for detecting spindles and K-complexes in detail.

Sleep spindles are brief bursts of neural oscillations (9–16Hz, 0.5–3s) generated by the interplay of the thalamic reticular nucleus and other thalamic nuclei during NREM sleep (N2 and N3 stages)~\cite{de2003sleep} (an example is shown in Figure~\ref{fig:signal_examples_eeg}). Like spindle, K-complex is a great hallmark of NREM sleep stage 2 and is often followed by a sleep spindle. K-complexes are generated in widespread cortical locations though they tend to predominate over the frontal parts of the brain. K-complex characteristics are very different from spindles. They consist of a brief negative sharp wave immediately followed by a positive component, creating slow-wave (0.8 Hz) and delta (1.6–4.0 Hz) oscillations of 0.5–3s in duration~\cite{cash2009human:kcomplex} (an example is shown in Figure~\ref{fig:signal_examples_eeg}). We train two different models of the same architecture for detection of spindles and K-complexes. Several classification algorithms were tested during this study, including random forests, neural networks, and support vector machines. We found the random forest classifier to provide the best performance.

\begin{figure}
\centering
\includegraphics[width=1\linewidth]{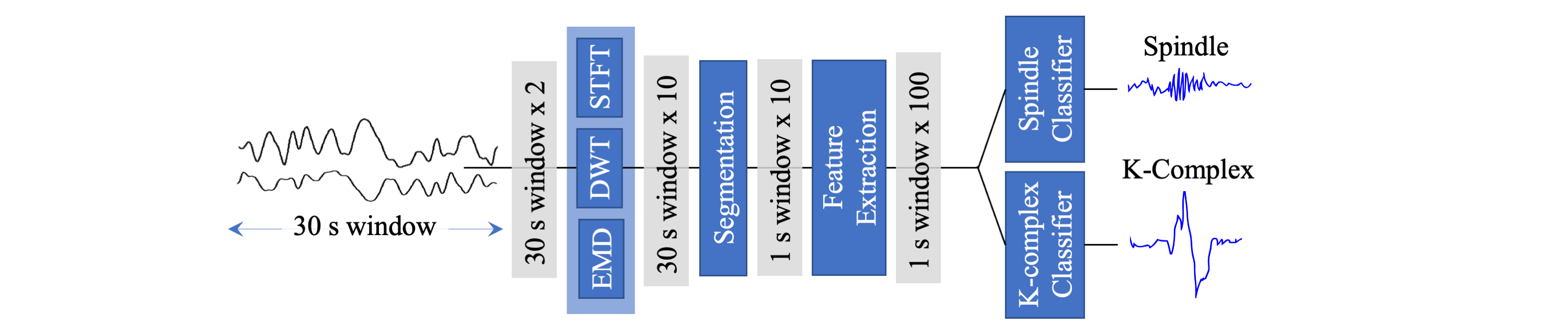}
\caption{\myname{} spindle and K-complex detection algorithm pipeline.}
\label{fig:eeg_pipeline}
\end{figure}

In order to prepare the data for our binary random forest classifier, we first apply a \nth{5} order butterworth filter with a passband of 0.5-35Hz to filter out noises from the EEG streams. Then, we extract features in both time and frequency domains. Our feature extraction pipeline has two steps (summarized in Figure~\ref{fig:eeg_pipeline}). First, we derive Short-time Fourier Transform (STFT), Discrete Wavelet Transform (DWT), and Empirical Mode Decomposition (EMD) features from 30-second epochs of the EEG stream and append them as additional channels to the original raw signal. In the second step, we extract statistical features from overlapping windows of the output signal of step 1. We empirically choose an input window size of 1s (125 samples) for both spindle and K-complex detection. In the following we describe the details of our feature extraction pipeline. 

\mypara{Short-time Fourier transform (STFT).}
In the literature, STFT is widely used as one of the main features for automatic spindle and K-complex detection~\cite{costa2012automatic, patti2014automated, yucelbacs2018automatic}. STFT computation algorithm divides the input vector into overlapping shorter segments of equal length and then Fourier transform is computed separately on each segment. The output is a 3-dimensional time-series containing information regarding the frequency content of each time segment. Given the different frequency characteristic of spindle and K-complex, we extract two separate features from STFT for each event. 
For spindle that has a frequency content of 9–16Hz, we extract 1) sum of the powers of frequency components of the signal in 10–12 Hz, and 2) sum of the powers of frequency components of the signal within 9–16 Hz of the signal. For K-complex, which has most of its frequency content around 0.8 Hz and within 1.6-4.0 Hz, we extract 1) sum of the powers of frequency components of the signal within 0.5–1 Hz , and 2) sum of the powers of frequency components of the signal within 1.5–4.5 Hz. 
We then up-sample the outputs to generate a time-series of the same length as the original EEG signal and augment them with the raw EEG stream as additional channels.

\mypara{Discrete Wavelet transform (DWT).} To further investigate the frequency content of the EEG time-series data, we use Discrete Wavelet transform (DWT)~\cite{shensa1992discrete}. DWT obtains low-frequency resolution and high-frequency information using long- and short-time windows, respectively. Because of this, DWT is appropriate for the analysis of the non-stationary signals such as EEG~\cite{yucelbacs2018automatic}. 2-level DWT and Daubechies-2 (db2) mother wavelet is used in our computation, and the second-level coefficients are taken as the output. The output is up-sampled and appended to the raw EEG and extracted STFT features.

\mypara{Empirical Mode Decomposition (EMD).}
Empirical mode decomposition (EMD) can be used for nonlinear and non-stationary signals such as EEG~\cite{huang1998empirical}. In this method, a signal is divided into substatements referred to as intrinsic mode functions (IMF) and subpiece (residue). The main idea behind EMD is to locally reconstruct a signal. The reconstruction is a sum of a local trend and detail. The local trend implies the low-frequency components of the signal (residual), while the local detail (IMF) is the high-frequency parts. According to this method, a signal is recursively separated step by step. Then a certain number of IMFs and residuals are obtained depending on the content of the signal. We choose the first IMF component as the output, and then append it to the output of the previous step. Y{\"u}celbac{\c s} et al.~\cite{yucelbacs2018automatic} has also utilized EMD for automatic spindle detection. In their solution, they apply binary thresholding on the first IMF component derived from the EMD of a single EEG stream to detect spindle events.

\mypara{Statistical feature extraction.} In this step, we chunk the output of the previous step, i.e. a multi-channel signal consisting of the raw EEG, extracted STFT features, DWT features, and the EMD components, into overlapping 1-second windows and then extract statistical features including maximum, mean, median, standard deviation, sum, energy, mean-crossing, interquartile range, \nth{10} percentile, and \nth{90} percentile. Finally in order to train our random forest classifier, we use SMOTE~\cite{smote} to balance the training dataset.

\begin{figure}
\centering
\begin{minipage}{.45\textwidth}
  \centering
  \includegraphics[width=\linewidth]{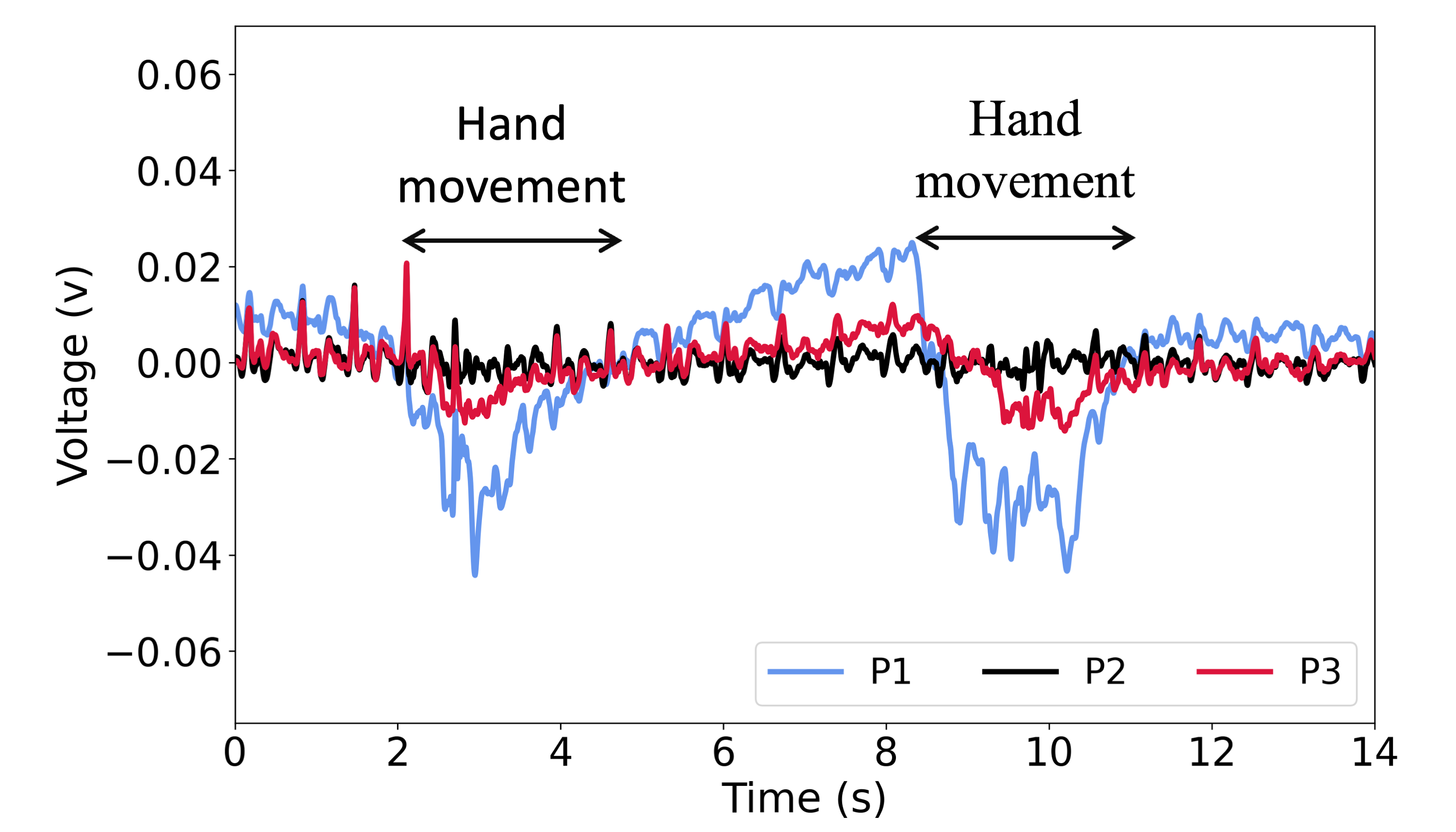}
  \captionof{figure}{DC-removed pressure patches signal are illustrated when the user is lying on their back and swinging their hand in time 2-4s and 9-11s. The periodic small spikes on the signals correspond to the heart beats.}
  \label{fig:hand_movement_ex}
\end{minipage}%
\hspace{0.2cm}
\begin{minipage}{.41\textwidth}
  \centering
  \includegraphics[width=\linewidth]{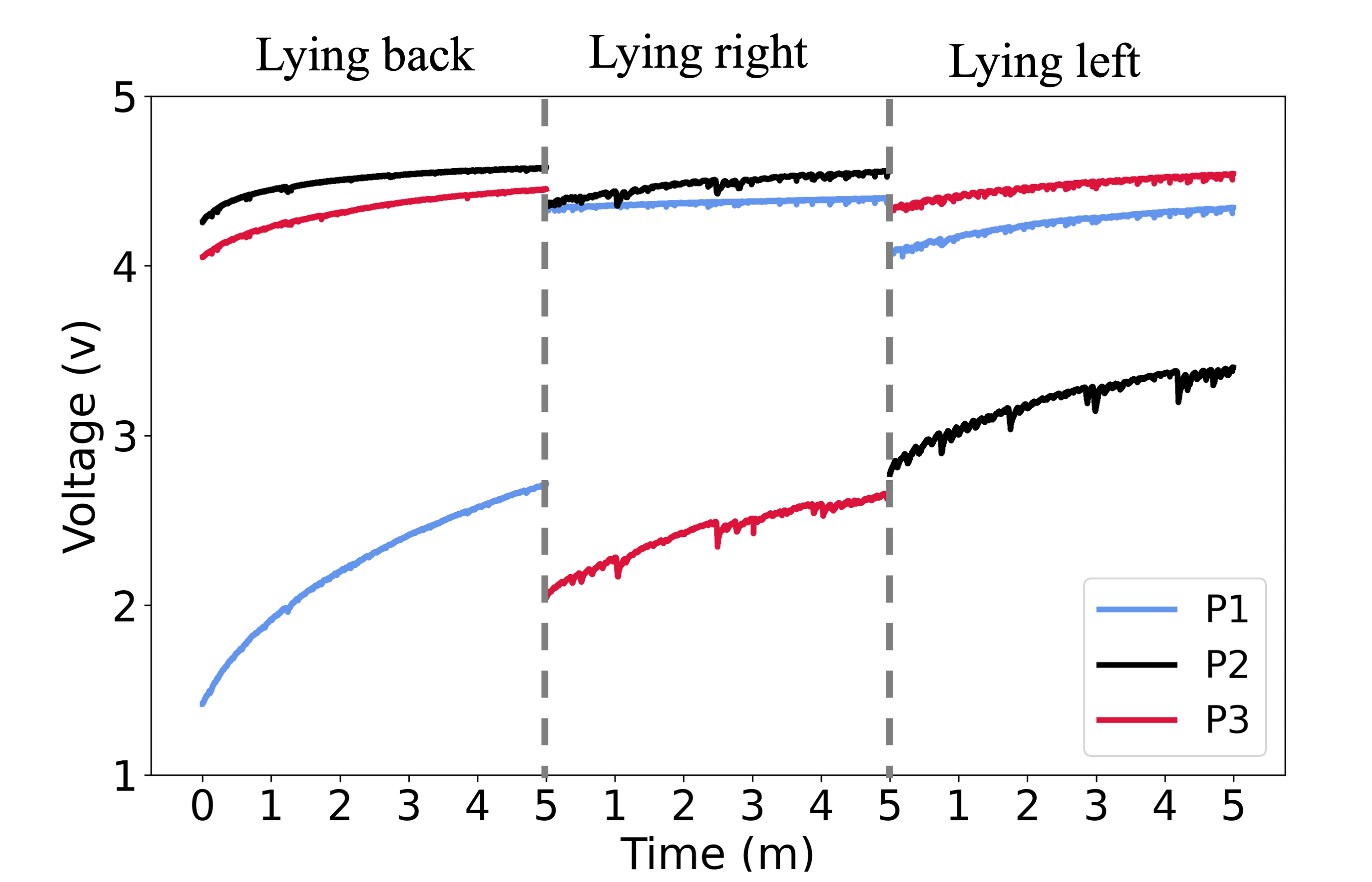}
  \captionof{figure}{Pressure patches baseline across postures. The slow oscillations on the signal correspond to the respiration while the more choppy ones represent the gross body movement.}
  \label{fig:posture}
\end{minipage}%
\end{figure}

\subsubsection{Gross Motor Movement and Posture Estimation}\label{sec:gross_body_posture_estimation}
\myname{} can easily measure pressure changes that reflect gross motor movement of the limbs. We illustrate by looking at voltage changes of the \myname{} pressure patches when a user moves their hand. Figure~\ref{fig:hand_movement_ex} shows the DC-removed pressure signal for the three patches when the user is lying on their back. As the user moves their hand during time 2-4s and 9-11s, the pressure signals voltage, especially for the pressed patch, changes dramatically. Please note that the small periodic spikes on the signals correspond to the heart beats.

The head posture information during sleep can be useful in understanding which posture leads to better or worse sleep quality. For example, lying on the back is usually not recommended for sleep apnea patients~\cite{ravesloot2013undervalued}. Figure~\ref{fig:posture} shows the baseline signal for the P1, P2, and P3 pressure patches when a user is lying in different sleep postures for a total of 15 minutes. As it can be seen, in each sleep posture, the pressed patch has the lowest baseline. This is aligned with our expectation as the voltage being measured via the voltage divider circuit (shown in Figure~\ref{fig:PhyMask_board_sch}a) is inversely proportional to the pressure, so lower voltage means higher pressure.

\subsubsection{Respiration and Heart Rate Estimation}\label{sec:HR_estimation}

In order to accurately estimate respiratory rate from the pressure sensors baseline signals (an example is shown in Figure~\ref{fig:signal_examples_resp}), we first apply FFT to find the frequency bin with the highest power resulted from the respiration signal. Then, we perform band-pass filtering based around the FFT peak to avoid counting fluctuations of the second harmonic. We then count the number of peaks in one-minute-long windows of the filtered time series to obtain the respiration rate. The estimated respiration rate is updated once every half-minute (the step size is 30s). The strongest respiration signal is obtained from the patch on which the head is resting, i.e. the most pressure is sensed on.

The estimation of the heart rate is more challenging as the cardiac signal captured by the pressure patches has very a low SNR due to the mixed-in involuntary head movements, and the noise in all frequency bins. Furthermore, the amplitude of the cardiac signal can also be affected by many factors including sleep posture and user’s physiological characteristics.

\begin{figure}
\centering
\includegraphics[width=0.7\linewidth]{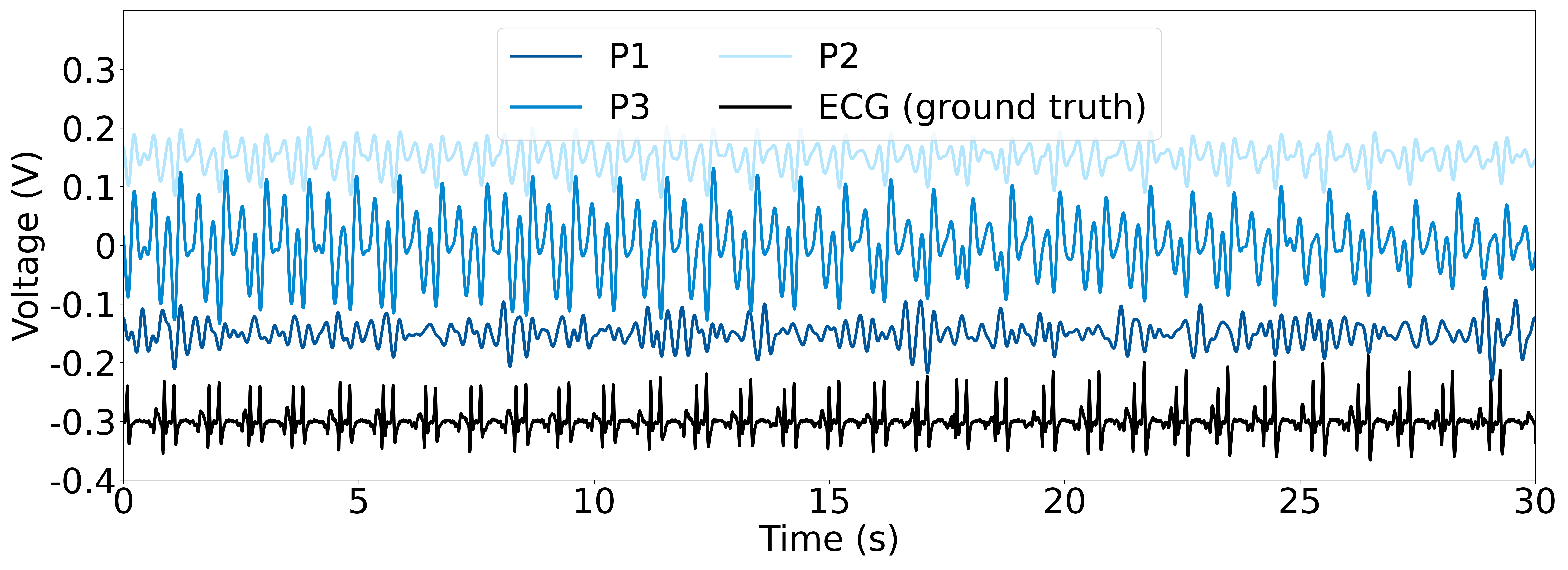}
\caption{Time series of the back (P1), left (P2), and right (P3) pressure patches signals along with ECG ground truth. While the user is lying on their back, the side patches capture the pulses stronger than the back patch.}
\label{fig:example_lying_back_strong_right}
\end{figure}

\begin{figure}
\centering
\includegraphics[width=0.9\linewidth]{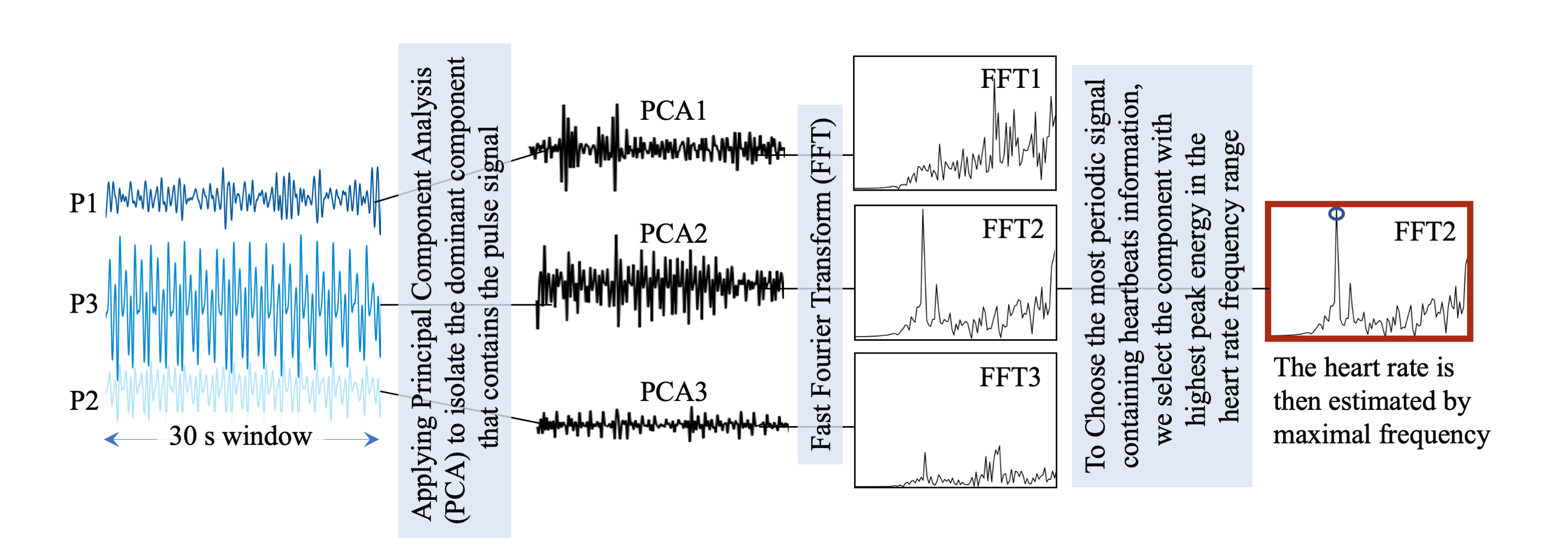}
\caption{\myname{} heart rate detection algorithm pipeline.}
\label{fig:hr_pipeline}
\end{figure}

Generally, the signal captured by the patch under pressure in each sleep posture has better resolution in capturing the small head movements related to cardiac activity. An example is shown in Figure~\ref{fig:signal_examples_bcg}, where the user is lying on their back and the back patch (P1) can clearly capture the heartbeats, better than the other two sensors. However, this assumption does not always hold true. Figure~\ref{fig:example_lying_back_strong_right} shows another example of when the user is lying on their back and this time, the back pressure signal (P1) is quite corrupted while the right (P3) and left (P2) signals clearly detect the heartbeats. This can happen due to several reasons. Having wrinkles in the pressed patch decreases the sensor sensitivity and results in poor and noisy signal. Also, for the users who naturally have strong superficial temporal artery pulse on the temple, the heartbeat signals captured from side patches are much stronger than the back patch that only detects the subtle head motion caused by the Newtonian reaction to the influx of blood at each beat. Therefore, we have two main challenges that we need to encounter when designing our algorithm: (1) the heartbeat signal is weak and usually interfered by many sources of noises, and (2) the signal quality of patches differs across different users and sleep postures. In the following, we describe our proposed heart rate detection algorithm summarized in Figure~\ref{fig:hr_pipeline}.

In order to block the DC baseline, respiration related frequency components, and higher frequency noises, we apply a \nth{5} order butterworth filter with a passband of 0.75-3Hz on three cardiac signals. Please note that a normal adult’s resting pulse rate falls within [0.75, 2] Hz, or [45, 120] beats/min. After eliminating some of the noise, we need to isolate the heartbeat component of the three signals. In order to do this, we use Principal Component Analysis (PCA) to isolate the dominant component that contains the pulse signal. To do this, we consider each pressure patch signal stream as a separate data point and use PCA to find a set of basis vectors along which the signal has the most variation. We then select a dimension on which to project the time-series to obtain the pulse signal.

The 3 filtered BCG signals at time $t$ are represented by $m_t = [y1(t), y2(t), y3(t)]$. PCA finds the principal axes of variation of the signal as the eigenvectors of the covariance matrix of the data $\Sigma{m}$, i.e.

\begin{equation}
\label{equ:PCA1}
\Sigma{m}\:\Phi{m} = \Phi{m}\:\Lambda{m} ~,
\end{equation}
where $\Lambda{m}$ denotes a diagonal matrix of the eigenvalues $\lambda{1}$, $\lambda{2}$, and $\lambda{3}$ corresponding to the eigenvectors in the columns of $\Phi{m}$, $\phi{1}$, $\phi{2}$, and $\phi{3}$. We finally obtain three 1-D 
signals $si(t)$ by projecting $m_t$ onto $\phi{i}$.
\begin{equation}
\label{equ:PCA2}
s_i(t) = \sum_t^T{m_t\phi{i}}
\end{equation}

We then select the most periodic component $s_p(t) = \argmax_{s_i} h(s_i(t))$ for heart rate estimation, where the function $h(.)$ calculates the periodicity of the input signal as the percentage of the spectral power accounted for by the frequency with maximal power to the total spectral power. The heart rate is then estimated as $\frac{60}{f_{pulse}}$ beats/min, where $f_{pulse}$ is the maximal frequency of $s_p(t)$.  Please note that we perform PCA on a 30-second window and slide it with a hop size of 10s.

\subsection{PhyMask Hardware and Data Acquisition}

We now turn to describing our hardware platform. PhyMask uses a single low-power and compact circuit board for filtering, amplification, digitization, and transmission of the raw EEG, EOG, cardiac, and respiratory signals collected from the biopotential electrodes and the pressure patches. There are three main challenges in designing such a board: first, each of these signals has different characteristics that require specific electronics circuitry design; second, all of the signal channels need to be collected at high-enough sampling rate and transmitted at the same time to ensure real-time and accurate tracking; third, the board should have very low power consumption enabling long-term usability of the system with a single battery charge. In the following, we explain the steps taken to tackle these challenges in the design of the PhyMask electronics.

\begin{figure}
\centering
\includegraphics[width=0.9\linewidth]{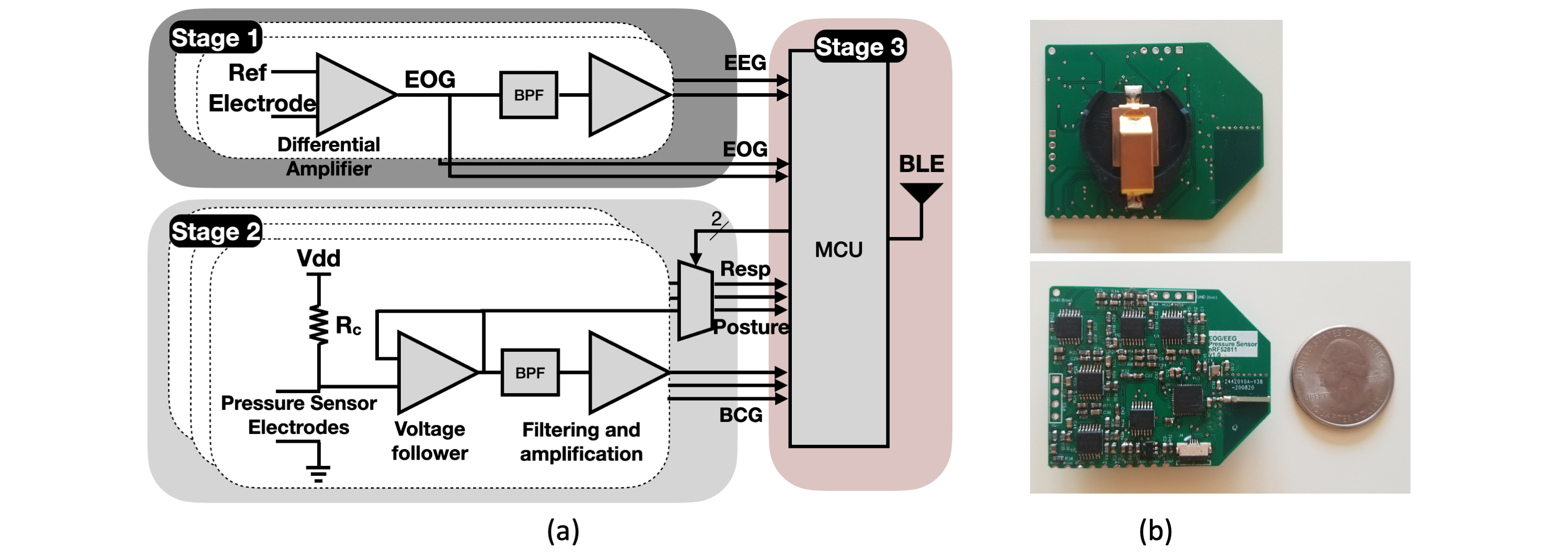}
\caption{PhyMask board. (a) Block diagram of analog circuit board. The board has three main stages responsible for biopotential signals sensing, pressure sensors sensing, and signals transmission. (b) Front and rear view where the battery should be placed.}
\label{fig:PhyMask_board_sch}
\end{figure}

As mentioned before, the board is connected to 3 biopotential electrodes (one reference and two sensors) and 3 pressure sensors. The biopotential electrodes provide the EOG and EEG signals while pressure sensors provide cardiorespiratory signals. Our board is composed of three main modules: 1) EEG and EOG signals acquisition, 2) cardiac and respiration signals acquisition, and 3) low-power wireless signal transmission. The board schematic is illustrated in Figure~\ref{fig:PhyMask_board_sch}a.

\mypara{Stage 1: EEG and EOG signals collection.} EEG and EOG signals have very different characteristics. EOG signals, representing eye movements, have magnitudes in the orders of 100s of $\mu Vs$ with a frequency range of about DC-100Hz. EEG signals however, are roughly three orders of magnitude weaker than the EOG signals and their frequency content during sleep can vary from 0.5Hz to about 30Hz.

To deal with this, we use two two-stage differential amplifiers each amplifying one biopotential signal with respect to the reference electrode. The gain of the first stage is about 30 and it outputs the EOG signal stream. We selected this level of amplification gain empirically such that the obtained EOG signal is clearly distinguishable, yet it does not saturate the next stage amplifier. This diminishes the need for a DC reject filter which consequently would have eliminated critical low-frequency contents of the EOG signal. In the second stage of the amplifier, we further amplify this signal by 2500 times and filter it to only pass through frequency content in the range of 1-50 Hz to acquire the EEG signal.

\mypara{Stage 2: BCG and respirations signals collection.} As discussed earlier, the fabric-based pressure patches are designed to capture the subtle ballistics and head movements caused by cardiorespiratory activities. But these two signals are very different --- the pressure response of respiration is stronger than the heartbeat, and the respiration signal contains very low frequency components (typically, below 1 Hz) whereas cardiac signal is in the 1-20 Hz frequency range. 

The second stage of the board receives the signal from three pressure sensors and after using a voltage divider to translate resistance changes into changes in voltage, we use a voltage follower to create a copy of the signal with low output impedance for each channel, suitable for digitization. The resultant three signals represent the posture/respiration signals of the user. To find the heartbeat instances in these signals, we then DC reject and amplify the signals up to 400 times to find the ballistics of the veins covered by the sensor.

\mypara{Stage 3: low-power wireless signal transmission.}
The last module of the board uses a low-power MCU with integrated BLE capability (nRF52811~\cite{nRF52811}) to wirelessly transmit the acquired data to the host device. The MCU is capable of converting a maximum of 8 simultaneous analog channels into digital signals. This is less than the 10 analog channels captured by \myname{}, i.e., 2 EEG, 2 EOG, 3 cardiac, and 3 respiration. A naive solution would be to sequentially sample each channel and effectively reduce the sampling rate of all channels symmetrically. However, given the fact that the dominant frequency content of respiration is placed lower in frequency spectrum compared to other desired signals, we placed a 3:1 analog MUX to output one of the three respiration signals in each sampling round. As a result, the effective sampling rate of the respiration signal is one-third of the other physiological signals. The decisions presented above result in sampling rate of 125 samples per second for EEG, EOG, and cardiac signals, while providing 42 samples per second for the respiration signal. These sampling rates are sufficient for reconstruction of the above signals given the fact that EEG, which contains the highest frequency content among all, requires a minimum sampling rate of 60 Hz to reach the Nyquist rate and the respiration signal requires a minimum sampling rate of 2 Hz.

Our choice of sampling rate is also limited by power consumption and hardware compatibility considerations. Higher sampling rate leads to higher transmission rate for the BLE module which greatly increases the power consumption; on the other hand, some host devices have lower bound restrictions for transmission interval in their BLE connection. In our setup, the transmission interval is set to 100 ms which makes the system capable of connecting with devices with more strict restrictions and leads to average power consumption of 3.3 mw. The power consumption of the analog processing components and the MUX is negligible in comparison with the MCU. This means that with a small battery of 250 mAh, 250/1.1 capacity, PhyMask can run up to 5 days. An image of the board, which has dimensions as small as 4 cm by 5 cm, is presented in Figure~\ref{fig:PhyMask_board_sch}b.

Once signals are captured and transmitted to the host device, we apply notch, median, and outlier filters in order to remove the powerline noise, and smooth out the signals. Then we filter each signal with a tailored band-pass butterworth filter (we choose \nth{5} order for its maximally flat passband). Examples of the \myname{} collected EEG, EOG, BCG, and respiration signals are shown in Figure~\ref{fig:signal_examples}. 

\section{Dataset Collection and Labeling}

In this section we explain the details of our user study and the ground truth labeling approach. All of these datasets were collected under Institutional Review Board approval.

\subsection{Benchmarking Dataset} \label{sec:benchmarking_dataset}

Our first data collection aims to evaluate \myname{}'s ability in detecting heart rate, respiration, sleep posture, gross body movement, and eye movement. For this, we asked 10 participants (average age of 27) to wear \myname{} and we recorded the output voltage in various stationary conditions. Participants varied in weight, 110-220 lb, and height, 5’1" to 6’. 4 out of 10 participants were females.

We asked our participants to lie down on their back, left, and right side, posing a sleep position that feels natural to them. We collected data for five minutes in each sleep posture. Then we demonstrated the participants a set of both smooth and jerky limb movements that are common during sleep (particularly for sleep disorders like periodic limb movement). We then repeated the first experiment and asked our participants to mimic these movements while in the sleep positions. The participants were also asked to perform random eye movements in each position for two minutes. The whole experiment led to a total of $\sim$30 minutes of recording from each user. Each recording consists of ten channels, six of which correspond to pressure sensing patches and four corresponding to the hydrogel fabric electrodes.

To assess the \myname{}'s performance, we collect ground truth measures of the target physiological signals. For heart rate, we use a three-channel ECG measurement (2 wrists and an ankle) using the AD8232 evaluation board~\cite{AD8232} (sampling rate of 200 Hz), and for respiration, we used a Go Direct respiration belt~\cite{respbelt} (sampling rate of 10 Hz).

\begin{figure}
\centering
\includegraphics[width=0.9\linewidth]{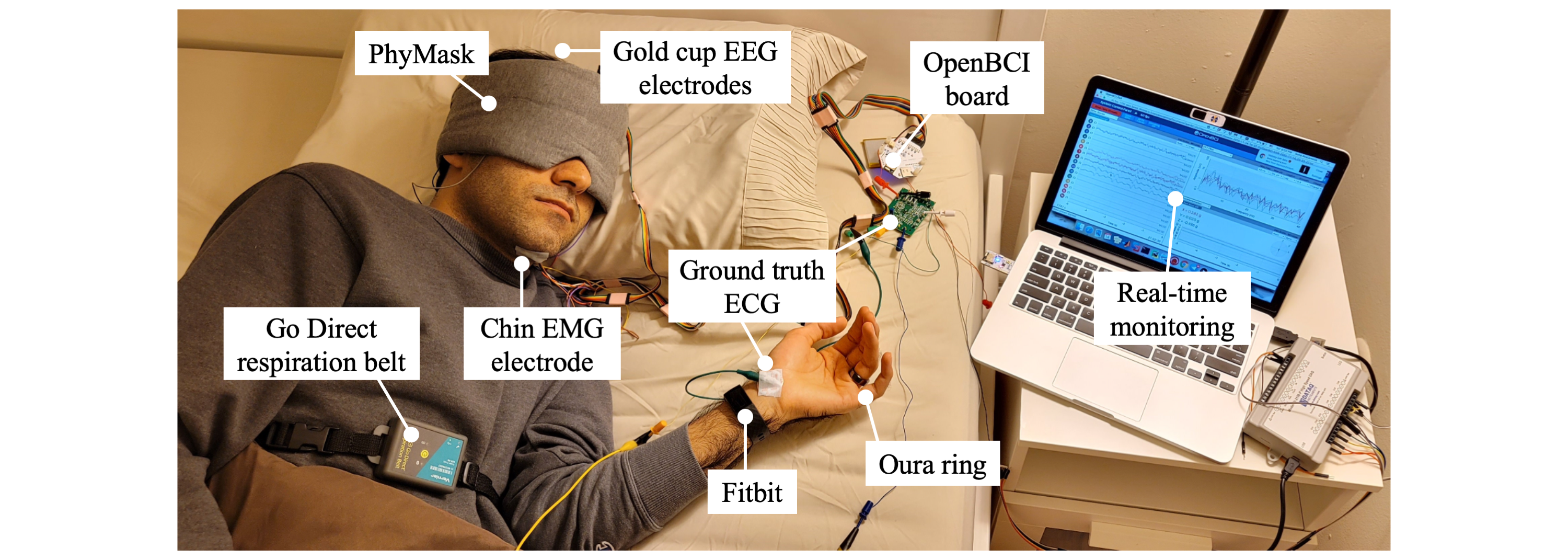}
\caption{Data collection setup.}
\label{fig:data_collection}
\end{figure}

\subsection{Sleep Dataset} \label{sec:sleep_dataset}
Our second data collection aims to validate PhyMask’s ability in measuring the brain waves (EEG) in a more uncontrolled and naturalistic setting and to evaluate biopotential signal quality over a long time of wear. For this, we asked a participant (Male, weight = 160lbs) to wear PhyMask for five nights during sleep and we collected the signals for total of 2118 minutes (about 7 hours per night). Due to the COVID-19 precaution, the experiment took place in participant’s home, providing him a safe environment to perform the study. In addition, it allowed us to collect more naturalistic sleep behavior.

To evaluate how \myname{} compares to commercially-available sleep tracking wearables, we asked our participant to wear a Fitbit Charge 2 and an Oura Ring (fit size 11). The Fitbit and Oura Ring are the dominant sleep tracking wearables in the market, and this comparison allows us to gauge the benefits that \myname{} can provide due to measuring the EEG signal in addition to the physiological signals collected by these wearables.

\begin{figure}
\centering
\includegraphics[width=0.5\linewidth]{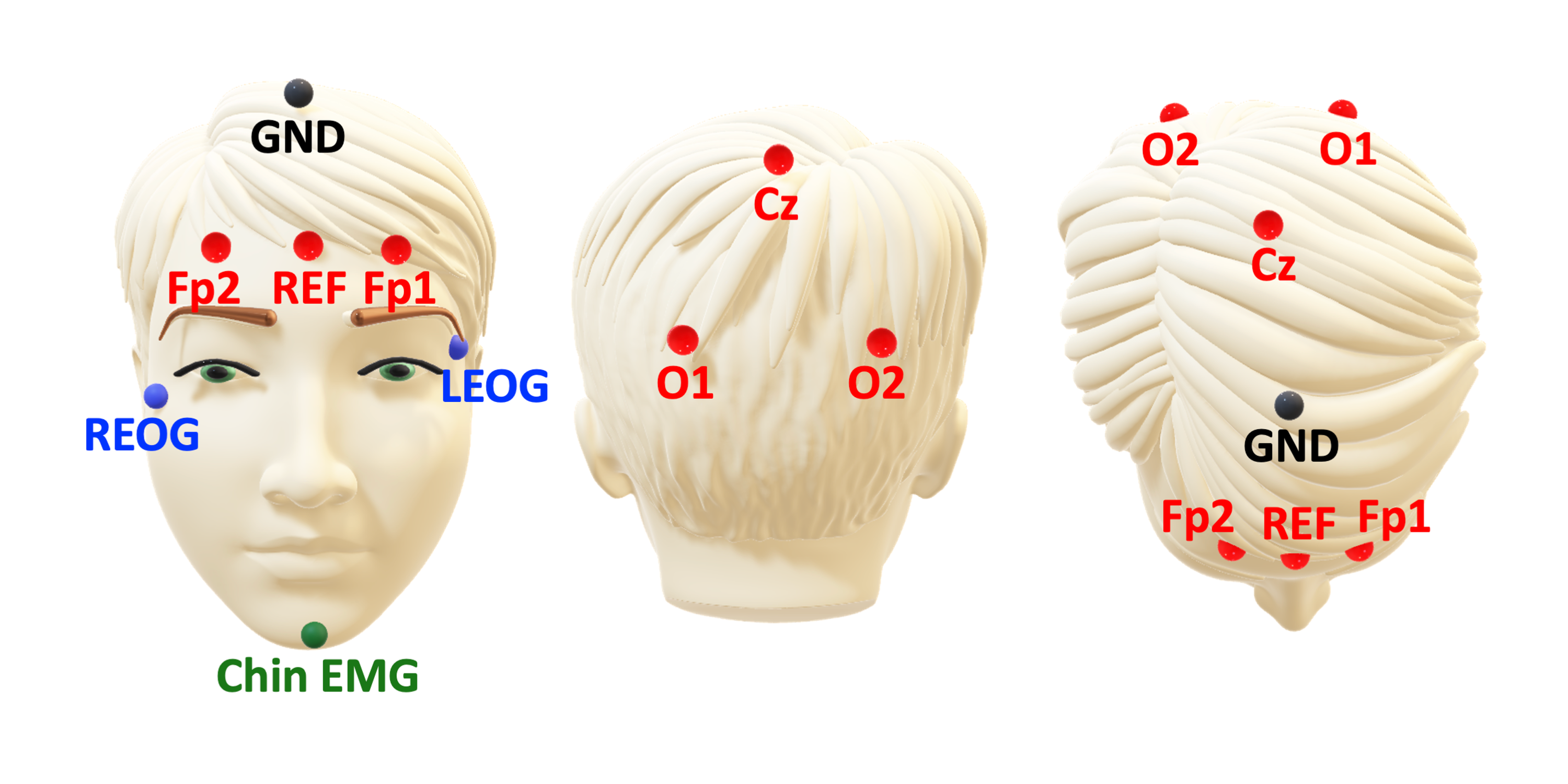}
\caption{The ground truth gold-cup electrodes’ placements used in our sleep monitoring study, based on the standard 10-20 EEG recording system.}
\label{fig:eeg_aasm}
\end{figure}

\mypara{Ground truth annotation.} For ground truth, the PSG data was collected simultaneously with data from \myname{}. PSG included electroencephalographic (EEG: Fp1, Fp2, Cz, O1, O2 sites), chin electromyographic (EMG), and electrooculographic (EOG) recordings performed according to American Academy of Sleep Medicine (AASM) rules~\cite{AASMwebsite} (the electrodes placement is shown in Figure~\ref{fig:eeg_aasm}). The electroconductive paste (Ten20~\cite{ten20}) were used to improve contact between the participant’s scalp and the gold cup electrodes (with 10 mm diameter~\cite{goldcup}). The eight channels of data were collected via an OpenBCI Cyton amplifier~\cite{cytondaisy}. In order to synchronize \myname{}'s EEG signals with the PSG setup, the two \myname{}'s EEG channels were also collected via the OpenBCI Cyton amplifier through the expansion Daisy Module~\cite{cytondaisy}. The OpenBCI board was wirelessly connected to a laptop by the USB Dongle. The raw EEG signals were sampled at a rate of 125Hz, passed through a 60Hz notch filer, a 0.5-50Hz bandpass filter, and finally processed through the use of a mean smoothing filter to mitigate movement artifacts. Figure~\ref{fig:data_collection} shows a user sleeping while wearing PhyMask, sleep tracking wearables, and all the ground truth devices.

Sleep stages (Wake, N1, N2, N3, REM sleep) were scored in 30-sec epochs by certified sleep experts according to sleep scoring guideline of AASM~\cite{berry2012aasm}, separately for both PSG and \myname{}. In addition, the sleep micro-events i.e. spindles and K-complexes, were annotated (2094 number of spindles and 1397 number of K-complexes). For this, a sleep specialist first mark the events with Embla RemLogic PSG software~\cite{emblascoring}, and then visually validate all the events.

\section{Evaluation}
In order to evaluate \myname{}, we break down the results into two subsections. In~\S\ref{sec:results:hydrogelelectrode}, we analyze the performance of our hydrogel electrodes in picking up the brain activity and eye movement signals. We present \myname{} results in detecting sleep stages as well as sleep micro-events (spindle and K-complex), and compare it with commercial wearable trackers. In~\S\ref{sec:results:pressurepatch}, we evaluate our pressure patches performance in estimating the head posture, gross body activity, and respiration and heart rate in various sleep postures.

\subsection{Evaluation of Hydrogel Electrodes in Measuring Biopotential Signals}\label{sec:results:hydrogelelectrode}

In this section, we explore our novel fabric hydrogel sensor's ability in measuring brain waves and eye movement patterns during the longitudinal sleep study. To the best of our knowledge, this is the first time to measure electroencephalogram (EEG) with all-fabric sensing elements during sleep. The validation of this sensor can potentially impact the next generation of EEG monitoring wearable devices.

\begin{figure}
\centering
\includegraphics[width=0.5\linewidth]{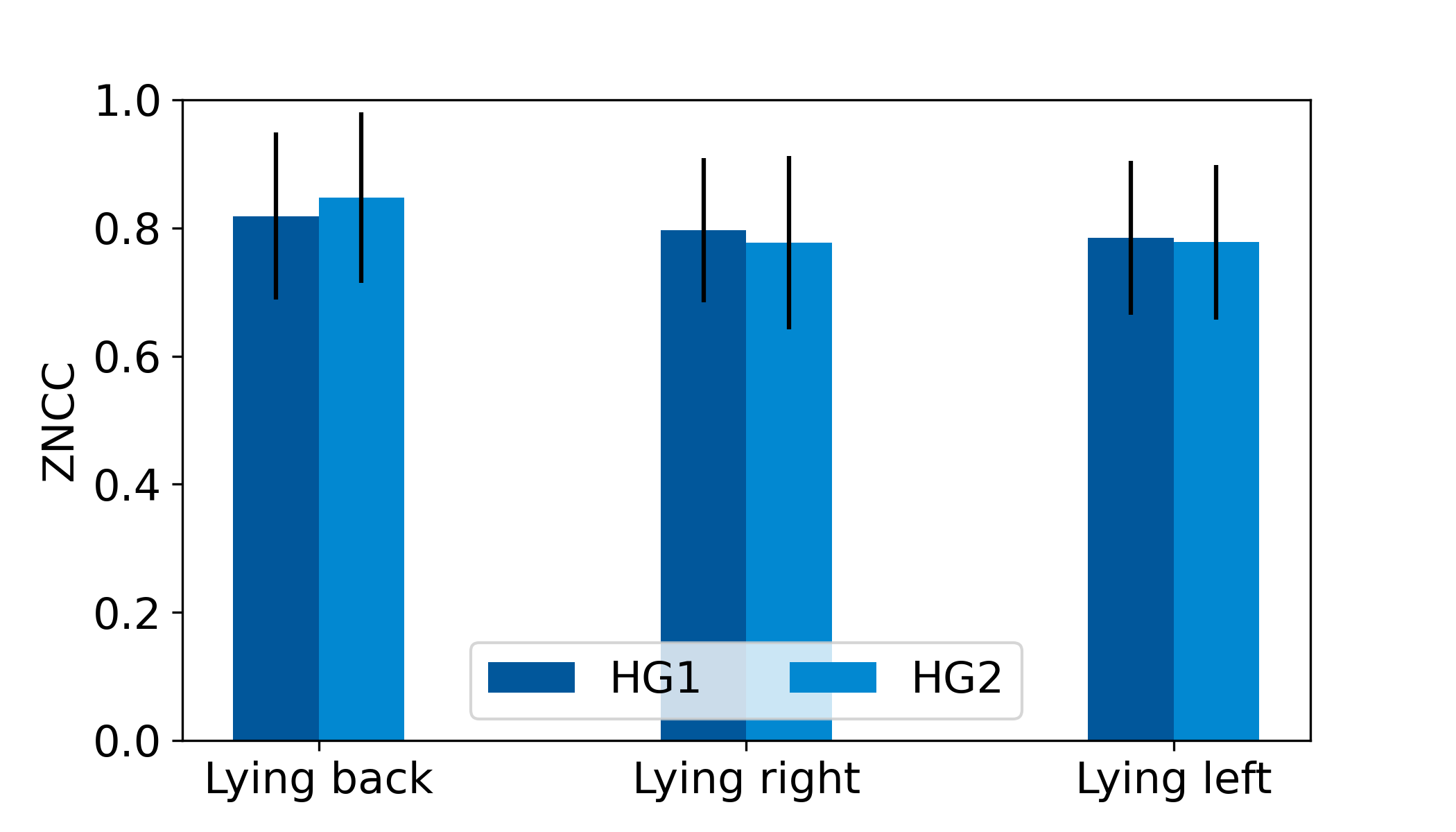}
\captionof{figure}{Zero-normalized cross-correlation values for the electrodes pairs, HG1/Fp1 and HG2/Fp2, across different sleep postures.}
\label{fig:EOG_ZNCC_posture}
\end{figure}

\subsubsection{Evaluating PhyMask Signal Integrity}
We start by evaluating the quality of the biopotential signal obtained from the textile electrodes. We use all data for this evaluation i.e. no data is discarded due to motion artifacts or other noise issues.

We evaluate EOG signal quality using Zero-Normalized Cross-Correlation (ZNCC) metric. We use the benchmarking dataset and report the ZNCC score between the signal captured by our hydrogel electrodes and the gold cup electrodes placed at the nearest standard location, i.e. HG1 with Fp1 and HG2 with Fp2 (The sensor placement is shown in Figure~\ref{fig:eeg_aasm}). ZNCC is defined as

\begin{equation}
\label{equ:NCC}
ZNCC = \frac{1}{n}\:\sum_{x}\frac{1}{\sigma_f \sigma_t} (f(x) - \mu_f) (t(x) - \mu_t)
\end{equation}
where $n$ is the length of $f(x)$ and $t(x)$, $\mu_f$ is the average of $f$, and $\sigma_f$ is the standard deviation of $f$.

Figure~\ref{fig:EOG_ZNCC_posture} illustrates the ZNCC score of 10 participants for lying back, right, and left scenarios. As can be seen the results are consistently good under different postures. This shows that the hydrogel electrodes precisely capture the low-frequency content of the EOG signal even when there is inconsistent pressure applied on the hydrogel electrodes caused by various sleep postures. We observe that the inconsistent pressure only affect the baseline which is removed by our band-pass filter (described in \S\ref{sec:eeg_analysis}).

We evaluate the EEG signal quality using the coherence measure. This metric is known to identify the level of coupling in cortical pathways given its sensitivity to signal phase difference. Maximum coherence occurs when the phase difference is fixed between two signals and a near zero coherence value indicates a random phase difference between signals over time~\cite{coherence1, coherence2}. Coherence is calculated as:

\begin{equation}
\label{equ:coherence}
C_{xy}(f) = \frac{\mid G_{xy}(f) \mid^2}{G_{xx}(f)\:G_{yy}(f)}
\end{equation}
where $G_{xy}(f)$ is the Cross-spectral density between $x$ and $y$, and $G_{xx}(f)$ is the autospectral density of $x$.
\begin{figure}
\centering
\includegraphics[width=1\linewidth]{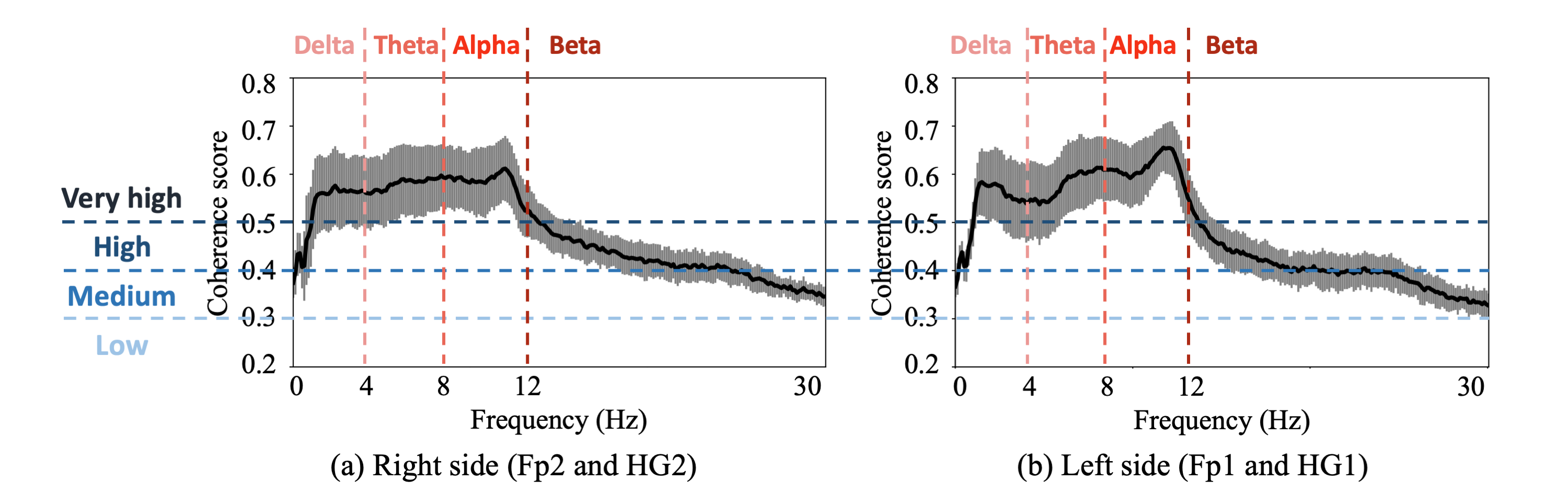}
\caption{The averaged coherence measures over each frequency domain is illustrated for two groups of signals.}
\label{fig:coherence}
\end{figure}

The goal is to evaluate how precise our electrodes can pick up the brain signals with respect to the gold standard EEG recording approach used in PSG, in an uncontrolled and naturalistic setting over a long time of wear. To understand this, we use the sleep dataset~\S\ref{sec:sleep_dataset} which includes over 2118 minutes (more than 35 hours) of night sleep EEG recordings and calculate the coherence for these two groups of signals, HG1/Fp1 and HG2/Fp2, over 30-second epochs in the frequency domain. The calculated coherence values for all epochs are then averaged over all five nights of recording. 

Figure~\ref{fig:coherence} illustrates the averaged coherence measures over each frequency domain. As suggested by previous studies in EEG signal analysis ~\cite{coherence1}, we label the coherence scores as following: low (0.2 to 0.3), medium (0.3 to 0.4), high (0.4 to 0.5), and very high (0.5 to 1). As it can be seen, the majority of frequency domains have very high coherence scores, confirming the maximum coupling between our fabric-based hydrogel electrodes and the standard gold-cup electrodes. Please note that for analysis of EEG signal during sleep, we are more interested in the Delta (0.5 to 4 Hz), Theta (4 to 7 Hz), and Alpha (8 to 12 Hz) sub-bands, as the Beta (12 to 25 Hz) is present when the person is awake with open eyes and is actively thinking~\cite{SATAPATHY20191}.

\begin{figure}
\centering
\includegraphics[width=0.55\linewidth]{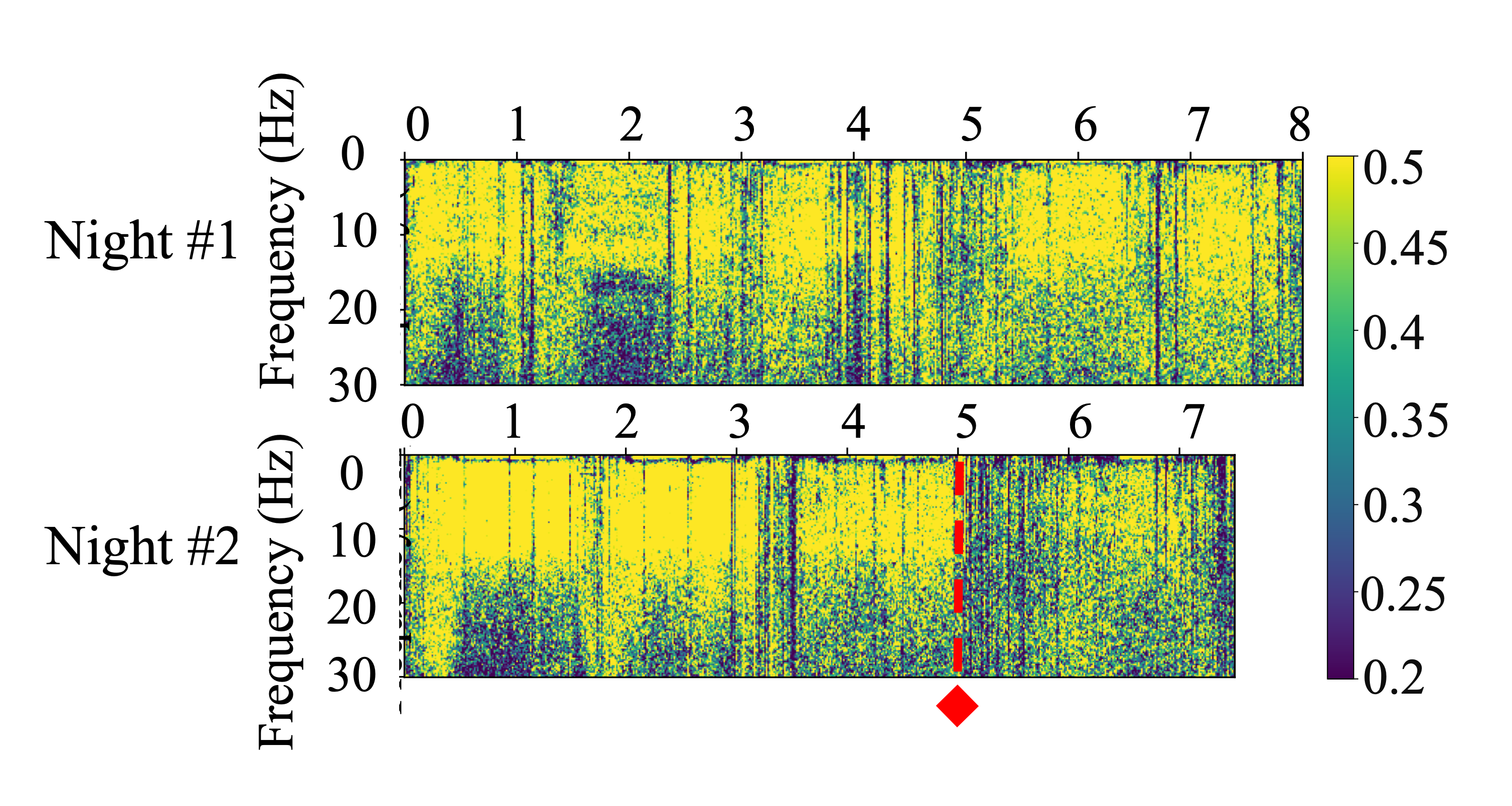}
\caption{Time series of the coherence score of the right hydrogel electrode/Fp2 pair for two different nights. While the coherence is consistently high for night \#1, for night \#2, it degrades after 5:00 due to sudden body motions.}
\label{fig:coherence_time_series}
\end{figure}

While Figure~\ref{fig:coherence} illustrates very high coherence for the two groups of signals, it is also important to understand whether the coupling changes as time progresses. In order to investigate this factor, we have plotted the coherence score of the HG2/Fp2 pair for two different full-night sleep sessions in Figure~\ref{fig:coherence_time_series}. For night \#1, the consistently high coherence score (above 0.4) in lower frequency band (0.5 to 12 Hz), which includes Delta, Theta, and Alpha waves, indicates good performance and high coupling between HG2 electrode and standard Fp2 electrode. However, for night \#2, the coherence degrades after time 5:00 (marked with a red arrow). The reason for this observation is that during that time period, the user had sudden body movements and roll-overs that resulted in \myname{}'s displacement (about 3cm shift to the right). We see that even with this level of displacement, the average coherence measure is around 0.4 which is well within the acceptable range~\cite{coherence1}.

\subsubsection{Sleep Spindle and K-complex Detection} 
In this section, we evaluate 1) how well \myname{} is able to detect spindles and K-complexes, and 2) how \myname{} compares to PSG-based spindle and K-complex detection; PSG has higher-dimension EEG signals captured by standard electrodes from various points on the scalp, so it provides a measure of the best-case performance.

\begin{table}
  \centering
  \caption{Comparing spindle detection of \myname{} versus PSG}
  \label{table:micro-sleep-spindle}
  \begin{tabular}{lccc}
  \toprule
  \textbf{Platform} & \textbf{Sensitivity} & \textbf{Specificity} & \textbf{Accuracy} \\ 
  \midrule
  \textbf{\myname{}} & 0.78    & 0.85  & 0.85 \\
  \textbf{PSG} & 0.80    & 0.86  & 0.85\\
  \bottomrule
  \end{tabular}
\end{table}

\begin{table}
  \centering
  \caption{Comparing K-complex detection of \myname{} versus PSG}
  \label{table:micro-sleep-kcomplex}
  \begin{tabular}{lccc}
  \toprule
  \textbf{Platform} & \textbf{Sensitivity} & \textbf{Specificity} & \textbf{Accuracy} \\ 
  \midrule
  \textbf{\myname{}} & 0.83    & 0.86  & 0.86 \\
  \textbf{PSG} & 0.85    & 0.89  & 0.89 \\
  \bottomrule
  \end{tabular}
\end{table}

In order to compare the performance of \myname{} with PSG in sleep micro-event detection, we train our classifiers explained in~\S\ref{sec:eeg_analysis} separately on \myname{} EEG data and the gold standard PSG. The results of running the two sets of classifiers using K-fold (k = 20) cross validation is summarized in Table~\ref{table:micro-sleep-spindle} and \ref{table:micro-sleep-kcomplex}. Due to the highly imbalanced nature of the data, we use sensitivity and specificity measures for evaluating the performance of the micro-event detectors as also common in the literature~\cite{ranjan2018fuzzy, schonwald2006benchmarking}.

\begin{equation}
\label{equ:metrics}
\begin{aligned}
Sensitivity = \frac{TP}{TP + FN} \:\:\:\:\:\:\:\:\:\:\:\:
Specificity = \frac{TN}{TN + FP} \:\:\:
\end{aligned}
\end{equation}

Our results are very promising and show that: (1) \myname{} accurately detects spindle and K-complex events with higher than 0.8 sensitivity, specificity, and accuracy scores, and (2) \myname{} performs as precise as PSG in spindle and K-complex detection. This means that \myname{} is capable of recording high fidelity EEG signals that contain the micro-sleep events such as spindles and K-complexes.

\subsubsection{Validation of \myname{} against Polysomnography for Sleep Stage Tracking} 

In this section, we evaluate the quality of the PhyMask signal against gold-standard PSG for the purpose of sleep stage tracking. In order to do this, both PhyMask and PSG sleep data (more than 35 hours of sleep data) have been annotated by sleep experts into five sleep stages, i.e., N1, N2, N3 (Deep sleep), REM, and Awake. Due to the low number of N1 stage instances in our dataset, we grouped both N1 and N2 stages as Light sleep. In order to quantify the level of agreement for annotated sleep stages between PhyMask and PSG, we use Cohen's kappa~\cite{cohen1960coefficient} defined as follows:

\begin{equation}
\label{equ:cohenskappa}
\kappa = \frac{p_o - p_e}{1 - p_e}
\end{equation}
where $p_o$ is the relative observed agreement among raters, and $p_e$ is the hypothetical probability of chance agreement, using the observed data to calculate the probabilities of each observer randomly seeing each category. If the raters are in complete agreement then $\kappa = 1$. If there is no agreement among the raters, $\kappa = 0$. It is possible for the statistic to be negative, which implies that there is no effective agreement between the two raters or the agreement is worse than random.

\begin{figure}
\centering
\begin{minipage}{.4\textwidth}
  \centering
  \includegraphics[width=\linewidth]{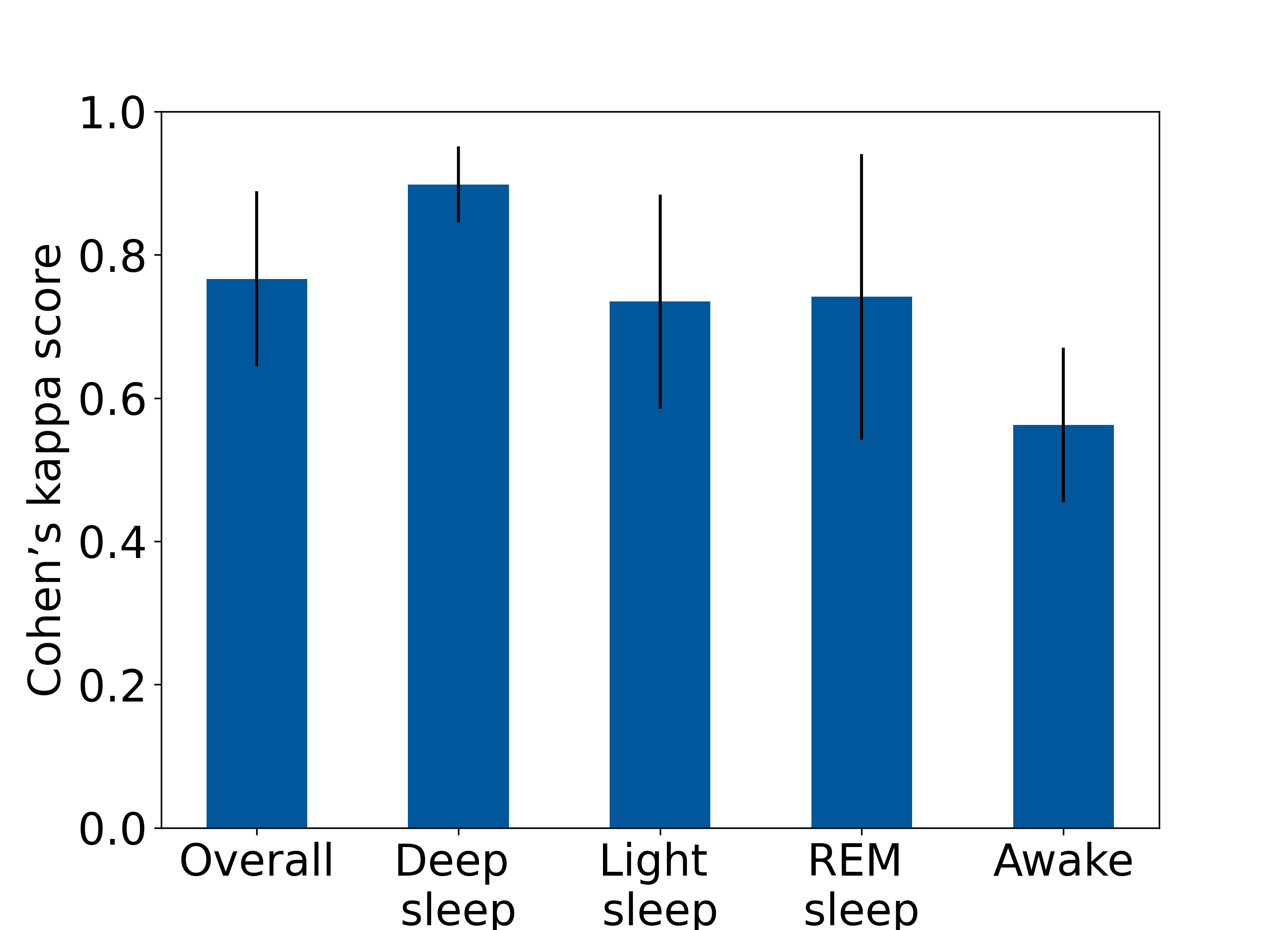}
  \captionof{figure}{Median cohen's kappa measures over all five sleep sessions for \myname{}.}
  \label{fig:cohen_kappa_score_PhyMask}
\end{minipage}%
\hspace{0.2cm}
\begin{minipage}{.45\textwidth}
  \centering
  \includegraphics[width=0.85\linewidth]{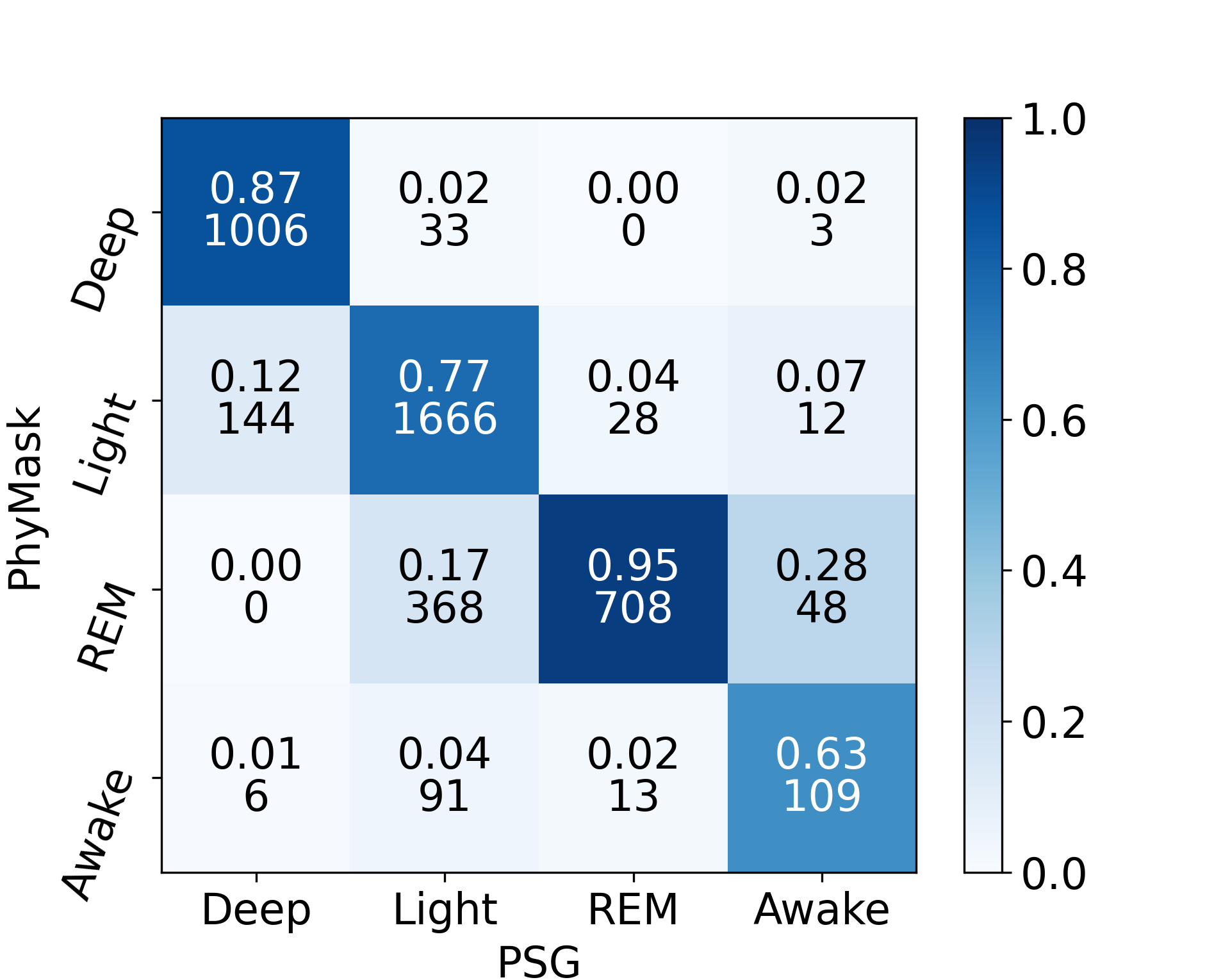}
  \captionof{figure}{The confusion matrix of \myname{} in sleep stage classification}
  \label{fig:cf_PhyMask}
\end{minipage}%
\end{figure}

It is well-known that there exists inevitable variability and disagreements between human experts on epoch ratings -- Wang et al.~\cite{wang2015evaluation} found kappas of 0.72-0.85 between two human raters and kappas of 0.82–0.85 between well-performing raters. Therefore, a Cohen's kappa of $0.8$ is considered to show strong agreement once we take into account such variabilities. Figure~\ref{fig:cohen_kappa_score_PhyMask} summarizes, for each sleep stage, the Cohen's kappa coefficients over all five sleep sessions (total number of epochs = 4236). The overall performance of \myname{} reaches the state of the art with a median Cohen's kappa of 0.77. For most of the sleep stages, \myname{} agrees very well with the PSG. This high level of agreement between \myname{} and PSG further validates the quality of the measured EEG signal and shows that \myname{} successfully captures information that is critical for sleep stage tracking.

The common misclassified stage in PhyMask data is Awake. In order to better understand when sleep stage misclassifications happen, we calculate the confusion matrix (Figure~\ref{fig:cf_PhyMask}), where each data point in the confusion matrix represents a 30-second epoch. As can be see, the awake stage is mostly confused with the REM stage. Since, these stages have overlapping frequency bands (13Hz-30Hz), the main distinguishing signal between the two is the EOG. During the REM stage, as its name suggests, fast rapid eye movements occur as opposed to slow rolling eye patterns that happen during the Awake/drowsiness stage~\cite{roebuck2013review_sleepfeatures}. \myname{} can only capture the horizontal eye movements; however, human raters are trained on conventional data that includes both horizontal and vertical eye movement patterns. We believe that this might have led to lower classification accuracy in this case, and we hypothesize that in the future, training a classifier on the \myname{} data should significantly help with this. 

\subsubsection{Comparison between \myname{} and Commercial Sleep Trackers} 

\begin{figure}
\centering
\includegraphics[width=1\linewidth]{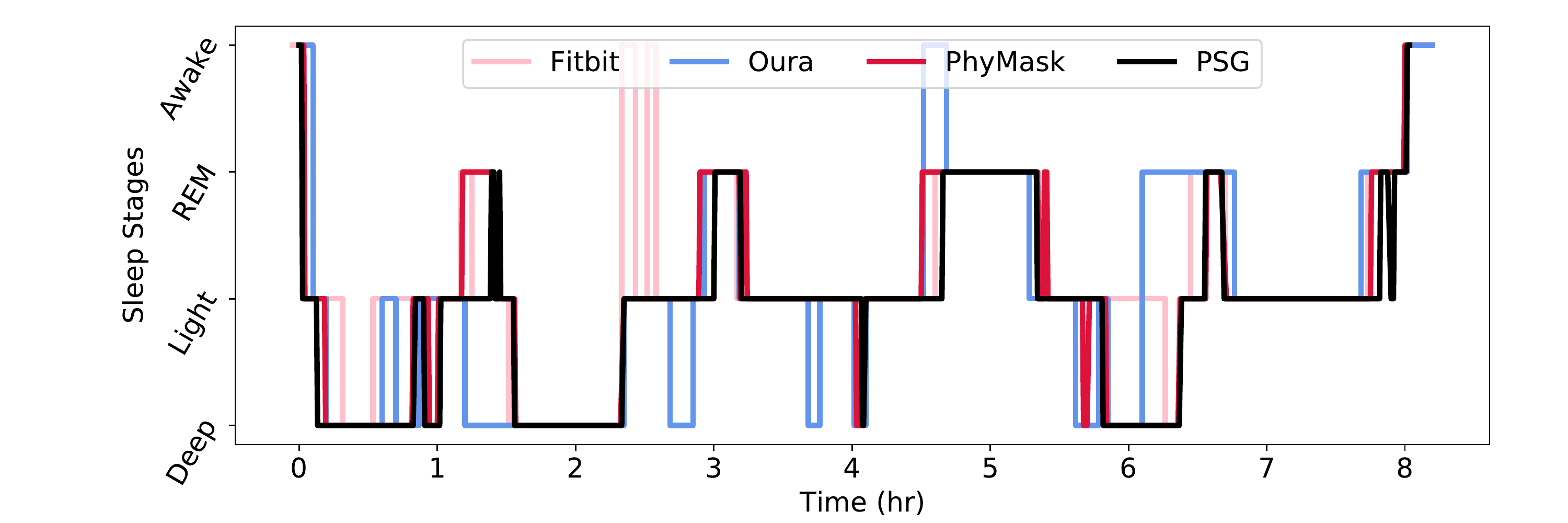}
\caption{The extracted sleep stages of one night for ground-truth (PSG), \myname{}, Oura, and Fitbit.}
\label{fig:night_stages}
\end{figure}

In this section we compare \myname{} with two other commercially-available sleep stage tracking devices i.e. Fitbit Charge 2 and Oura Ring. Figure~\ref{fig:night_stages} illustrates the extracted sleep stages of one night from the three devices with respect to the PSG ground-truth system. 

We compare the accuracy, precision, recall, and f1 score measures of PhyMask, Oura, and Fitbit, calculated as

\begin{equation}
\label{equ:metrics}
\begin{aligned}
Accuracy = \frac{TP + TN}{TP + TN + FP + FN} \:\:\:
Precision = \frac{TP}{TP + FP} \:\:\:
Recall = \frac{TP}{TP + FN} \:\:\:
F1\:score = \frac{2TP}{2TP + FP + FN} \:\:\:
\end{aligned}
\end{equation}
where TP (true positive) is the number of actual positive epochs which are correctly classified; TN (true negative) is the number of actual negative epochs which are correctly classified; FP (false positive) is the number of actual negative epochs which are incorrectly classified as positive; FN (false negative) is the number of actual positive epochs which are incorrectly classified as negative.

\begin{figure}
\centering
\includegraphics[width=0.5\linewidth]{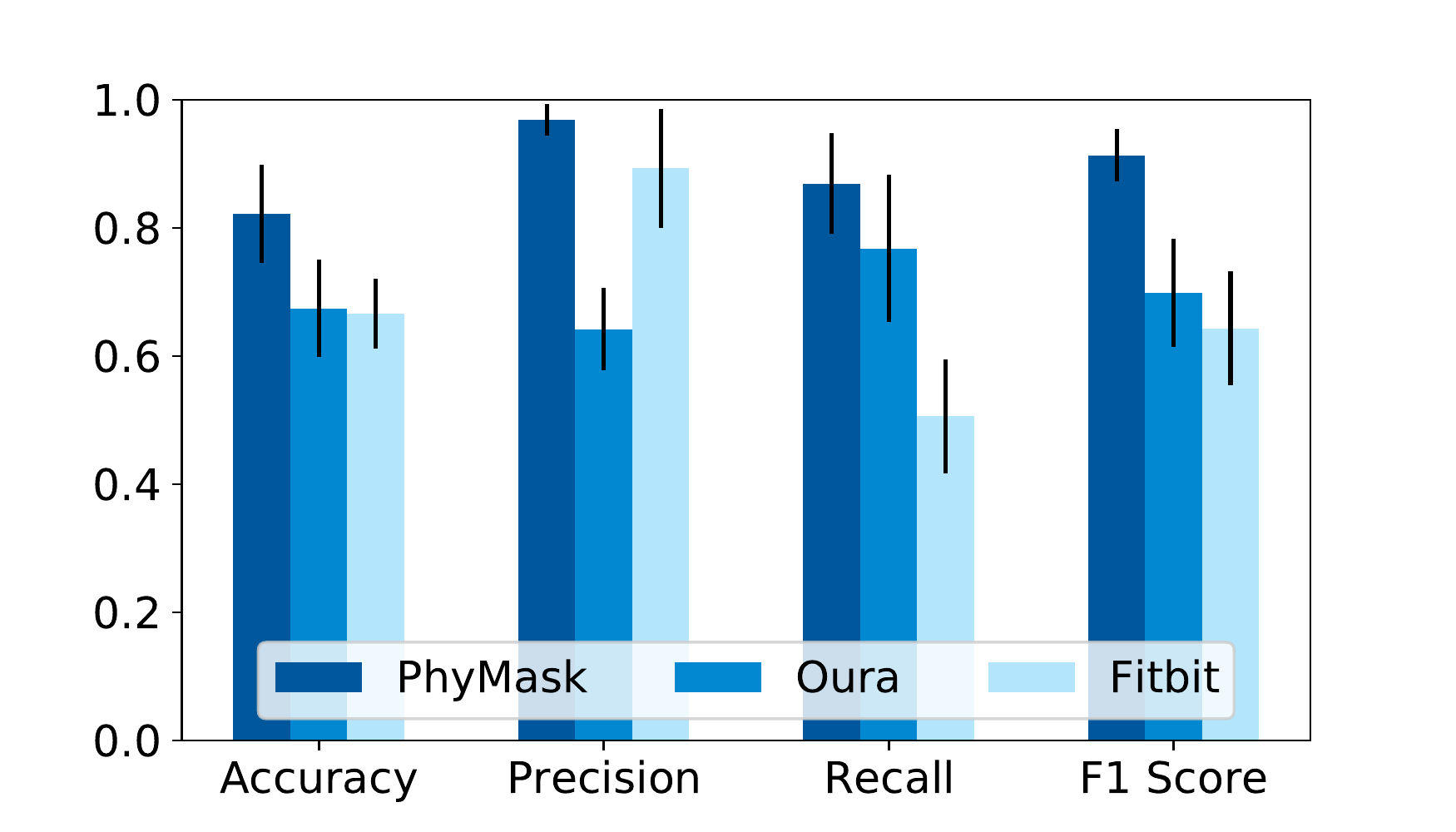}
\caption{Accuracy, precision, recall, and F1 score of sleep staging for five sleep sessions over all of the sleep stage categories.}
\label{fig:fscores}
\end{figure}

Figure~\ref{fig:fscores} summarizes the results for all of the sleep stages over all nights data. \myname{} outperforms the other two devices in terms of all four metrics, which demonstrates the advantage that \myname{} offers in accurate sleep stage tracking by leveraging its capability in recording the EEG signal.

\begin{figure}
     \centering
     \begin{subfigure}[b]{0.49\textwidth}
         \centering
         \includegraphics[width=0.82\textwidth]{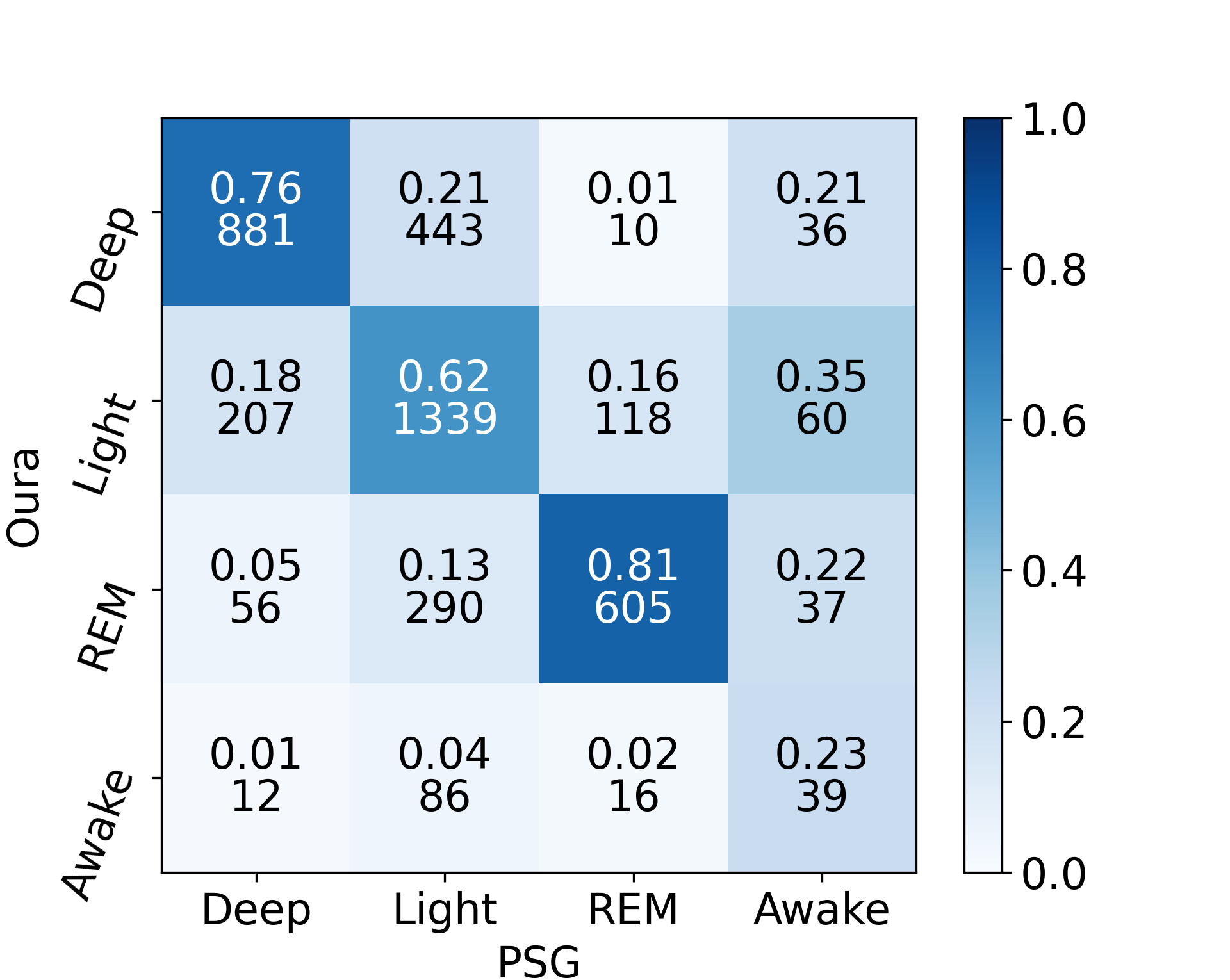}
         \caption{Oura Ring}
         \label{fig:confusion_matrix_oura}
     \end{subfigure}
     \hfill
     \begin{subfigure}[b]{0.49\textwidth}
         \centering
         \includegraphics[width=0.82\textwidth]{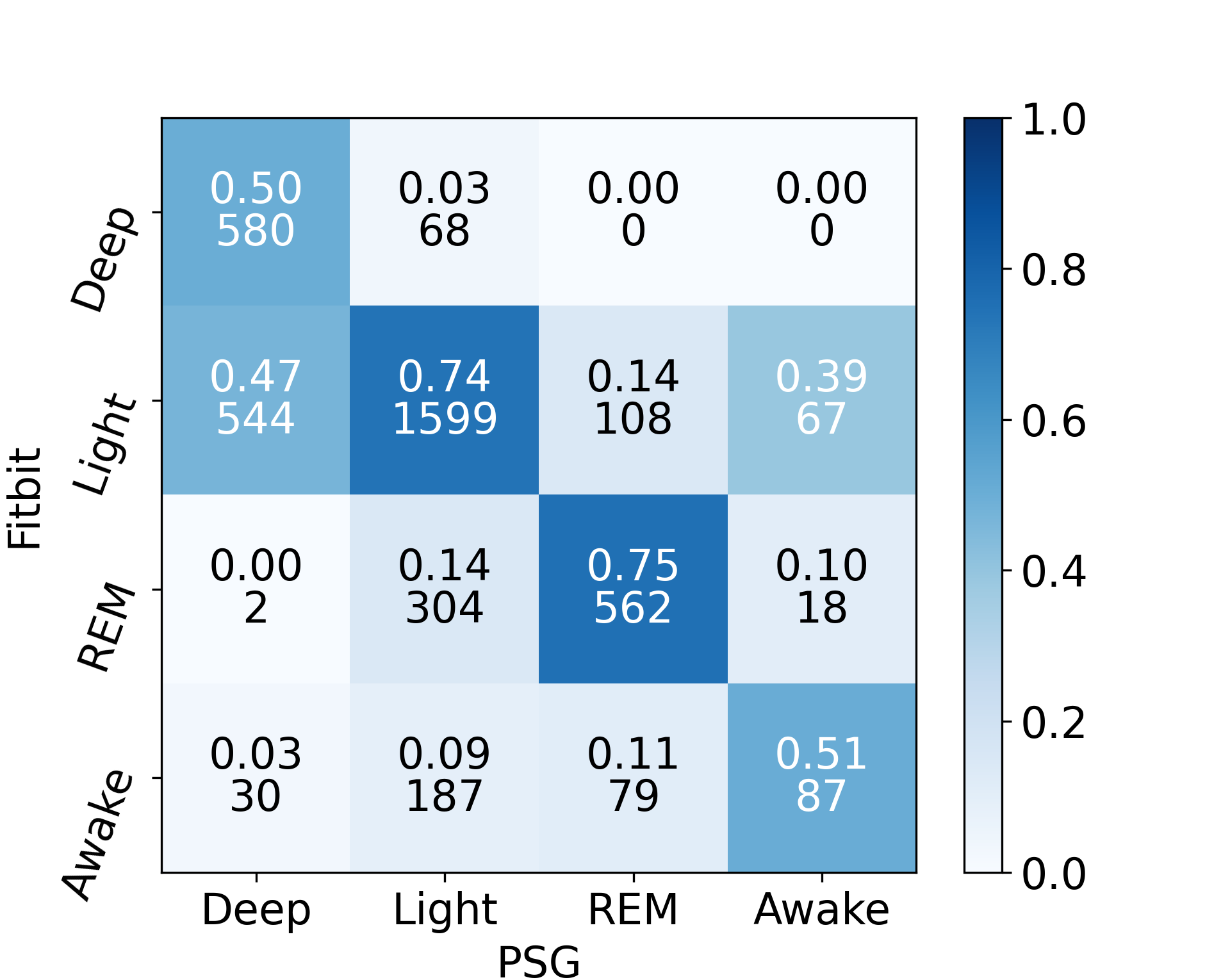}
         \caption{Fitbit}
         \label{fig:confusion_matrix_fitbit}
     \end{subfigure}
        \caption{Sleep stage classification results: the confusion matrices show the performance of Oura and Fitbit.}
        \label{fig:confusion_matrix}
\end{figure}

We also calculate the cohen's kappa measures and confusion matrices for Oura and Fitbit (Figure~\ref{fig:confusion_matrix}). Oura (median kappa = 0.54) and Fitbit (median kappa = 0.51) have much lower kappas than \myname{} (median kappa = 0.77). Low accuracy of these sleep trackers in detecting sleep stages has also been observed by previous studies~\cite{oura_psg, fitbit_psg}. Since we do not have access to the raw data captured by Oura and Fitbit sleep tracking algorithms, we cannot fully analyze their behaviour. However, if we look at the confusion matrices, we find that Oura has a very low recall for the Awake stage. This is because Oura does not provide the micro-awake events after sleep onset in their provided sleep hypnogram on the dashboard. They instead seem to be reported as the closest sleep stage and that is why the predicted labels for Awake events by Oura are distributed almost evenly among all the stages. Fitbit, on the other hand, has a noticeable bias towards classifying sleep sessions as the Light sleep stage. Figure~\ref{fig:confusion_matrix_fitbit} shows that nearly half of the Deep sleep stage epochs are wrongly classified as Light stage. This is an example where leveraging the EEG signal can greatly boost the performance as these two stages can easily be differentiated based on their frequency content (Deep: 0.5-4 Hz and Light: 4-8 Hz).

\subsection{Evaluation of Pressure Patches in Measuring Physiological Parameters}\label{sec:results:pressurepatch}

In this section, we evaluate the performance of our novel pressure-based method for detecting the physiological variables of interest. For this purpose we use our benchmarking dataset~\ref{sec:benchmarking_dataset}. 

\subsubsection{Head posture and Gross Body Movement Estimation}
\hfill

\mypara{Gross body movement.} As discussed earlier, the \myname{} pressure patches are designed to be sensitive to capture the subtle head movements caused by heart pulses and respiration. Therefore, the gross body movement, which is an important marker for sleep tracking and sleep disorder diagnosis, can easily be detected by the embedded pressure sensors. 

\begin{figure}
     \centering
     \begin{subfigure}[b]{0.49\textwidth}
         \centering
         \includegraphics[width=\textwidth]{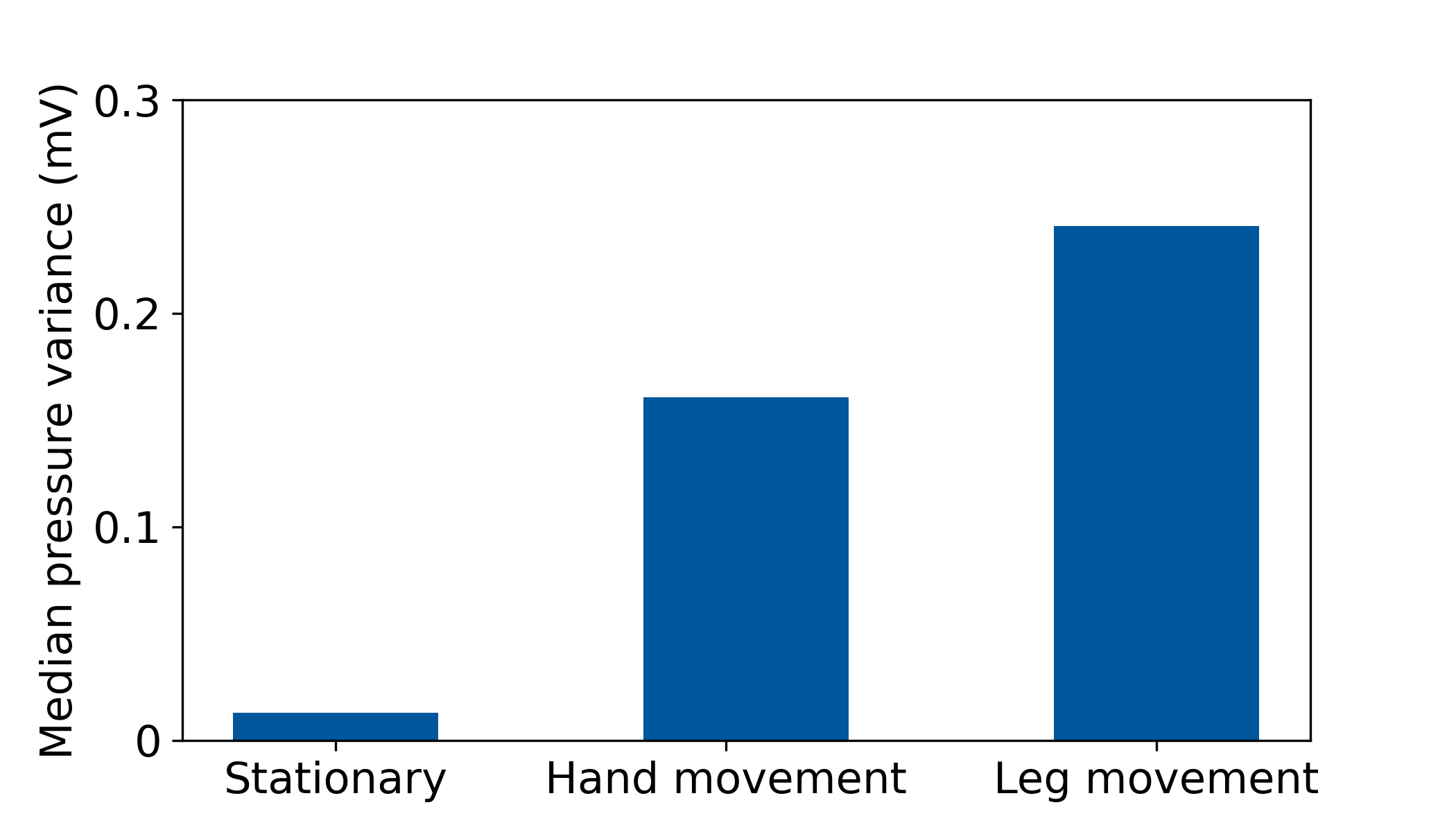}
         \caption{The median variance of baseline signal voltage of the pressure patches for stationary, hand movement, and leg movement scenarios across all sleep postures and participants.}
         \label{fig:body_movement_bar}
     \end{subfigure}
     \hfill
     \begin{subfigure}[b]{0.49\textwidth}
         \centering
         \includegraphics[width=0.8\textwidth]{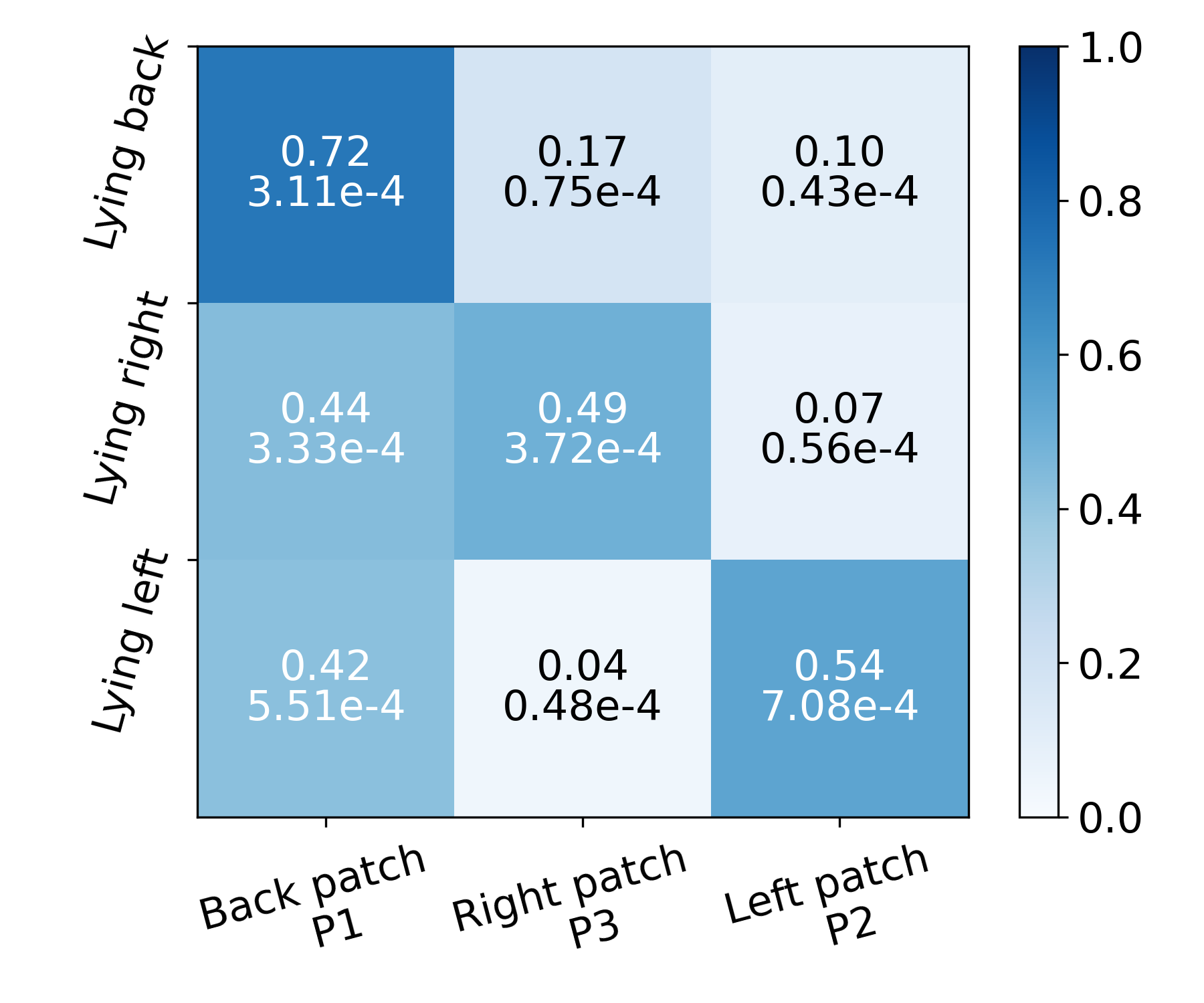}
         \caption{The median variance of each patch baseline voltage during limb movement periods over each sleep posture across all participants.}
         \label{fig:body_movement_cf}
     \end{subfigure}
        \caption{Analysis of baseline voltage changes of the pressure patches for limb movement scenarios in different sleep postures. We can see that pressure patches easily capture information on gross body movement.}
        \label{fig:three graphs}
\end{figure}

In order to evaluate this, we asked our participants to mimic both smooth and jerky limb movements that are common during sleep in each sleep posture (\S\ref{sec:benchmarking_dataset}). The median variance of baseline signal voltage of the three patches for stationary, hand movement, and leg movement scenarios across all sleep postures and over 10 participants are illustrated in Figure~\ref{fig:body_movement_bar}. As can be seen, the variance difference between the limb movement scenarios and the stationary scenario is very distinct. We can also see that the signal corresponding to the hand movement scenario has lower variance compared to the leg movement scenario since people in general use more force to move their legs as opposed to their hands due to the legs physical characteristics such as weight and length which results in applying more pressure on the patches. As a result, even naive thresholding on the variance of the pressure patch signals can identify different types of gross body movement events.

In order to understand how gross motor activity affect each pressure patch in different sleep postures, we calculate the variance of each patch baseline voltage during limb movement periods over each sleep posture. The median voltage across all participants is shown in Figure~\ref{fig:body_movement_cf}. As can be seen, the under-pressure patch has the highest variance meaning the higher sensitivity to detect the body movements. An interesting point is that in both lying on right and left scenarios, the variance of the back patch is as high as the pressed patch. This is because, based on our observation, the participants tend to swing their whole body a bit back and forth while doing the activities, specifically during the leg movements. These motions greatly affect the back patch which results in high variance.  

\begin{figure}
\centering
\includegraphics[width=0.5\linewidth]{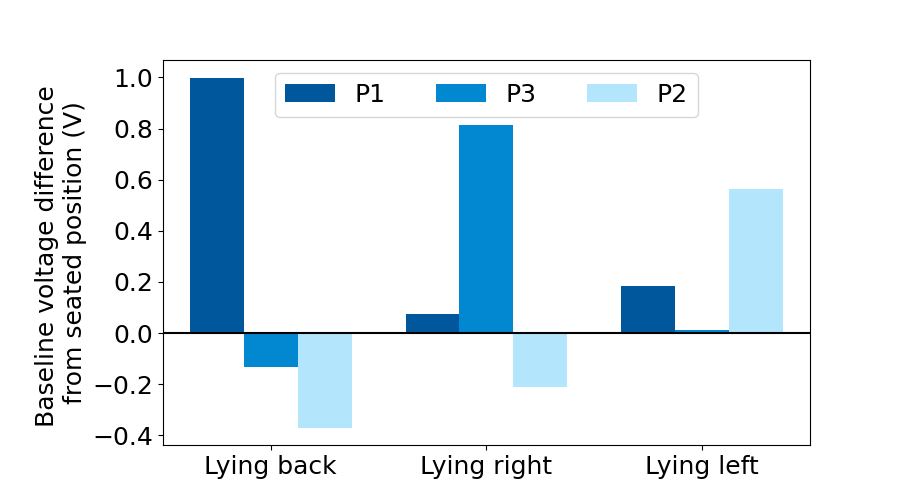}
\caption{The mean pressure patches voltage changes across all participants for each sleep posture.}
\label{fig:posture_result}
\end{figure}

\mypara{Head posture.} In addition to gross motor activity, \myname{} can also provide information about head posture. This information can be useful in understanding which posture leads to better or worse sleep. Head posture can be unambiguously estimated using \myname{} by using the signal baseline across pressure patches (shown in Figure~\ref{fig:posture}). Figure~\ref{fig:posture_result} summarizes the voltage changes averaged across all participants for each sleep posture. In each posture, the voltage change is calculated for all the patches (back, right, and left) with the following formula,

\begin{equation}
  \Delta V_p = V_{p0}^s - V_p
\end{equation}

where $V_{p0}^s$ is the mean baseline voltage of patch $p$ when the participant is in a seated position and there is no pressure applied to the pressure patch, and $V_p$ represents the mean baseline voltage of patch $p$ in the specified sleep posture. 

As expected, the pressed pressure patch in each sleep posture has the highest voltage difference comparing with other pressure patches. We find that a simple decision tree can easily identify posture with 100\% accuracy across all subjects. 

Note that the negative voltage occurs when the pressure applied to the corresponding patch in a specific sleep posture is lower than what it is in the seated position. One example is the right and left patches in the \emph{lying on back} sleep posture (Figure~\ref{fig:posture_result}). This is quite intuitive, since in the seated position the right and left pressure patches are resting on the ears and more pressure is applied to them comparing to when the person is lying on back and the side patches are sort-of loose.

\subsubsection{Respiration Rate Estimation}
\begin{figure}
\centering
\begin{minipage}{.45\textwidth}
  \centering
  \includegraphics[width=\linewidth]{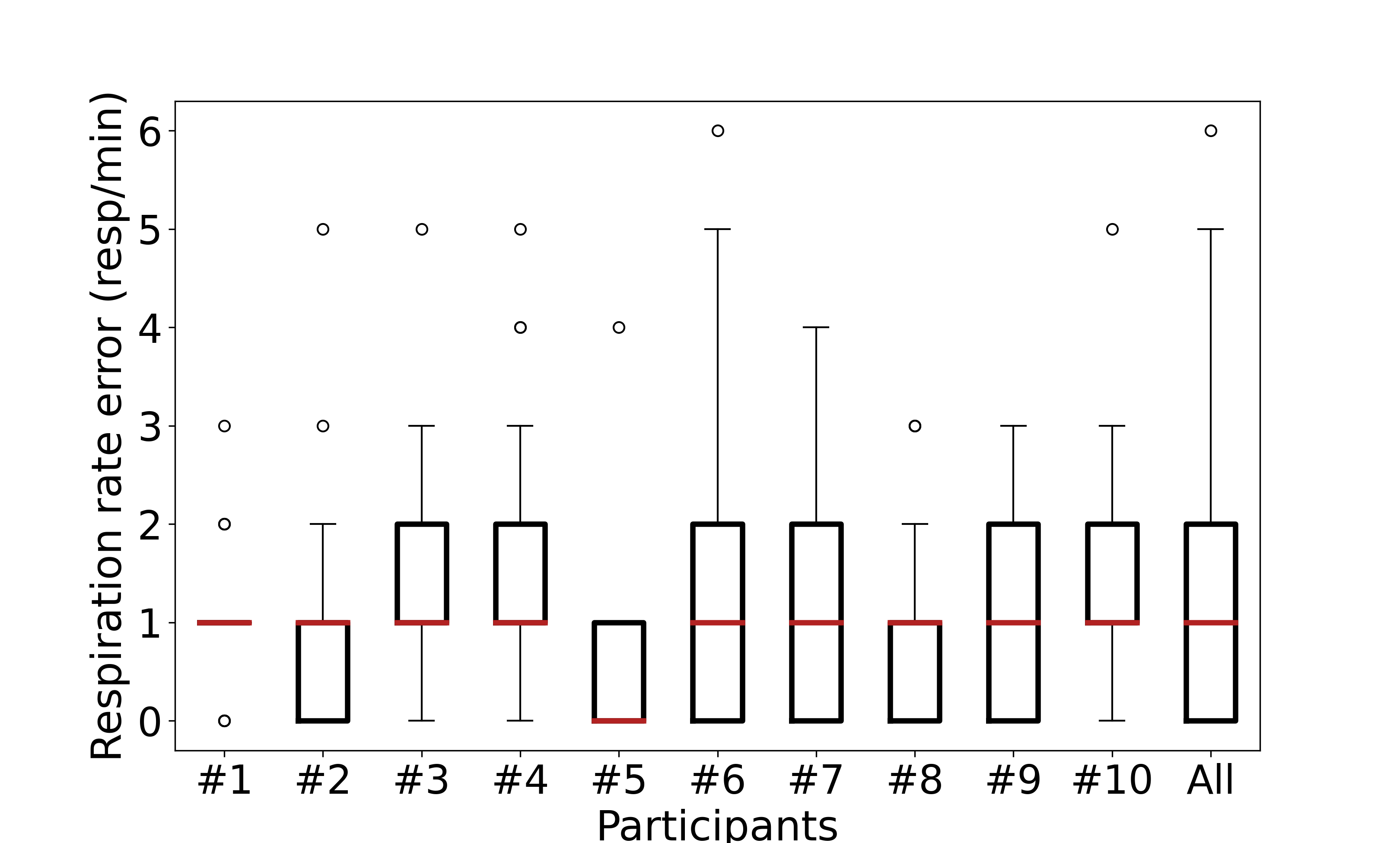}
  \captionof{figure}{Performance of \myname{} in estimating breathing rate across all participants.}
  \label{fig:performance_respiration}
\end{minipage}%
\hspace{0.2cm}
\begin{minipage}{.45\textwidth}
  \centering
  \includegraphics[width=\linewidth]{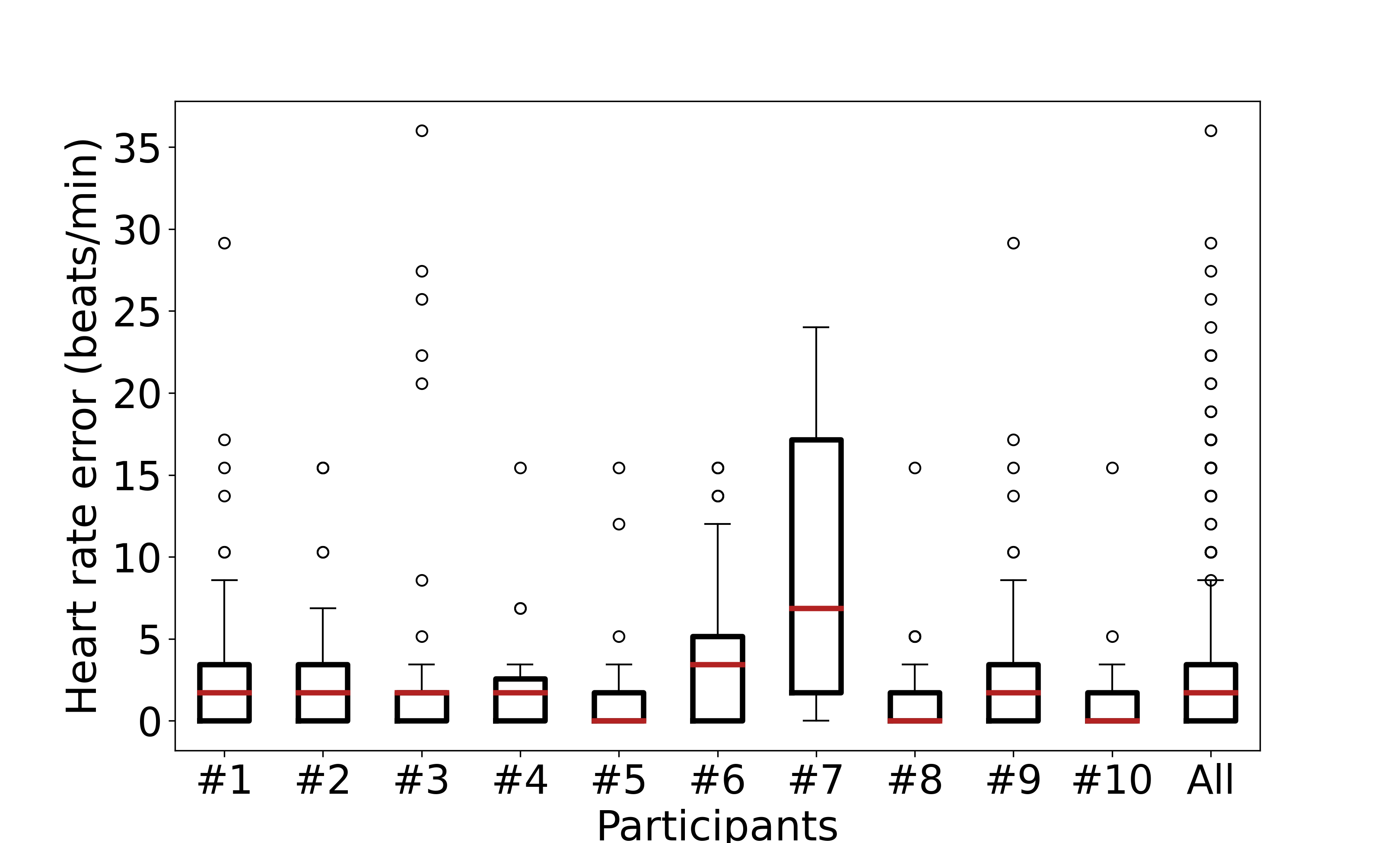}
  \captionof{figure}{Performance of \myname{} in estimating heart rate across all participants.}
  \label{fig:performance_hr}
\end{minipage}%
\end{figure}

Figure~\ref{fig:performance_respiration} shows the performance of \myname{} in estimating the breathing rate for all of the participants across three sleep postures. We see that respiration can be accurately captured by the \myname{} pressure sensor simply by looking at periodic pressure changes on the head. The respiration metrics are very good and the median error is generally about 1 resp/minute. 

\subsubsection{Heart Rate Estimation} The heart rate estimation error is also very good. As can be seen in Figure~\ref{fig:performance_hr}, the median heart rate error over all participants across all the sleep postures is about 1.7 beats/min, that is within the acceptable error margin (= 5 bpm) for heart rate measurement~\cite{hassan2017heart}. The only participant with high error is \#7. Based on the ground truth values, we find that this participant has generally weaker heart beats that result in more subtle head oscillations. We believe that in the future, we can further increase the sensitivity of our pressure patches in order to better detect these small movements. 

\begin{figure}
\centering
\begin{minipage}{.45\textwidth}
  \centering
  \includegraphics[width=0.8\linewidth]{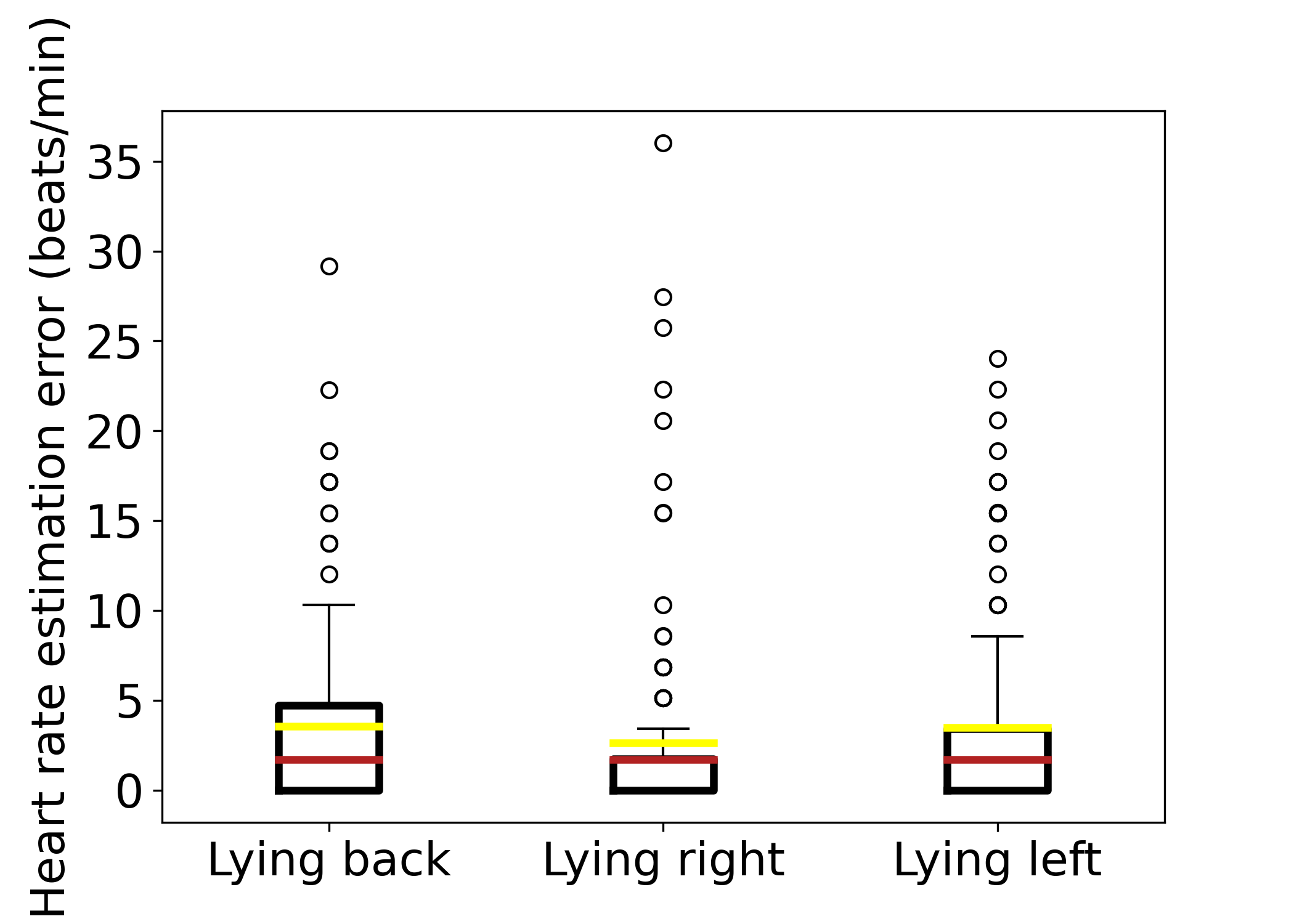}
  \captionof{figure}{Performance of \myname{} in estimating heart rate in different sleep postures across all participants.}
  \label{fig:performance_hr_postures}
\end{minipage}%
\hspace{0.2cm}
\begin{minipage}{.45\textwidth}
  \centering
  \includegraphics[width=0.88\linewidth]{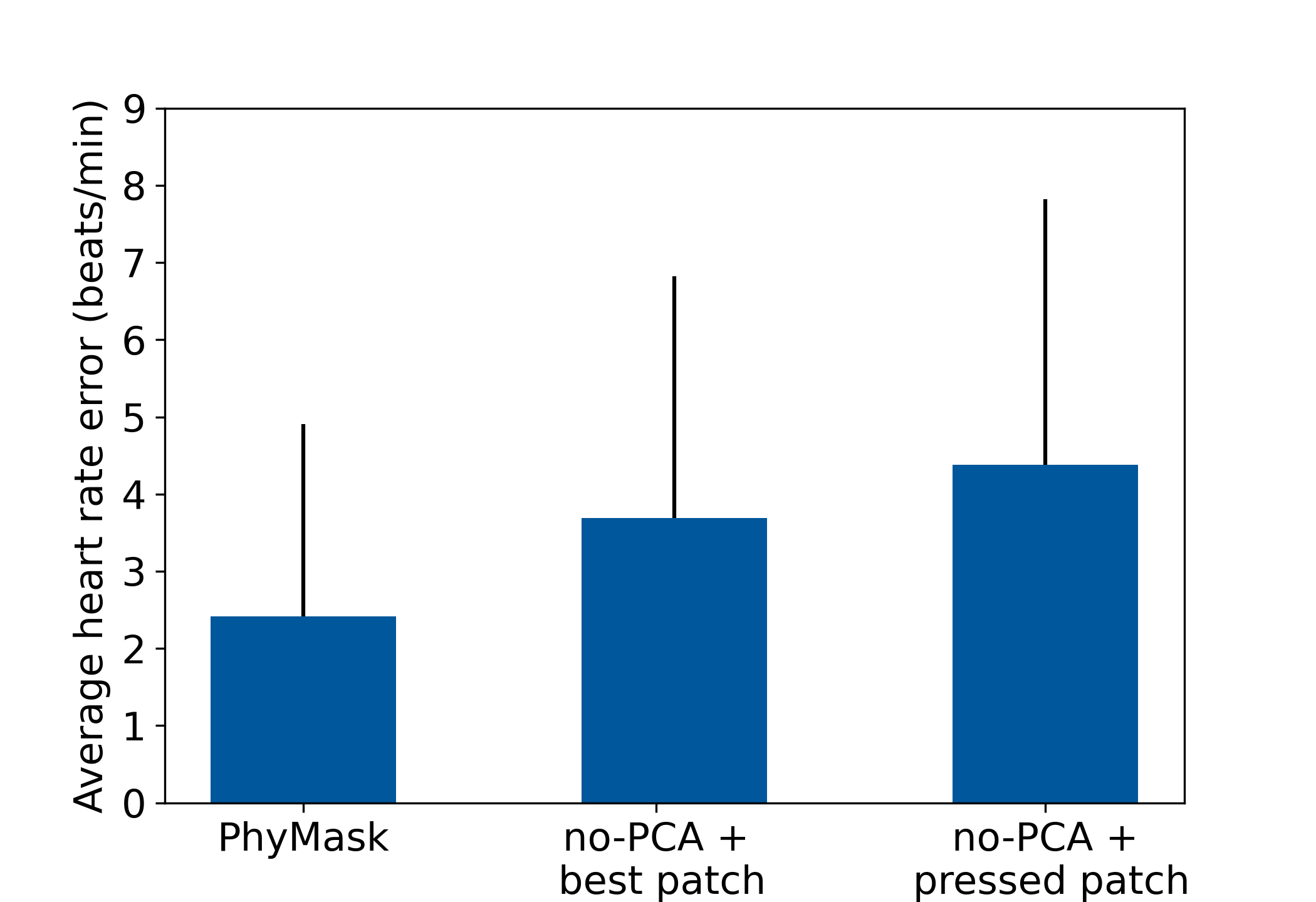}
  \captionof{figure}{Comparing the contributions of PCA and best component detection steps in overall heart rate estimation performance.}
  \label{fig:comp_hr_models}
\end{minipage}%
\end{figure}

In Figure~\ref{fig:performance_hr_postures}, we breakdown the \myname{} performance in estimating heart rate for different sleep postures across all the participants. The results are very promising and shows that \myname{} is robust in all the scenarios.

\mypara{Breaking down the contributions of the heart rate estimation model components.} As described in \S\ref{sec:HR_estimation}, in order to measure heart rate, we first apply PCA on three pressure patches and then detect the best component that holds the majority of the frequency content corresponding to the heart rate. But one question is how much each of these steps contribute to the overall heart rate estimation performance. To understand this, we compare three variants of our model. First we have the full model, \myname{}. For the second model, ``no-PCA + best patch'', we remove the PCA step that is responsible for decomposing the patch signals into sub-signals in order to isolate pulse. Therefore, the best component detection algorithm is directly applied to the pressure patches, and hence the heart rate is derived from the best selected patch. Lastly, we look at the performance, when we directly choose the highest peak in the pressed pressure patch derived fft as the estimated heart rate. This model can help us better understand whether the pressed patch is the best patch for accurate heart rate estimation and how effective our best component detection algorithm is. We call this model ``no-PCA + pressed patch''.

Figure~\ref{fig:comp_hr_models} summarizes the results. We see that removing PCA (``no-PCA + best patch'') increases the mean heart rate error about 1.3 beats/minute. This demonstrates that PCA is successful in isolating heart beats across the three patches. In the case of ``no-PCA + pressed patch'', where we also remove the best component detection step and directly measure the heart rate from the pressed pressure patch, the mean heart rate error increases about 1 beat/minute. This validates that the under-pressure patch in each sleep posture is not necessarily the most sensitive one. This is also consistent with our observation as described in \S\ref{sec:HR_estimation}.

\subsection{Comfort of wear}

\begin{figure}
\centering
\includegraphics[width=1\linewidth]{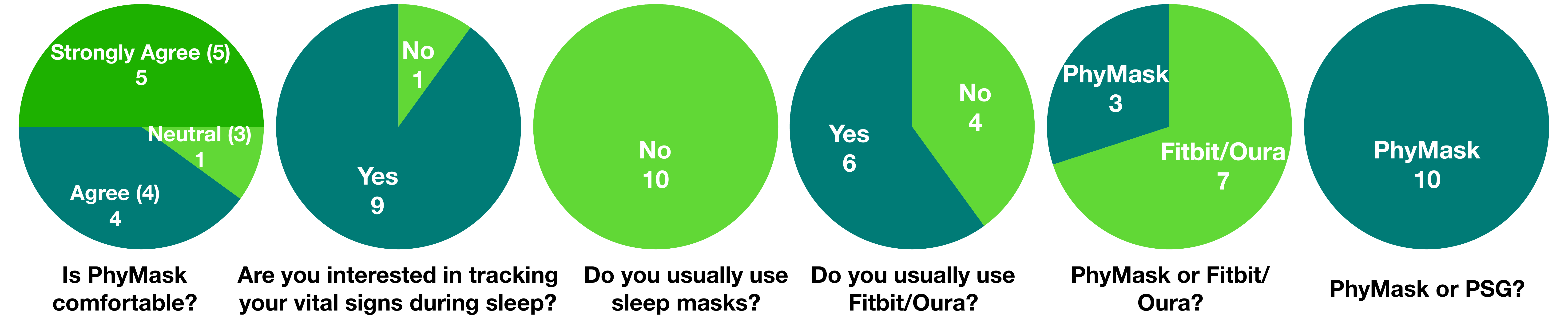}
\caption{Summary of subjective assessment report.}
\label{fig:Questionnaire}
\end{figure}

At the end of our user study, we asked our participants to fill out a questionnaire regarding the \myname{} comfort and their preferences. The questions and the users’ answers are presented in Figure~\ref{fig:Questionnaire}. The results show that the majority found \myname{} quite comfortable. 9/10 users were interested in tracking their physiological signals during the sleep. None of the users had used sleep masks before, while 6/10 users had been using wearable tracking devices on a daily basis. While all the participants preferred \myname{} over PSG, only 3 participants preferred \myname{} over Fitbit/Oura, none of the whom owned any wearable tracking devices. An explanation for this can be that participants who already are used to wearing smart wearables, are also less willing to change their habit.

\section{Conclusion}

In conclusion, we introduce a new all-textile system, the \myname{}, for monitoring a number of important sleep signals including EEG, EOG, respiration, heart rate, and gross body movement. Despite a plethora of commercial and research prototype sleep tracking solutions, we lack a reliable and comfortable solution that can continually monitor the whole range of sleep markers that are useful for clinical-grade sleep monitoring without impacting sleep. \myname{} bridges this gap and shows that all of these signals can be monitored by solely using soft textile-based sensors by leveraging a combination of soft hydrogel electrodes and sensitive textile-based pressure sensors. Such a system can be useful to enable high-quality clinical-grade sleep monitoring at home. Our results are very promising and demonstrate that we can measure a number of advanced sleep markers such as spindles and K-complexes in addition to sleep stages. We can also provide robust measures of physiological features from an unconventional place, i.e., head, with high accuracy across different sleep postures.

\section{Acknowledgements}
This work was funded by the National Science Foundation under agreement CSR Medium 1763524. S.R. and D. G. acknowledge support from the National Science Foundation under agreements 1815347 and 1839999. S.Z.H. and T.L.A. also thank the David and Lucile Packard Foundation for providing partial support.

\bibliographystyle{ACM-Reference-Format}
\bibliography{bibliography}


\begin{thebibliography}{92}


\ifx \showCODEN    \undefined \def \showCODEN     #1{\unskip}     \fi
\ifx \showDOI      \undefined \def \showDOI       #1{#1}\fi
\ifx \showISBNx    \undefined \def \showISBNx     #1{\unskip}     \fi
\ifx \showISBNxiii \undefined \def \showISBNxiii  #1{\unskip}     \fi
\ifx \showISSN     \undefined \def \showISSN      #1{\unskip}     \fi
\ifx \showLCCN     \undefined \def \showLCCN      #1{\unskip}     \fi
\ifx \shownote     \undefined \def \shownote      #1{#1}          \fi
\ifx \showarticletitle \undefined \def \showarticletitle #1{#1}   \fi
\ifx \showURL      \undefined \def \showURL       {\relax}        \fi
\providecommand\bibfield[2]{#2}
\providecommand\bibinfo[2]{#2}
\providecommand\natexlab[1]{#1}
\providecommand\showeprint[2][]{arXiv:#2}

\bibitem[\protect\citeauthoryear{??}{AAS}{2020}]%
        {AASMwebsite}
 \bibinfo{year}{2020}\natexlab{}.
\newblock \bibinfo{title}{American Academy of Sleep Medicine}.
\newblock
  \bibinfo{howpublished}{\url{https://aasm.org/clinical-resources/practice-standards/}}.
\newblock


\bibitem[\protect\citeauthoryear{??}{cyt}{2020}]%
        {cytondaisy}
 \bibinfo{year}{2020}\natexlab{}.
\newblock \bibinfo{title}{Cyton + Daisy Biosensing Boards (16-Channels)}.
\newblock
  \bibinfo{howpublished}{\url{https://shop.openbci.com/collections/frontpage/products/cyton-daisy-biosensing-boards-16-channel?variant=38959256526}}.
\newblock


\bibitem[\protect\citeauthoryear{??}{emb}{2020}]%
        {emblascoring}
 \bibinfo{year}{2020}\natexlab{}.
\newblock \bibinfo{title}{Embla RemLogic PSG Software for sleep and
  micro-events annotations}.
\newblock
  \bibinfo{howpublished}{\url{https://neuro.natus.com/products-services/embla-remlogic-software}}.
\newblock


\bibitem[\protect\citeauthoryear{??}{res}{2020}]%
        {respbelt}
 \bibinfo{year}{2020}\natexlab{}.
\newblock \bibinfo{title}{Go Direct Respiration Belt, Vernier}.
\newblock
  \bibinfo{howpublished}{\url{https://www.vernier.com/product/go-direct-respiration-belt/}}.
\newblock


\bibitem[\protect\citeauthoryear{??}{gol}{2020}]%
        {goldcup}
 \bibinfo{year}{2020}\natexlab{}.
\newblock \bibinfo{title}{Gold Cup Electrodes}.
\newblock
  \bibinfo{howpublished}{\url{https://shop.openbci.com/products/openbci-gold-cup-electrodes?variant=9056028163}}.
\newblock


\bibitem[\protect\citeauthoryear{??}{Mur}{2020}]%
        {Murata}
 \bibinfo{year}{2020}\natexlab{}.
\newblock \bibinfo{title}{Murata bcg mems sensor.}
\newblock
\newblock
\newblock
\shownote{\url{https://www.mouser.com/new/Murata/murata-bcg-mems-sensor/}.}


\bibitem[\protect\citeauthoryear{??}{neu}{2020}]%
        {neuroon}
 \bibinfo{year}{2020}\natexlab{}.
\newblock \bibinfo{title}{Neuroon Open - Sleep tracking solution}.
\newblock \bibinfo{howpublished}{\url{https://neuroon.com}}.
\newblock


\bibitem[\protect\citeauthoryear{??}{AD8}{2020}]%
        {AD8232}
 \bibinfo{year}{2020}\natexlab{}.
\newblock \bibinfo{title}{Single Lead Heart Rate Monitor AD8232}.
\newblock
  \bibinfo{howpublished}{\url{https://www.analog.com/en/products/ad8232.html}}.
\newblock


\bibitem[\protect\citeauthoryear{??}{ten}{2020}]%
        {ten20}
 \bibinfo{year}{2020}\natexlab{}.
\newblock \bibinfo{title}{Ten20 EEG Conductive Paste}.
\newblock
  \bibinfo{howpublished}{\url{https://bio-medical.com/ten20-eeg-conductive-paste8oz-jars-3-pack.html}}.
\newblock


\bibitem[\protect\citeauthoryear{??}{Act}{2021}]%
        {Actiwatch}
 \bibinfo{year}{2021}\natexlab{}.
\newblock \bibinfo{title}{{Actiwatch Spectrum Activity monitor | Philips
  Healthcare}}.
\newblock
\newblock
\urldef\tempurl%
\url{https://www.usa.philips.com/healthcare/product/HC1046964/actiwatch-spectrum-activity-monitor}
\showURL{%
\tempurl}


\bibitem[\protect\citeauthoryear{??}{bed}{2021}]%
        {beddit}
 \bibinfo{year}{2021}\natexlab{}.
\newblock \bibinfo{title}{{Beddit Sleep Monitor}}.
\newblock
\newblock
\urldef\tempurl%
\url{https://www.beddit.com/}
\showURL{%
\tempurl}


\bibitem[\protect\citeauthoryear{??}{dre}{2021}]%
        {dreem}
 \bibinfo{year}{2021}\natexlab{}.
\newblock \bibinfo{title}{{Dreem 2 - Sleep, finally.}}
\newblock
\newblock
\urldef\tempurl%
\url{https://dreem.com/}
\showURL{%
\tempurl}


\bibitem[\protect\citeauthoryear{??}{fit}{2021}]%
        {fitbit}
 \bibinfo{year}{2021}\natexlab{}.
\newblock \bibinfo{title}{{Fitbit official site for activity trackers.}}
\newblock
\newblock
\urldef\tempurl%
\url{https://www.fitbit.com/us/home}
\showURL{%
\tempurl}


\bibitem[\protect\citeauthoryear{??}{viv}{2021}]%
        {vivofit}
 \bibinfo{year}{2021}\natexlab{}.
\newblock \bibinfo{title}{{Garmin v{\'{i}}vofit{\textregistered} 4 | Fitness
  Activity Tracker}}.
\newblock
\newblock
\urldef\tempurl%
\url{https://buy.garmin.com/en-US/US/p/582444}
\showURL{%
\tempurl}


\bibitem[\protect\citeauthoryear{??}{mus}{2021}]%
        {muse}
 \bibinfo{year}{2021}\natexlab{}.
\newblock \bibinfo{title}{{Muse™ - Meditation Made Easy with the Muse
  Headband}}.
\newblock
\newblock
\urldef\tempurl%
\url{https://choosemuse.com/}
\showURL{%
\tempurl}


\bibitem[\protect\citeauthoryear{??}{nRF}{2021}]%
        {nRF52811}
 \bibinfo{year}{2021}\natexlab{}.
\newblock \bibinfo{title}{{nRF52811 - Bluetooth 5.2 SoC - nordicsemi.com}}.
\newblock
\newblock
\urldef\tempurl%
\url{https://www.nordicsemi.com/Products/Low-power-short-range-wireless/nRF52811}
\showURL{%
\tempurl}


\bibitem[\protect\citeauthoryear{??}{our}{2021}]%
        {oura}
 \bibinfo{year}{2021}\natexlab{}.
\newblock \bibinfo{title}{{Oura Ring: Accurate Health Information Accessible to
  Everyone}}.
\newblock
\newblock
\urldef\tempurl%
\url{https://ouraring.com/}
\showURL{%
\tempurl}


\bibitem[\protect\citeauthoryear{??}{Phi}{2021}]%
        {Phillips-Smart-Sleep}
 \bibinfo{year}{2021}\natexlab{}.
\newblock \bibinfo{title}{Philips SmartSleep Deep Sleep Headband}.
\newblock
\newblock
\newblock
\shownote{\url{https://www.usa.philips.com/c-e/smartsleep/deep-sleep-headband.html}.}


\bibitem[\protect\citeauthoryear{??}{app}{2021}]%
        {applewatch}
 \bibinfo{year}{2021}\natexlab{}.
\newblock \bibinfo{title}{{Watch - Apple}}.
\newblock
\newblock
\urldef\tempurl%
\url{https://www.apple.com/watch/}
\showURL{%
\tempurl}


\bibitem[\protect\citeauthoryear{??}{Bra}{2021}]%
        {Brainbit}
 \bibinfo{year}{2021}\natexlab{}.
\newblock \bibinfo{title}{{Wearable Smart EEG headband | BrainBit}}.
\newblock
\newblock
\urldef\tempurl%
\url{https://brainbit.com/}
\showURL{%
\tempurl}


\bibitem[\protect\citeauthoryear{??}{WHO}{2021}]%
        {WHOOP}
 \bibinfo{year}{2021}\natexlab{}.
\newblock \bibinfo{title}{{WHOOP - The World's Most Powerful Fitness
  Membership.}}
\newblock
\newblock
\urldef\tempurl%
\url{https://www.whoop.com/}
\showURL{%
\tempurl}


\bibitem[\protect\citeauthoryear{Acharya, Hani, Cheek, Thirumala, and
  Tsuchida}{Acharya et~al\mbox{.}}{2016}]%
        {acharya2016american}
\bibfield{author}{\bibinfo{person}{Jayant~N Acharya}, \bibinfo{person}{Abeer~J
  Hani}, \bibinfo{person}{Janna Cheek}, \bibinfo{person}{Parthasarathy
  Thirumala}, {and} \bibinfo{person}{Tammy~N Tsuchida}.}
  \bibinfo{year}{2016}\natexlab{}.
\newblock \showarticletitle{American Clinical Neurophysiology Society guideline
  2: guidelines for standard electrode position nomenclature}.
\newblock \bibinfo{journal}{\emph{The Neurodiagnostic Journal}}
  \bibinfo{volume}{56}, \bibinfo{number}{4} (\bibinfo{year}{2016}),
  \bibinfo{pages}{245--252}.
\newblock


\bibitem[\protect\citeauthoryear{Adib, Mao, Kabelac, Katabi, and Miller}{Adib
  et~al\mbox{.}}{2015}]%
        {fadel:2015}
\bibfield{author}{\bibinfo{person}{Fadel Adib}, \bibinfo{person}{Hongzi Mao},
  \bibinfo{person}{Zachary Kabelac}, \bibinfo{person}{Dina Katabi}, {and}
  \bibinfo{person}{Robert~C. Miller}.} \bibinfo{year}{2015}\natexlab{}.
\newblock \showarticletitle{Smart Homes That Monitor Breathing and Heart Rate}.
  In \bibinfo{booktitle}{\emph{Proceedings of the 33rd Annual ACM Conference on
  Human Factors in Computing Systems}} (Seoul, Republic of Korea)
  \emph{(\bibinfo{series}{CHI ’15})}. \bibinfo{publisher}{Association for
  Computing Machinery}, \bibinfo{address}{New York, NY, USA},
  \bibinfo{pages}{837–846}.
\newblock
\showISBNx{9781450331456}
\urldef\tempurl%
\url{https://doi.org/10.1145/2702123.2702200}
\showDOI{\tempurl}


\bibitem[\protect\citeauthoryear{Alba, Sclabassi, Sun, and Cui}{Alba
  et~al\mbox{.}}{2010}]%
        {alba2010novel}
\bibfield{author}{\bibinfo{person}{Nicolas~Alexander Alba},
  \bibinfo{person}{Robert~J Sclabassi}, \bibinfo{person}{Mingui Sun}, {and}
  \bibinfo{person}{Xinyan~Tracy Cui}.} \bibinfo{year}{2010}\natexlab{}.
\newblock \showarticletitle{Novel hydrogel-based preparation-free EEG
  electrode}.
\newblock \bibinfo{journal}{\emph{IEEE transactions on neural systems and
  rehabilitation engineering}} \bibinfo{volume}{18}, \bibinfo{number}{4}
  (\bibinfo{year}{2010}), \bibinfo{pages}{415--423}.
\newblock


\bibitem[\protect\citeauthoryear{Altevogt, Colten, et~al\mbox{.}}{Altevogt
  et~al\mbox{.}}{2006}]%
        {altevogt2006sleep}
\bibfield{author}{\bibinfo{person}{Bruce~M Altevogt}, \bibinfo{person}{Harvey~R
  Colten}, {et~al\mbox{.}}} \bibinfo{year}{2006}\natexlab{}.
\newblock \bibinfo{booktitle}{\emph{Sleep disorders and sleep deprivation: an
  unmet public health problem}}.
\newblock \bibinfo{publisher}{National Academies Press}.
\newblock


\bibitem[\protect\citeauthoryear{Andreoni, Standoli, and Perego}{Andreoni
  et~al\mbox{.}}{2016}]%
        {andreoni2016defining}
\bibfield{author}{\bibinfo{person}{Giuseppe Andreoni},
  \bibinfo{person}{Carlo~Emilio Standoli}, {and} \bibinfo{person}{Paolo
  Perego}.} \bibinfo{year}{2016}\natexlab{}.
\newblock \showarticletitle{Defining requirements and related methods for
  designing sensorized garments}.
\newblock \bibinfo{journal}{\emph{Sensors}} \bibinfo{volume}{16},
  \bibinfo{number}{6} (\bibinfo{year}{2016}), \bibinfo{pages}{769}.
\newblock


\bibitem[\protect\citeauthoryear{{Balakrishnan}, {Durand}, and
  {Guttag}}{{Balakrishnan} et~al\mbox{.}}{2013}]%
        {Balakrishnan:2013}
\bibfield{author}{\bibinfo{person}{G. {Balakrishnan}}, \bibinfo{person}{F.
  {Durand}}, {and} \bibinfo{person}{J. {Guttag}}.}
  \bibinfo{year}{2013}\natexlab{}.
\newblock \showarticletitle{Detecting Pulse from Head Motions in Video}. In
  \bibinfo{booktitle}{\emph{2013 IEEE Conference on Computer Vision and Pattern
  Recognition}}. \bibinfo{pages}{3430--3437}.
\newblock
\showISSN{1063-6919}
\urldef\tempurl%
\url{https://doi.org/10.1109/CVPR.2013.440}
\showDOI{\tempurl}


\bibitem[\protect\citeauthoryear{Bartula, Tigges, and Muehlsteff}{Bartula
  et~al\mbox{.}}{2013}]%
        {bartula2013camera}
\bibfield{author}{\bibinfo{person}{Marek Bartula}, \bibinfo{person}{Timo
  Tigges}, {and} \bibinfo{person}{Jens Muehlsteff}.}
  \bibinfo{year}{2013}\natexlab{}.
\newblock \showarticletitle{Camera-based system for contactless monitoring of
  respiration}. In \bibinfo{booktitle}{\emph{2013 35th Annual International
  Conference of the IEEE Engineering in Medicine and Biology Society (EMBC)}}.
  IEEE, \bibinfo{pages}{2672--2675}.
\newblock


\bibitem[\protect\citeauthoryear{Berry, Brooks, Gamaldo, Harding, Marcus,
  Vaughn, et~al\mbox{.}}{Berry et~al\mbox{.}}{2012}]%
        {berry2012aasm}
\bibfield{author}{\bibinfo{person}{Richard~B Berry}, \bibinfo{person}{Rita
  Brooks}, \bibinfo{person}{Charlene~E Gamaldo}, \bibinfo{person}{Susan~M
  Harding}, \bibinfo{person}{C Marcus}, \bibinfo{person}{Bradley~V Vaughn},
  {et~al\mbox{.}}} \bibinfo{year}{2012}\natexlab{}.
\newblock \showarticletitle{The AASM manual for the scoring of sleep and
  associated events}.
\newblock \bibinfo{journal}{\emph{Rules, Terminology and Technical
  Specifications, Darien, Illinois, American Academy of Sleep Medicine}}
  \bibinfo{volume}{176} (\bibinfo{year}{2012}), \bibinfo{pages}{2012}.
\newblock


\bibitem[\protect\citeauthoryear{Cash, Halgren, Dehghani, Rossetti, Thesen,
  Wang, Devinsky, Kuzniecky, Doyle, Madsen, et~al\mbox{.}}{Cash
  et~al\mbox{.}}{2009}]%
        {cash2009human:kcomplex}
\bibfield{author}{\bibinfo{person}{Sydney~S Cash}, \bibinfo{person}{Eric
  Halgren}, \bibinfo{person}{Nima Dehghani}, \bibinfo{person}{Andrea~O
  Rossetti}, \bibinfo{person}{Thomas Thesen}, \bibinfo{person}{ChunMao Wang},
  \bibinfo{person}{Orrin Devinsky}, \bibinfo{person}{Ruben Kuzniecky},
  \bibinfo{person}{Werner Doyle}, \bibinfo{person}{Joseph~R Madsen},
  {et~al\mbox{.}}} \bibinfo{year}{2009}\natexlab{}.
\newblock \showarticletitle{The human K-complex represents an isolated cortical
  down-state}.
\newblock \bibinfo{journal}{\emph{Science}} \bibinfo{volume}{324},
  \bibinfo{number}{5930} (\bibinfo{year}{2009}), \bibinfo{pages}{1084--1087}.
\newblock


\bibitem[\protect\citeauthoryear{Chawla, Bowyer, Hall, and Kegelmeyer}{Chawla
  et~al\mbox{.}}{2002}]%
        {smote}
\bibfield{author}{\bibinfo{person}{Nitesh~V Chawla}, \bibinfo{person}{Kevin~W
  Bowyer}, \bibinfo{person}{Lawrence~O Hall}, {and} \bibinfo{person}{W~Philip
  Kegelmeyer}.} \bibinfo{year}{2002}\natexlab{}.
\newblock \showarticletitle{SMOTE: synthetic minority over-sampling technique}.
\newblock \bibinfo{journal}{\emph{Journal of artificial intelligence research}}
   \bibinfo{volume}{16} (\bibinfo{year}{2002}), \bibinfo{pages}{321--357}.
\newblock


\bibitem[\protect\citeauthoryear{Cohen}{Cohen}{1960}]%
        {cohen1960coefficient}
\bibfield{author}{\bibinfo{person}{Jacob Cohen}.}
  \bibinfo{year}{1960}\natexlab{}.
\newblock \showarticletitle{A coefficient of agreement for nominal scales}.
\newblock \bibinfo{journal}{\emph{Educational and psychological measurement}}
  \bibinfo{volume}{20}, \bibinfo{number}{1} (\bibinfo{year}{1960}),
  \bibinfo{pages}{37--46}.
\newblock


\bibitem[\protect\citeauthoryear{Costa, Ortigueira, Batista, and Paiva}{Costa
  et~al\mbox{.}}{2012}]%
        {costa2012automatic}
\bibfield{author}{\bibinfo{person}{Joao Costa}, \bibinfo{person}{Manuel
  Ortigueira}, \bibinfo{person}{Arnaldo Batista}, {and} \bibinfo{person}{T
  Paiva}.} \bibinfo{year}{2012}\natexlab{}.
\newblock \showarticletitle{An Automatic Sleep Spindle detector based on WT,
  STFT and WMSD}.
\newblock \bibinfo{journal}{\emph{International Journal of Biomedical and
  Biological Engineering}} \bibinfo{volume}{6}, \bibinfo{number}{8}
  (\bibinfo{year}{2012}), \bibinfo{pages}{397--400}.
\newblock


\bibitem[\protect\citeauthoryear{De~Gennaro and Ferrara}{De~Gennaro and
  Ferrara}{2003}]%
        {de2003sleep}
\bibfield{author}{\bibinfo{person}{Luigi De~Gennaro} {and}
  \bibinfo{person}{Michele Ferrara}.} \bibinfo{year}{2003}\natexlab{}.
\newblock \bibinfo{title}{Sleep spindles: an overview}.
\newblock
\newblock


\bibitem[\protect\citeauthoryear{de~Zambotti, Goldstone, Claudatos, Colrain,
  and Baker}{de~Zambotti et~al\mbox{.}}{2018}]%
        {fitbit_psg}
\bibfield{author}{\bibinfo{person}{Massimiliano de Zambotti},
  \bibinfo{person}{Aimee Goldstone}, \bibinfo{person}{Stephanie Claudatos},
  \bibinfo{person}{Ian~M. Colrain}, {and} \bibinfo{person}{Fiona~C. Baker}.}
  \bibinfo{year}{2018}\natexlab{}.
\newblock \showarticletitle{A validation study of Fitbit Charge 2™ compared
  with polysomnography in adults}.
\newblock \bibinfo{journal}{\emph{Chronobiology International}}
  \bibinfo{volume}{35}, \bibinfo{number}{4} (\bibinfo{year}{2018}),
  \bibinfo{pages}{465--476}.
\newblock
\urldef\tempurl%
\url{https://doi.org/10.1080/07420528.2017.1413578}
\showDOI{\tempurl}
\showeprint{https://doi.org/10.1080/07420528.2017.1413578}
\newblock
\shownote{PMID: 29235907.}


\bibitem[\protect\citeauthoryear{de~Zambotti, Rosas, Colrain, and
  Baker}{de~Zambotti et~al\mbox{.}}{2019}]%
        {oura_psg}
\bibfield{author}{\bibinfo{person}{Massimiliano de Zambotti},
  \bibinfo{person}{Leonardo Rosas}, \bibinfo{person}{Ian~M. Colrain}, {and}
  \bibinfo{person}{Fiona~C. Baker}.} \bibinfo{year}{2019}\natexlab{}.
\newblock \showarticletitle{The Sleep of the Ring: Comparison of the ŌURA
  Sleep Tracker Against Polysomnography}.
\newblock \bibinfo{journal}{\emph{Behavioral Sleep Medicine}}
  \bibinfo{volume}{17}, \bibinfo{number}{2} (\bibinfo{year}{2019}),
  \bibinfo{pages}{124--136}.
\newblock
\urldef\tempurl%
\url{https://doi.org/10.1080/15402002.2017.1300587}
\showDOI{\tempurl}
\showeprint{https://doi.org/10.1080/15402002.2017.1300587}
\newblock
\shownote{PMID: 28323455.}


\bibitem[\protect\citeauthoryear{Deng, Dong, Wang, Fang, Liu, Yu, Liu, and
  Chen}{Deng et~al\mbox{.}}{2018}]%
        {deng2018design}
\bibfield{author}{\bibinfo{person}{Fei Deng}, \bibinfo{person}{Jianwu Dong},
  \bibinfo{person}{Xiangyu Wang}, \bibinfo{person}{Ying Fang},
  \bibinfo{person}{Yu Liu}, \bibinfo{person}{Zhaofei Yu}, \bibinfo{person}{Jing
  Liu}, {and} \bibinfo{person}{Feng Chen}.} \bibinfo{year}{2018}\natexlab{}.
\newblock \showarticletitle{Design and implementation of a noncontact sleep
  monitoring system using infrared cameras and motion sensor}.
\newblock \bibinfo{journal}{\emph{IEEE Transactions on Instrumentation and
  Measurement}} \bibinfo{volume}{67}, \bibinfo{number}{7}
  (\bibinfo{year}{2018}), \bibinfo{pages}{1555--1563}.
\newblock


\bibitem[\protect\citeauthoryear{El~Helou, Navarro, Depienne, Fedirko, LeGuern,
  Baulac, An-Gourfinkel, and Adam}{El~Helou et~al\mbox{.}}{2008}]%
        {el2008k}
\bibfield{author}{\bibinfo{person}{J El~Helou}, \bibinfo{person}{V Navarro},
  \bibinfo{person}{C Depienne}, \bibinfo{person}{E Fedirko}, \bibinfo{person}{E
  LeGuern}, \bibinfo{person}{M Baulac}, \bibinfo{person}{I An-Gourfinkel},
  {and} \bibinfo{person}{C Adam}.} \bibinfo{year}{2008}\natexlab{}.
\newblock \showarticletitle{K-complex-induced seizures in autosomal dominant
  nocturnal frontal lobe epilepsy}.
\newblock \bibinfo{journal}{\emph{Clinical neurophysiology}}
  \bibinfo{volume}{119}, \bibinfo{number}{10} (\bibinfo{year}{2008}),
  \bibinfo{pages}{2201--2204}.
\newblock


\bibitem[\protect\citeauthoryear{Ferman, Boeve, Smith, Silber, Kokmen,
  Petersen, and Ivnik}{Ferman et~al\mbox{.}}{1999}]%
        {ferman1999rem}
\bibfield{author}{\bibinfo{person}{Tanis~J Ferman}, \bibinfo{person}{BF Boeve},
  \bibinfo{person}{GE Smith}, \bibinfo{person}{MH Silber}, \bibinfo{person}{E
  Kokmen}, \bibinfo{person}{RC Petersen}, {and} \bibinfo{person}{RJ Ivnik}.}
  \bibinfo{year}{1999}\natexlab{}.
\newblock \showarticletitle{REM sleep behavior disorder and dementia: cognitive
  differences when compared with AD}.
\newblock \bibinfo{journal}{\emph{Neurology}} \bibinfo{volume}{52},
  \bibinfo{number}{5} (\bibinfo{year}{1999}), \bibinfo{pages}{951--951}.
\newblock


\bibitem[\protect\citeauthoryear{Ghaderi, Moradkhani, Haghighatfard, Akrami,
  Khayyer, and Balc{\i}}{Ghaderi et~al\mbox{.}}{2018}]%
        {coherence1}
\bibfield{author}{\bibinfo{person}{Amir~Hossein Ghaderi},
  \bibinfo{person}{Shadi Moradkhani}, \bibinfo{person}{Arvin Haghighatfard},
  \bibinfo{person}{Fatemeh Akrami}, \bibinfo{person}{Zahra Khayyer}, {and}
  \bibinfo{person}{Fuat Balc{\i}}.} \bibinfo{year}{2018}\natexlab{}.
\newblock \showarticletitle{Time estimation and beta segregation: An EEG study
  and graph theoretical approach}.
\newblock \bibinfo{journal}{\emph{PLoS One}} \bibinfo{volume}{13},
  \bibinfo{number}{4} (\bibinfo{year}{2018}), \bibinfo{pages}{e0195380}.
\newblock


\bibitem[\protect\citeauthoryear{Golparvar and Yapici}{Golparvar and
  Yapici}{2018}]%
        {golparvar2018electrooculography}
\bibfield{author}{\bibinfo{person}{Ata~Jedari Golparvar} {and}
  \bibinfo{person}{Murat~Kaya Yapici}.} \bibinfo{year}{2018}\natexlab{}.
\newblock \showarticletitle{Electrooculography by wearable graphene textiles}.
\newblock \bibinfo{journal}{\emph{IEEE Sensors Journal}} \bibinfo{volume}{18},
  \bibinfo{number}{21} (\bibinfo{year}{2018}), \bibinfo{pages}{8971--8978}.
\newblock


\bibitem[\protect\citeauthoryear{Hassan, Malik, Fofi, Saad, Karasfi, Ali, and
  Meriaudeau}{Hassan et~al\mbox{.}}{2017}]%
        {hassan2017heart}
\bibfield{author}{\bibinfo{person}{Mohamed~Abul Hassan},
  \bibinfo{person}{Aamir~Saeed Malik}, \bibinfo{person}{David Fofi},
  \bibinfo{person}{Naufal Saad}, \bibinfo{person}{Babak Karasfi},
  \bibinfo{person}{Yasir~Salih Ali}, {and} \bibinfo{person}{Fabrice
  Meriaudeau}.} \bibinfo{year}{2017}\natexlab{}.
\newblock \showarticletitle{Heart rate estimation using facial video: A
  review}.
\newblock \bibinfo{journal}{\emph{Biomedical Signal Processing and Control}}
  \bibinfo{volume}{38} (\bibinfo{year}{2017}), \bibinfo{pages}{346--360}.
\newblock


\bibitem[\protect\citeauthoryear{Hennies, Ralph, Kempkes, Cousins, and
  Lewis}{Hennies et~al\mbox{.}}{2016}]%
        {hennies2016sleep}
\bibfield{author}{\bibinfo{person}{Nora Hennies}, \bibinfo{person}{Matthew
  A~Lambon Ralph}, \bibinfo{person}{Marleen Kempkes}, \bibinfo{person}{James~N
  Cousins}, {and} \bibinfo{person}{Penelope~A Lewis}.}
  \bibinfo{year}{2016}\natexlab{}.
\newblock \showarticletitle{Sleep spindle density predicts the effect of prior
  knowledge on memory consolidation}.
\newblock \bibinfo{journal}{\emph{Journal of Neuroscience}}
  \bibinfo{volume}{36}, \bibinfo{number}{13} (\bibinfo{year}{2016}),
  \bibinfo{pages}{3799--3810}.
\newblock


\bibitem[\protect\citeauthoryear{Homayounfar, Kiaghadi, Ganesan, and
  Andrew}{Homayounfar et~al\mbox{.}}{2021}]%
        {PressION}
\bibfield{author}{\bibinfo{person}{S.~Zohreh Homayounfar}, \bibinfo{person}{Ali
  Kiaghadi}, \bibinfo{person}{Deepak Ganesan}, {and} \bibinfo{person}{Trisha~L.
  Andrew}.} \bibinfo{year}{2021}\natexlab{}.
\newblock \showarticletitle{{PressION}: An All-Fabric Piezoionic Pressure
  Sensor for Extracting Physiological Metrics in Both Static and Dynamic
  Contexts}.
\newblock \bibinfo{journal}{\emph{Journal of the Electrochemical Society}}
  \bibinfo{volume}{168}, \bibinfo{number}{1} (\bibinfo{date}{jan}
  \bibinfo{year}{2021}), \bibinfo{pages}{017515}.
\newblock
\urldef\tempurl%
\url{https://doi.org/10.1149/1945-7111/abdc65}
\showDOI{\tempurl}


\bibitem[\protect\citeauthoryear{Homayounfar, Rostaminia, Kiaghadi, Chen,
  Alexander, Ganesan, and Andrew}{Homayounfar et~al\mbox{.}}{2020}]%
        {chesma}
\bibfield{author}{\bibinfo{person}{S.~Zohreh Homayounfar},
  \bibinfo{person}{Soha Rostaminia}, \bibinfo{person}{Ali Kiaghadi},
  \bibinfo{person}{Xingda Chen}, \bibinfo{person}{Emerson~T. Alexander},
  \bibinfo{person}{Deepak Ganesan}, {and} \bibinfo{person}{Trisha~L. Andrew}.}
  \bibinfo{year}{2020}\natexlab{}.
\newblock \showarticletitle{Multimodal Smart Eyewear for Longitudinal Eye
  Movement Tracking}.
\newblock \bibinfo{journal}{\emph{Matter}} \bibinfo{volume}{3},
  \bibinfo{number}{4} (\bibinfo{year}{2020}), \bibinfo{pages}{1275--1293}.
\newblock
\showISSN{2590-2385}
\urldef\tempurl%
\url{https://doi.org/10.1016/j.matt.2020.07.030}
\showDOI{\tempurl}


\bibitem[\protect\citeauthoryear{Huang, Shen, Long, Wu, Shih, Zheng, Yen, Tung,
  and Liu}{Huang et~al\mbox{.}}{1998}]%
        {huang1998empirical}
\bibfield{author}{\bibinfo{person}{Norden~E Huang}, \bibinfo{person}{Zheng
  Shen}, \bibinfo{person}{Steven~R Long}, \bibinfo{person}{Manli~C Wu},
  \bibinfo{person}{Hsing~H Shih}, \bibinfo{person}{Quanan Zheng},
  \bibinfo{person}{Nai-Chyuan Yen}, \bibinfo{person}{Chi~Chao Tung}, {and}
  \bibinfo{person}{Henry~H Liu}.} \bibinfo{year}{1998}\natexlab{}.
\newblock \showarticletitle{The empirical mode decomposition and the Hilbert
  spectrum for nonlinear and non-stationary time series analysis}.
\newblock \bibinfo{journal}{\emph{Proceedings of the Royal Society of London.
  Series A: mathematical, physical and engineering sciences}}
  \bibinfo{volume}{454}, \bibinfo{number}{1971} (\bibinfo{year}{1998}),
  \bibinfo{pages}{903--995}.
\newblock


\bibitem[\protect\citeauthoryear{Javaheri, Storfer-Isser, Rosen, and
  Redline}{Javaheri et~al\mbox{.}}{2008}]%
        {javaheri2008sleep}
\bibfield{author}{\bibinfo{person}{Sogol Javaheri}, \bibinfo{person}{Amy
  Storfer-Isser}, \bibinfo{person}{Carol~L Rosen}, {and} \bibinfo{person}{Susan
  Redline}.} \bibinfo{year}{2008}\natexlab{}.
\newblock \showarticletitle{Sleep quality and elevated blood pressure in
  adolescents}.
\newblock \bibinfo{journal}{\emph{Circulation}} \bibinfo{volume}{118},
  \bibinfo{number}{10} (\bibinfo{year}{2008}), \bibinfo{pages}{1034--1040}.
\newblock


\bibitem[\protect\citeauthoryear{Jegou, Schabus, Gosseries, Dahmen, Albouy,
  Desseilles, Sterpenich, Phillips, Maquet, Grova, et~al\mbox{.}}{Jegou
  et~al\mbox{.}}{2019}]%
        {jegou2019cortical}
\bibfield{author}{\bibinfo{person}{Aude Jegou}, \bibinfo{person}{Manuel
  Schabus}, \bibinfo{person}{Olivia Gosseries}, \bibinfo{person}{Brigitte
  Dahmen}, \bibinfo{person}{Genevi{\`e}ve Albouy}, \bibinfo{person}{Martin
  Desseilles}, \bibinfo{person}{Virginie Sterpenich},
  \bibinfo{person}{Christophe Phillips}, \bibinfo{person}{Pierre Maquet},
  \bibinfo{person}{Christophe Grova}, {et~al\mbox{.}}}
  \bibinfo{year}{2019}\natexlab{}.
\newblock \showarticletitle{Cortical reactivations during sleep spindles
  following declarative learning}.
\newblock \bibinfo{journal}{\emph{Neuroimage}}  \bibinfo{volume}{195}
  (\bibinfo{year}{2019}), \bibinfo{pages}{104--112}.
\newblock


\bibitem[\protect\citeauthoryear{Jia, Bonde, Li, Xu, Wang, Zhang, Howard, and
  Zhang}{Jia et~al\mbox{.}}{2017a}]%
        {Zhenhua:2017:sleepsharedbed}
\bibfield{author}{\bibinfo{person}{Zhenhua Jia}, \bibinfo{person}{Amelie
  Bonde}, \bibinfo{person}{Sugang Li}, \bibinfo{person}{Chenren Xu},
  \bibinfo{person}{Jingxian Wang}, \bibinfo{person}{Yanyong Zhang},
  \bibinfo{person}{Richard~E. Howard}, {and} \bibinfo{person}{Pei Zhang}.}
  \bibinfo{year}{2017}\natexlab{a}.
\newblock \showarticletitle{Monitoring a Person's Heart Rate and Respiratory
  Rate on a Shared Bed Using Geophones}. In
  \bibinfo{booktitle}{\emph{Proceedings of the 15th ACM Conference on Embedded
  Network Sensor Systems}} (Delft, Netherlands) \emph{(\bibinfo{series}{SenSys
  '17})}. \bibinfo{publisher}{Association for Computing Machinery},
  \bibinfo{address}{New York, NY, USA}, Article \bibinfo{articleno}{6},
  \bibinfo{numpages}{14}~pages.
\newblock
\showISBNx{9781450354592}
\urldef\tempurl%
\url{https://doi.org/10.1145/3131672.3131679}
\showDOI{\tempurl}


\bibitem[\protect\citeauthoryear{Jia, Bonde, Li, Xu, Wang, Zhang, Howard, and
  Zhang}{Jia et~al\mbox{.}}{2017b}]%
        {jia:2017}
\bibfield{author}{\bibinfo{person}{Zhenhua Jia}, \bibinfo{person}{Amelie
  Bonde}, \bibinfo{person}{Sugang Li}, \bibinfo{person}{Chenren Xu},
  \bibinfo{person}{Jingxian Wang}, \bibinfo{person}{Yanyong Zhang},
  \bibinfo{person}{Richard~E. Howard}, {and} \bibinfo{person}{Pei Zhang}.}
  \bibinfo{year}{2017}\natexlab{b}.
\newblock \showarticletitle{Monitoring a Person’s Heart Rate and Respiratory
  Rate on a Shared Bed Using Geophones}. In
  \bibinfo{booktitle}{\emph{Proceedings of the 15th ACM Conference on Embedded
  Network Sensor Systems}} (Delft, Netherlands) \emph{(\bibinfo{series}{SenSys
  ’17})}. \bibinfo{publisher}{Association for Computing Machinery},
  \bibinfo{address}{New York, NY, USA}, Article \bibinfo{articleno}{6},
  \bibinfo{numpages}{14}~pages.
\newblock
\showISBNx{9781450354592}
\urldef\tempurl%
\url{https://doi.org/10.1145/3131672.3131679}
\showDOI{\tempurl}


\bibitem[\protect\citeauthoryear{Kam, Parekh, Sharma, Andrade, Lewin, Castillo,
  Bubu, Chua, Miller, Mullins, et~al\mbox{.}}{Kam et~al\mbox{.}}{2019}]%
        {kam2019sleep}
\bibfield{author}{\bibinfo{person}{Korey Kam}, \bibinfo{person}{Ankit Parekh},
  \bibinfo{person}{Ram~A Sharma}, \bibinfo{person}{Andreia Andrade},
  \bibinfo{person}{Monica Lewin}, \bibinfo{person}{Bresne Castillo},
  \bibinfo{person}{Omonigho~M Bubu}, \bibinfo{person}{Nicholas~J Chua},
  \bibinfo{person}{Margo~D Miller}, \bibinfo{person}{Anna~E Mullins},
  {et~al\mbox{.}}} \bibinfo{year}{2019}\natexlab{}.
\newblock \showarticletitle{Sleep oscillation-specific associations with
  Alzheimer’s disease CSF biomarkers: novel roles for sleep spindles and
  tau}.
\newblock \bibinfo{journal}{\emph{Molecular neurodegeneration}}
  \bibinfo{volume}{14}, \bibinfo{number}{1} (\bibinfo{year}{2019}),
  \bibinfo{pages}{10}.
\newblock


\bibitem[\protect\citeauthoryear{Kang, Kang, Ko, Park, Mariani, and Weng}{Kang
  et~al\mbox{.}}{2017}]%
        {kang2017validity}
\bibfield{author}{\bibinfo{person}{Seung-Gul Kang}, \bibinfo{person}{Jae~Myeong
  Kang}, \bibinfo{person}{Kwang-Pil Ko}, \bibinfo{person}{Seon-Cheol Park},
  \bibinfo{person}{Sara Mariani}, {and} \bibinfo{person}{Jia Weng}.}
  \bibinfo{year}{2017}\natexlab{}.
\newblock \showarticletitle{Validity of a commercial wearable sleep tracker in
  adult insomnia disorder patients and good sleepers}.
\newblock \bibinfo{journal}{\emph{Journal of psychosomatic research}}
  \bibinfo{volume}{97} (\bibinfo{year}{2017}), \bibinfo{pages}{38--44}.
\newblock


\bibitem[\protect\citeauthoryear{Kasasbeh, Chi, and Krishnaswamy}{Kasasbeh
  et~al\mbox{.}}{2006}]%
        {kasasbeh2006inflammatory}
\bibfield{author}{\bibinfo{person}{E Kasasbeh}, \bibinfo{person}{David~S Chi},
  {and} \bibinfo{person}{G Krishnaswamy}.} \bibinfo{year}{2006}\natexlab{}.
\newblock \showarticletitle{Inflammatory aspects of sleep apnea and their
  cardiovascular consequences}.
\newblock \bibinfo{journal}{\emph{Southern medical journal}}
  \bibinfo{volume}{99}, \bibinfo{number}{1} (\bibinfo{year}{2006}),
  \bibinfo{pages}{58--68}.
\newblock


\bibitem[\protect\citeauthoryear{Kiaghadi, Homayounfar, Gummeson, Andrew, and
  Ganesan}{Kiaghadi et~al\mbox{.}}{2019}]%
        {Kiaghadi:2019}
\bibfield{author}{\bibinfo{person}{Ali Kiaghadi},
  \bibinfo{person}{Seyedeh~Zohreh Homayounfar}, \bibinfo{person}{Jeremy
  Gummeson}, \bibinfo{person}{Trisha Andrew}, {and} \bibinfo{person}{Deepak
  Ganesan}.} \bibinfo{year}{2019}\natexlab{}.
\newblock \showarticletitle{Phyjama: Physiological Sensing via Fiber-Enhanced
  Pyjamas}.
\newblock \bibinfo{journal}{\emph{Proc. ACM Interact. Mob. Wearable Ubiquitous
  Technol.}} \bibinfo{volume}{3}, \bibinfo{number}{3}, Article
  \bibinfo{articleno}{89} (\bibinfo{date}{Sept.} \bibinfo{year}{2019}),
  \bibinfo{numpages}{29}~pages.
\newblock
\urldef\tempurl%
\url{https://doi.org/10.1145/3351247}
\showDOI{\tempurl}


\bibitem[\protect\citeauthoryear{Kim, Lee, Yeom, Lee, Kim, and Kim}{Kim
  et~al\mbox{.}}{2020}]%
        {kim2020wearable}
\bibfield{author}{\bibinfo{person}{Sung-Woo Kim}, \bibinfo{person}{Kwangmuk
  Lee}, \bibinfo{person}{Junyeong Yeom}, \bibinfo{person}{Tae-Hoon Lee},
  \bibinfo{person}{Don-Han Kim}, {and} \bibinfo{person}{Jae~Joon Kim}.}
  \bibinfo{year}{2020}\natexlab{}.
\newblock \showarticletitle{Wearable Multi-Biosignal Analysis Integrated
  Interface With Direct Sleep-Stage Classification}.
\newblock \bibinfo{journal}{\emph{IEEE Access}}  \bibinfo{volume}{8}
  (\bibinfo{year}{2020}), \bibinfo{pages}{46131--46140}.
\newblock


\bibitem[\protect\citeauthoryear{Knutson, Ryden, Mander, and
  Van~Cauter}{Knutson et~al\mbox{.}}{2006}]%
        {knutson2006role}
\bibfield{author}{\bibinfo{person}{Kristen~L Knutson},
  \bibinfo{person}{Armand~M Ryden}, \bibinfo{person}{Bryce~A Mander}, {and}
  \bibinfo{person}{Eve Van~Cauter}.} \bibinfo{year}{2006}\natexlab{}.
\newblock \showarticletitle{Role of sleep duration and quality in the risk and
  severity of type 2 diabetes mellitus}.
\newblock \bibinfo{journal}{\emph{Archives of internal medicine}}
  \bibinfo{volume}{166}, \bibinfo{number}{16} (\bibinfo{year}{2006}),
  \bibinfo{pages}{1768--1774}.
\newblock


\bibitem[\protect\citeauthoryear{Li, Yadollahi, and Taati}{Li
  et~al\mbox{.}}{2016}]%
        {li2016noncontact}
\bibfield{author}{\bibinfo{person}{Michael~H Li}, \bibinfo{person}{Azadeh
  Yadollahi}, {and} \bibinfo{person}{Babak Taati}.}
  \bibinfo{year}{2016}\natexlab{}.
\newblock \showarticletitle{Noncontact vision-based cardiopulmonary monitoring
  in different sleeping positions}.
\newblock \bibinfo{journal}{\emph{IEEE journal of biomedical and health
  informatics}} \bibinfo{volume}{21}, \bibinfo{number}{5}
  (\bibinfo{year}{2016}), \bibinfo{pages}{1367--1375}.
\newblock


\bibitem[\protect\citeauthoryear{Liang, Kuo, Lee, Lin, Liu, Chen, Cherng, and
  Shaw}{Liang et~al\mbox{.}}{2015a}]%
        {eogmask}
\bibfield{author}{\bibinfo{person}{Sheng-Fu Liang}, \bibinfo{person}{Chih-En
  Kuo}, \bibinfo{person}{Yi-Chieh Lee}, \bibinfo{person}{Wen-Chieh Lin},
  \bibinfo{person}{Yen-Chen Liu}, \bibinfo{person}{Peng-Yu Chen},
  \bibinfo{person}{Fu-Yin Cherng}, {and} \bibinfo{person}{Fu Shaw}.}
  \bibinfo{year}{2015}\natexlab{a}.
\newblock \showarticletitle{Development of an EOG-Based Automatic
  Sleep-Monitoring Eye Mask}.
\newblock \bibinfo{journal}{\emph{IEEE Transactions on Instrumentation and
  Measurement}}  \bibinfo{volume}{64} (\bibinfo{date}{11}
  \bibinfo{year}{2015}), \bibinfo{pages}{1--1}.
\newblock
\urldef\tempurl%
\url{https://doi.org/10.1109/TIM.2015.2433652}
\showDOI{\tempurl}


\bibitem[\protect\citeauthoryear{Liang, Kuo, Lee, Lin, Liu, Chen, Cherng, and
  Shaw}{Liang et~al\mbox{.}}{2015b}]%
        {liang2015development}
\bibfield{author}{\bibinfo{person}{Sheng-Fu Liang}, \bibinfo{person}{Chin-En
  Kuo}, \bibinfo{person}{Yi-Chieh Lee}, \bibinfo{person}{Wen-Chieh Lin},
  \bibinfo{person}{Yen-Chen Liu}, \bibinfo{person}{Peng-Yu Chen},
  \bibinfo{person}{Fu-Yin Cherng}, {and} \bibinfo{person}{Fu-Zen Shaw}.}
  \bibinfo{year}{2015}\natexlab{b}.
\newblock \showarticletitle{Development of an EOG-based automatic
  sleep-monitoring eye mask}.
\newblock \bibinfo{journal}{\emph{IEEE Transactions on Instrumentation and
  Measurement}} \bibinfo{volume}{64}, \bibinfo{number}{11}
  (\bibinfo{year}{2015}), \bibinfo{pages}{2977--2985}.
\newblock


\bibitem[\protect\citeauthoryear{Lin, Liao, Liu, Wang, Lin, and Chang}{Lin
  et~al\mbox{.}}{2011}]%
        {lin2011novel}
\bibfield{author}{\bibinfo{person}{Chin-Teng Lin}, \bibinfo{person}{Lun-De
  Liao}, \bibinfo{person}{Yu-Hang Liu}, \bibinfo{person}{I-Jan Wang},
  \bibinfo{person}{Bor-Shyh Lin}, {and} \bibinfo{person}{Jyh-Yeong Chang}.}
  \bibinfo{year}{2011}\natexlab{}.
\newblock \showarticletitle{Novel dry polymer foam electrodes for long-term EEG
  measurement}.
\newblock \bibinfo{journal}{\emph{IEEE Transactions on Biomedical Engineering}}
  \bibinfo{volume}{58}, \bibinfo{number}{5} (\bibinfo{year}{2011}),
  \bibinfo{pages}{1200--1207}.
\newblock


\bibitem[\protect\citeauthoryear{Liu, Wang, Chen, Yang, Chen, and Cheng}{Liu
  et~al\mbox{.}}{2015a}]%
        {Liu:2015}
\bibfield{author}{\bibinfo{person}{Jian Liu}, \bibinfo{person}{Yan Wang},
  \bibinfo{person}{Yingying Chen}, \bibinfo{person}{Jie Yang},
  \bibinfo{person}{Xu Chen}, {and} \bibinfo{person}{Jerry Cheng}.}
  \bibinfo{year}{2015}\natexlab{a}.
\newblock \showarticletitle{Tracking Vital Signs During Sleep Leveraging
  Off-the-Shelf WiFi}. In \bibinfo{booktitle}{\emph{Proceedings of the 16th ACM
  International Symposium on Mobile Ad Hoc Networking and Computing}}
  (Hangzhou, China) \emph{(\bibinfo{series}{MobiHoc '15})}.
  \bibinfo{publisher}{Association for Computing Machinery},
  \bibinfo{address}{New York, NY, USA}, \bibinfo{pages}{267–276}.
\newblock
\showISBNx{9781450334891}
\urldef\tempurl%
\url{https://doi.org/10.1145/2746285.2746303}
\showDOI{\tempurl}


\bibitem[\protect\citeauthoryear{Liu, Wang, Chen, Yang, Chen, and Cheng}{Liu
  et~al\mbox{.}}{2015b}]%
        {liu2015tracking}
\bibfield{author}{\bibinfo{person}{Jian Liu}, \bibinfo{person}{Yan Wang},
  \bibinfo{person}{Yingying Chen}, \bibinfo{person}{Jie Yang},
  \bibinfo{person}{Xu Chen}, {and} \bibinfo{person}{Jerry Cheng}.}
  \bibinfo{year}{2015}\natexlab{b}.
\newblock \showarticletitle{Tracking vital signs during sleep leveraging
  off-the-shelf wifi}. In \bibinfo{booktitle}{\emph{Proceedings of the 16th ACM
  International Symposium on Mobile Ad Hoc Networking and Computing}}.
  \bibinfo{pages}{267--276}.
\newblock


\bibitem[\protect\citeauthoryear{Malinowska, Durka, Blinowska, Szelenberger,
  and Wakarow}{Malinowska et~al\mbox{.}}{2006}]%
        {malinowska2006micro}
\bibfield{author}{\bibinfo{person}{Urszula Malinowska},
  \bibinfo{person}{Piotr~J Durka}, \bibinfo{person}{Katarzyna~J Blinowska},
  \bibinfo{person}{Waldemar Szelenberger}, {and} \bibinfo{person}{Andrzej
  Wakarow}.} \bibinfo{year}{2006}\natexlab{}.
\newblock \showarticletitle{Micro-and macrostructure of sleep EEG}.
\newblock \bibinfo{journal}{\emph{IEEE engineering in medicine and biology
  magazine}} \bibinfo{volume}{25}, \bibinfo{number}{4} (\bibinfo{year}{2006}),
  \bibinfo{pages}{26--31}.
\newblock


\bibitem[\protect\citeauthoryear{Martinez and Stiefelhagen}{Martinez and
  Stiefelhagen}{2012}]%
        {martinez2012breath}
\bibfield{author}{\bibinfo{person}{Manuel Martinez} {and}
  \bibinfo{person}{Rainer Stiefelhagen}.} \bibinfo{year}{2012}\natexlab{}.
\newblock \showarticletitle{Breath rate monitoring during sleep using near-IR
  imagery and PCA}. In \bibinfo{booktitle}{\emph{Proceedings of the 21st
  International Conference on Pattern Recognition (ICPR2012)}}. IEEE,
  \bibinfo{pages}{3472--3475}.
\newblock


\bibitem[\protect\citeauthoryear{Miller, Lastella, Scanlan, Bellenger, Halson,
  Roach, and Sargent}{Miller et~al\mbox{.}}{2020}]%
        {whoop:evaluation:2020}
\bibfield{author}{\bibinfo{person}{Dean~J. Miller}, \bibinfo{person}{Michele
  Lastella}, \bibinfo{person}{Aaron~T. Scanlan}, \bibinfo{person}{Clint
  Bellenger}, \bibinfo{person}{Shona~L. Halson}, \bibinfo{person}{Gregory~D.
  Roach}, {and} \bibinfo{person}{Charli Sargent}.}
  \bibinfo{year}{2020}\natexlab{}.
\newblock \showarticletitle{A validation study of the WHOOP strap against
  polysomnography to assess sleep}.
\newblock \bibinfo{journal}{\emph{Journal of Sports Sciences}}
  \bibinfo{volume}{38}, \bibinfo{number}{22} (\bibinfo{year}{2020}),
  \bibinfo{pages}{2631--2636}.
\newblock
\urldef\tempurl%
\url{https://doi.org/10.1080/02640414.2020.1797448}
\showDOI{\tempurl}
\showeprint{https://doi.org/10.1080/02640414.2020.1797448}
\newblock
\shownote{PMID: 32713257.}


\bibitem[\protect\citeauthoryear{Moreno-Pino, Porras-Segovia,
  L{\'o}pez-Esteban, Art{\'e}s, and Baca-Garc{\'\i}a}{Moreno-Pino
  et~al\mbox{.}}{2019}]%
        {moreno2019validation}
\bibfield{author}{\bibinfo{person}{Fernando Moreno-Pino},
  \bibinfo{person}{Alejandro Porras-Segovia}, \bibinfo{person}{Pilar
  L{\'o}pez-Esteban}, \bibinfo{person}{Antonio Art{\'e}s}, {and}
  \bibinfo{person}{Enrique Baca-Garc{\'\i}a}.} \bibinfo{year}{2019}\natexlab{}.
\newblock \showarticletitle{Validation of Fitbit charge 2 and Fitbit Alta HR
  against polysomnography for assessing sleep in adults with obstructive sleep
  apnea}.
\newblock \bibinfo{journal}{\emph{Journal of Clinical Sleep Medicine}}
  \bibinfo{volume}{15}, \bibinfo{number}{11} (\bibinfo{year}{2019}),
  \bibinfo{pages}{1645--1653}.
\newblock


\bibitem[\protect\citeauthoryear{Niu, Zhang, Xiong, Li, Yi, and Zhang}{Niu
  et~al\mbox{.}}{2018}]%
        {Niu:2018}
\bibfield{author}{\bibinfo{person}{Kai Niu}, \bibinfo{person}{Fusang Zhang},
  \bibinfo{person}{Jie Xiong}, \bibinfo{person}{Xiang Li},
  \bibinfo{person}{Enze Yi}, {and} \bibinfo{person}{Daqing Zhang}.}
  \bibinfo{year}{2018}\natexlab{}.
\newblock \showarticletitle{Boosting Fine-Grained Activity Sensing by Embracing
  Wireless Multipath Effects}. In \bibinfo{booktitle}{\emph{Proceedings of the
  14th International Conference on Emerging Networking EXperiments and
  Technologies}} (Heraklion, Greece) \emph{(\bibinfo{series}{CoNEXT ’18})}.
  \bibinfo{publisher}{Association for Computing Machinery},
  \bibinfo{address}{New York, NY, USA}, \bibinfo{pages}{139–151}.
\newblock
\showISBNx{9781450360807}
\urldef\tempurl%
\url{https://doi.org/10.1145/3281411.3281425}
\showDOI{\tempurl}


\bibitem[\protect\citeauthoryear{Nochino, Ohno, Kato, Taniike, and
  Okada}{Nochino et~al\mbox{.}}{2019}]%
        {nochino2019sleep}
\bibfield{author}{\bibinfo{person}{Teruaki Nochino}, \bibinfo{person}{Yuko
  Ohno}, \bibinfo{person}{Takafumi Kato}, \bibinfo{person}{Masako Taniike},
  {and} \bibinfo{person}{Shima Okada}.} \bibinfo{year}{2019}\natexlab{}.
\newblock \showarticletitle{Sleep stage estimation method using a camera for
  home use}.
\newblock \bibinfo{journal}{\emph{Biomedical engineering letters}}
  \bibinfo{volume}{9}, \bibinfo{number}{2} (\bibinfo{year}{2019}),
  \bibinfo{pages}{257--265}.
\newblock


\bibitem[\protect\citeauthoryear{Pani, Dess{\`\i}, Saenz-Cogollo, Barabino,
  Fraboni, and Bonfiglio}{Pani et~al\mbox{.}}{2015}]%
        {pani2015fully}
\bibfield{author}{\bibinfo{person}{Danilo Pani}, \bibinfo{person}{Alessia
  Dess{\`\i}}, \bibinfo{person}{Jose~F Saenz-Cogollo},
  \bibinfo{person}{Gianluca Barabino}, \bibinfo{person}{Beatrice Fraboni},
  {and} \bibinfo{person}{Annalisa Bonfiglio}.} \bibinfo{year}{2015}\natexlab{}.
\newblock \showarticletitle{Fully textile, PEDOT: PSS based electrodes for
  wearable ECG monitoring systems}.
\newblock \bibinfo{journal}{\emph{IEEE Transactions on Biomedical Engineering}}
  \bibinfo{volume}{63}, \bibinfo{number}{3} (\bibinfo{year}{2015}),
  \bibinfo{pages}{540--549}.
\newblock


\bibitem[\protect\citeauthoryear{Paradiso, Loriga, and Taccini}{Paradiso
  et~al\mbox{.}}{2005}]%
        {paradiso2005wearable}
\bibfield{author}{\bibinfo{person}{Rita Paradiso}, \bibinfo{person}{Giannicola
  Loriga}, {and} \bibinfo{person}{Nicola Taccini}.}
  \bibinfo{year}{2005}\natexlab{}.
\newblock \showarticletitle{A wearable health care system based on knitted
  integrated sensors}.
\newblock \bibinfo{journal}{\emph{IEEE transactions on Information Technology
  in biomedicine}} \bibinfo{volume}{9}, \bibinfo{number}{3}
  (\bibinfo{year}{2005}), \bibinfo{pages}{337--344}.
\newblock


\bibitem[\protect\citeauthoryear{Patti, Chaparro-Vargas, and Cvetkovic}{Patti
  et~al\mbox{.}}{2014}]%
        {patti2014automated}
\bibfield{author}{\bibinfo{person}{Chanakya~Reddy Patti},
  \bibinfo{person}{Ramiro Chaparro-Vargas}, {and} \bibinfo{person}{Dean
  Cvetkovic}.} \bibinfo{year}{2014}\natexlab{}.
\newblock \showarticletitle{Automated sleep spindle detection using novel EEG
  features and mixture models}. In \bibinfo{booktitle}{\emph{2014 36th Annual
  International Conference of the IEEE Engineering in Medicine and Biology
  Society}}. IEEE, \bibinfo{pages}{2221--2224}.
\newblock


\bibitem[\protect\citeauthoryear{Rahman, Adams, Ravichandran, Zhang, Patel,
  Kientz, and Choudhury}{Rahman et~al\mbox{.}}{2015}]%
        {Rahman:2015}
\bibfield{author}{\bibinfo{person}{Tauhidur Rahman},
  \bibinfo{person}{Alexander~T. Adams}, \bibinfo{person}{Ruth~Vinisha
  Ravichandran}, \bibinfo{person}{Mi Zhang}, \bibinfo{person}{Shwetak~N.
  Patel}, \bibinfo{person}{Julie~A. Kientz}, {and} \bibinfo{person}{Tanzeem
  Choudhury}.} \bibinfo{year}{2015}\natexlab{}.
\newblock \showarticletitle{DoppleSleep: A Contactless Unobtrusive Sleep
  Sensing System Using Short-Range Doppler Radar}. In
  \bibinfo{booktitle}{\emph{Proceedings of the 2015 ACM International Joint
  Conference on Pervasive and Ubiquitous Computing}} (Osaka, Japan)
  \emph{(\bibinfo{series}{UbiComp ’15})}. \bibinfo{publisher}{Association for
  Computing Machinery}, \bibinfo{address}{New York, NY, USA},
  \bibinfo{pages}{39–50}.
\newblock
\showISBNx{9781450335744}
\urldef\tempurl%
\url{https://doi.org/10.1145/2750858.2804280}
\showDOI{\tempurl}


\bibitem[\protect\citeauthoryear{Ranjan, Arya, Fernandes, Sravya, and
  Jain}{Ranjan et~al\mbox{.}}{2018}]%
        {ranjan2018fuzzy}
\bibfield{author}{\bibinfo{person}{Rakesh Ranjan}, \bibinfo{person}{Rajeev
  Arya}, \bibinfo{person}{Steven~Lawrence Fernandes}, \bibinfo{person}{Erukonda
  Sravya}, {and} \bibinfo{person}{Vinay Jain}.}
  \bibinfo{year}{2018}\natexlab{}.
\newblock \showarticletitle{A fuzzy neural network approach for automatic
  K-complex detection in sleep EEG signal}.
\newblock \bibinfo{journal}{\emph{Pattern Recognition Letters}}
  \bibinfo{volume}{115} (\bibinfo{year}{2018}), \bibinfo{pages}{74--83}.
\newblock


\bibitem[\protect\citeauthoryear{Ravesloot, Van~Maanen, Dun, and
  De~Vries}{Ravesloot et~al\mbox{.}}{2013}]%
        {ravesloot2013undervalued}
\bibfield{author}{\bibinfo{person}{MJL Ravesloot}, \bibinfo{person}{JP
  Van~Maanen}, \bibinfo{person}{L Dun}, {and} \bibinfo{person}{N De~Vries}.}
  \bibinfo{year}{2013}\natexlab{}.
\newblock \showarticletitle{The undervalued potential of positional therapy in
  position-dependent snoring and obstructive sleep apnea—a review of the
  literature}.
\newblock \bibinfo{journal}{\emph{Sleep and Breathing}} \bibinfo{volume}{17},
  \bibinfo{number}{1} (\bibinfo{year}{2013}), \bibinfo{pages}{39--49}.
\newblock


\bibitem[\protect\citeauthoryear{Roebuck, Monasterio, Gederi, Osipov, Behar,
  Malhotra, Penzel, and Clifford}{Roebuck et~al\mbox{.}}{2013}]%
        {roebuck2013review_sleepfeatures}
\bibfield{author}{\bibinfo{person}{A Roebuck}, \bibinfo{person}{V Monasterio},
  \bibinfo{person}{E Gederi}, \bibinfo{person}{M Osipov}, \bibinfo{person}{J
  Behar}, \bibinfo{person}{A Malhotra}, \bibinfo{person}{T Penzel}, {and}
  \bibinfo{person}{GD Clifford}.} \bibinfo{year}{2013}\natexlab{}.
\newblock \showarticletitle{A review of signals used in sleep analysis}.
\newblock \bibinfo{journal}{\emph{Physiological measurement}}
  \bibinfo{volume}{35}, \bibinfo{number}{1} (\bibinfo{year}{2013}),
  \bibinfo{pages}{R1}.
\newblock


\bibitem[\protect\citeauthoryear{Rostaminia, Lamson, Maji, Rahman, and
  Ganesan}{Rostaminia et~al\mbox{.}}{2019}]%
        {wince}
\bibfield{author}{\bibinfo{person}{Soha Rostaminia}, \bibinfo{person}{Alexander
  Lamson}, \bibinfo{person}{Subhransu Maji}, \bibinfo{person}{Tauhidur Rahman},
  {and} \bibinfo{person}{Deepak Ganesan}.} \bibinfo{year}{2019}\natexlab{}.
\newblock \showarticletitle{W!NCE: Unobtrusive Sensing of Upper Facial Action
  Units with EOG-Based Eyewear}.
\newblock \bibinfo{journal}{\emph{Proc. ACM Interact. Mob. Wearable Ubiquitous
  Technol.}} \bibinfo{volume}{3}, \bibinfo{number}{1}, Article
  \bibinfo{articleno}{23} (\bibinfo{date}{March} \bibinfo{year}{2019}),
  \bibinfo{numpages}{26}~pages.
\newblock
\urldef\tempurl%
\url{https://doi.org/10.1145/3314410}
\showDOI{\tempurl}


\bibitem[\protect\citeauthoryear{Sahito, Sun, Arbab, Qadir, and Jeong}{Sahito
  et~al\mbox{.}}{2015}]%
        {sahito2015graphene}
\bibfield{author}{\bibinfo{person}{Iftikhar~Ali Sahito},
  \bibinfo{person}{Kyung~Chul Sun}, \bibinfo{person}{Alvira~Ayoub Arbab},
  \bibinfo{person}{Muhammad~Bilal Qadir}, {and} \bibinfo{person}{Sung~Hoon
  Jeong}.} \bibinfo{year}{2015}\natexlab{}.
\newblock \showarticletitle{Graphene coated cotton fabric as textile structured
  counter electrode for DSSC}.
\newblock \bibinfo{journal}{\emph{Electrochimica Acta}}  \bibinfo{volume}{173}
  (\bibinfo{year}{2015}), \bibinfo{pages}{164--171}.
\newblock


\bibitem[\protect\citeauthoryear{Satapathy, Dehuri, Jagadev, and
  Mishra}{Satapathy et~al\mbox{.}}{2019}]%
        {SATAPATHY20191}
\bibfield{author}{\bibinfo{person}{Sandeep~Kumar Satapathy},
  \bibinfo{person}{Satchidananda Dehuri}, \bibinfo{person}{Alok~Kumar Jagadev},
  {and} \bibinfo{person}{Shruti Mishra}.} \bibinfo{year}{2019}\natexlab{}.
\newblock \showarticletitle{Chapter 1 - Introduction}.
\newblock In \bibinfo{booktitle}{\emph{EEG Brain Signal Classification for
  Epileptic Seizure Disorder Detection}},
  \bibfield{editor}{\bibinfo{person}{Sandeep~Kumar Satapathy},
  \bibinfo{person}{Satchidananda Dehuri}, \bibinfo{person}{Alok~Kumar Jagadev},
  {and} \bibinfo{person}{Shruti Mishra}} (Eds.). \bibinfo{publisher}{Academic
  Press}, \bibinfo{pages}{1--25}.
\newblock
\showISBNx{978-0-12-817426-5}
\urldef\tempurl%
\url{https://doi.org/10.1016/B978-0-12-817426-5.00001-6}
\showDOI{\tempurl}


\bibitem[\protect\citeauthoryear{Sch{\"o}nwald, Emerson, Rossatto, Chaves, and
  Gerhardt}{Sch{\"o}nwald et~al\mbox{.}}{2006}]%
        {schonwald2006benchmarking}
\bibfield{author}{\bibinfo{person}{Suzana~V Sch{\"o}nwald}, \bibinfo{person}{L
  Emerson}, \bibinfo{person}{Roberto Rossatto}, \bibinfo{person}{M{\'a}rcia~LF
  Chaves}, {and} \bibinfo{person}{G{\"u}nther~JL Gerhardt}.}
  \bibinfo{year}{2006}\natexlab{}.
\newblock \showarticletitle{Benchmarking matching pursuit to find sleep
  spindles}.
\newblock \bibinfo{journal}{\emph{Journal of neuroscience methods}}
  \bibinfo{volume}{156}, \bibinfo{number}{1-2} (\bibinfo{year}{2006}),
  \bibinfo{pages}{314--321}.
\newblock


\bibitem[\protect\citeauthoryear{Schwartz, Kohler, and Karatinos}{Schwartz
  et~al\mbox{.}}{2005}]%
        {schwartz2005symptoms}
\bibfield{author}{\bibinfo{person}{Daniel~J Schwartz},
  \bibinfo{person}{William~C Kohler}, {and} \bibinfo{person}{Gillian
  Karatinos}.} \bibinfo{year}{2005}\natexlab{}.
\newblock \showarticletitle{Symptoms of depression in individuals with
  obstructive sleep apnea may be amenable to treatment with continuous positive
  airway pressure}.
\newblock \bibinfo{journal}{\emph{Chest}} \bibinfo{volume}{128},
  \bibinfo{number}{3} (\bibinfo{year}{2005}), \bibinfo{pages}{1304--1309}.
\newblock


\bibitem[\protect\citeauthoryear{Schwarz, {\AA}kerstedt, Lindberg, Gruber,
  Fischer, and Theorell-Hagl{\"o}w}{Schwarz et~al\mbox{.}}{2017}]%
        {schwarz2017age}
\bibfield{author}{\bibinfo{person}{Johanna~FA Schwarz},
  \bibinfo{person}{Torbj{\"o}rn {\AA}kerstedt}, \bibinfo{person}{Eva Lindberg},
  \bibinfo{person}{Georg Gruber}, \bibinfo{person}{H{\aa}kan Fischer}, {and}
  \bibinfo{person}{Jenny Theorell-Hagl{\"o}w}.}
  \bibinfo{year}{2017}\natexlab{}.
\newblock \showarticletitle{Age affects sleep microstructure more than sleep
  macrostructure}.
\newblock \bibinfo{journal}{\emph{Journal of sleep research}}
  \bibinfo{volume}{26}, \bibinfo{number}{3} (\bibinfo{year}{2017}),
  \bibinfo{pages}{277--287}.
\newblock


\bibitem[\protect\citeauthoryear{Shensa}{Shensa}{1992}]%
        {shensa1992discrete}
\bibfield{author}{\bibinfo{person}{Mark~J Shensa}.}
  \bibinfo{year}{1992}\natexlab{}.
\newblock \showarticletitle{The discrete wavelet transform: wedding the a trous
  and Mallat algorithms}.
\newblock \bibinfo{journal}{\emph{IEEE Transactions on signal processing}}
  \bibinfo{volume}{40}, \bibinfo{number}{10} (\bibinfo{year}{1992}),
  \bibinfo{pages}{2464--2482}.
\newblock


\bibitem[\protect\citeauthoryear{Shu, Xu, and Xu}{Shu et~al\mbox{.}}{2019}]%
        {shu2019multilayer}
\bibfield{author}{\bibinfo{person}{Lin Shu}, \bibinfo{person}{Tianyuan Xu},
  {and} \bibinfo{person}{Xiangmin Xu}.} \bibinfo{year}{2019}\natexlab{}.
\newblock \showarticletitle{Multilayer sweat-absorbable textile electrode for
  EEG measurement in forehead site}.
\newblock \bibinfo{journal}{\emph{IEEE Sensors Journal}} \bibinfo{volume}{19},
  \bibinfo{number}{15} (\bibinfo{year}{2019}), \bibinfo{pages}{5995--6005}.
\newblock


\bibitem[\protect\citeauthoryear{Staner, Cornette, Maurice, Viardot, Bon, Haba,
  Staner, Luthringer, Muzet, and Macher}{Staner et~al\mbox{.}}{2003}]%
        {staner2003sleep}
\bibfield{author}{\bibinfo{person}{Luc Staner},
  \bibinfo{person}{Fran{\c{c}}oise Cornette}, \bibinfo{person}{Damien Maurice},
  \bibinfo{person}{Geoffrey Viardot}, \bibinfo{person}{Olivier~Le Bon},
  \bibinfo{person}{Jos{\'e} Haba}, \bibinfo{person}{Corinne Staner},
  \bibinfo{person}{R{\'e}my Luthringer}, \bibinfo{person}{Alain Muzet}, {and}
  \bibinfo{person}{Jean-Paul Macher}.} \bibinfo{year}{2003}\natexlab{}.
\newblock \showarticletitle{Sleep microstructure around sleep onset
  differentiates major depressive insomnia from primary insomnia}.
\newblock \bibinfo{journal}{\emph{Journal of sleep research}}
  \bibinfo{volume}{12}, \bibinfo{number}{4} (\bibinfo{year}{2003}),
  \bibinfo{pages}{319--330}.
\newblock


\bibitem[\protect\citeauthoryear{Taheri}{Taheri}{2006}]%
        {taheri2006link}
\bibfield{author}{\bibinfo{person}{Shahrad Taheri}.}
  \bibinfo{year}{2006}\natexlab{}.
\newblock \showarticletitle{The link between short sleep duration and obesity:
  we should recommend more sleep to prevent obesity}.
\newblock \bibinfo{journal}{\emph{Archives of disease in childhood}}
  \bibinfo{volume}{91}, \bibinfo{number}{11} (\bibinfo{year}{2006}),
  \bibinfo{pages}{881--884}.
\newblock


\bibitem[\protect\citeauthoryear{Terzano and Parrino}{Terzano and
  Parrino}{2000}]%
        {terzano2000origin}
\bibfield{author}{\bibinfo{person}{Mario~Giovanni Terzano} {and}
  \bibinfo{person}{Liborio Parrino}.} \bibinfo{year}{2000}\natexlab{}.
\newblock \showarticletitle{Origin and significance of the cyclic alternating
  pattern (CAP)}.
\newblock \bibinfo{journal}{\emph{Sleep medicine reviews}} \bibinfo{volume}{4},
  \bibinfo{number}{1} (\bibinfo{year}{2000}), \bibinfo{pages}{101--123}.
\newblock


\bibitem[\protect\citeauthoryear{Thatcher}{Thatcher}{2012}]%
        {coherence2}
\bibfield{author}{\bibinfo{person}{Robert~W. Thatcher}.}
  \bibinfo{year}{2012}\natexlab{}.
\newblock \showarticletitle{Coherence, Phase Differences, Phase Shift, and
  Phase Lock in EEG/ERP Analyses}.
\newblock \bibinfo{journal}{\emph{Developmental Neuropsychology}}
  \bibinfo{volume}{37}, \bibinfo{number}{6} (\bibinfo{year}{2012}),
  \bibinfo{pages}{476--496}.
\newblock
\urldef\tempurl%
\url{https://doi.org/10.1080/87565641.2011.619241}
\showDOI{\tempurl}
\showeprint{https://doi.org/10.1080/87565641.2011.619241}
\newblock
\shownote{PMID: 22889341.}


\bibitem[\protect\citeauthoryear{Vogels, Van~Gastel, Wang, and De~Haan}{Vogels
  et~al\mbox{.}}{2018}]%
        {vogels2018fully}
\bibfield{author}{\bibinfo{person}{Tom Vogels}, \bibinfo{person}{Mark
  Van~Gastel}, \bibinfo{person}{Wenjin Wang}, {and} \bibinfo{person}{Gerard
  De~Haan}.} \bibinfo{year}{2018}\natexlab{}.
\newblock \showarticletitle{Fully-automatic camera-based pulse-oximetry during
  sleep}. In \bibinfo{booktitle}{\emph{Proceedings of the IEEE Conference on
  Computer Vision and Pattern Recognition Workshops}}.
  \bibinfo{pages}{1349--1357}.
\newblock


\bibitem[\protect\citeauthoryear{Wang, Loparo, Kelly, and Kaplan}{Wang
  et~al\mbox{.}}{2015}]%
        {wang2015evaluation}
\bibfield{author}{\bibinfo{person}{Ying Wang}, \bibinfo{person}{Kenneth~A
  Loparo}, \bibinfo{person}{Monica~R Kelly}, {and} \bibinfo{person}{Richard~F
  Kaplan}.} \bibinfo{year}{2015}\natexlab{}.
\newblock \showarticletitle{Evaluation of an automated single-channel sleep
  staging algorithm}.
\newblock \bibinfo{journal}{\emph{Nature and science of sleep}}
  \bibinfo{volume}{7} (\bibinfo{year}{2015}), \bibinfo{pages}{101}.
\newblock


\bibitem[\protect\citeauthoryear{Wennberg, Wu, Rosenberg, and Spira}{Wennberg
  et~al\mbox{.}}{2017}]%
        {wennberg2017sleep}
\bibfield{author}{\bibinfo{person}{Alexandra~MV Wennberg},
  \bibinfo{person}{Mark~N Wu}, \bibinfo{person}{Paul~B Rosenberg}, {and}
  \bibinfo{person}{Adam~P Spira}.} \bibinfo{year}{2017}\natexlab{}.
\newblock \showarticletitle{Sleep disturbance, cognitive decline, and dementia:
  a review}. In \bibinfo{booktitle}{\emph{Seminars in neurology}},
  Vol.~\bibinfo{volume}{37}. Thieme Medical Publishers,
  \bibinfo{pages}{395--406}.
\newblock


\bibitem[\protect\citeauthoryear{Yu, Wu, Liou, Lee, and Hung}{Yu
  et~al\mbox{.}}{2012}]%
        {yu2012multiparameter}
\bibfield{author}{\bibinfo{person}{Meng-Chieh Yu}, \bibinfo{person}{Huan Wu},
  \bibinfo{person}{Jia-Ling Liou}, \bibinfo{person}{Ming-Sui Lee}, {and}
  \bibinfo{person}{Yi-Ping Hung}.} \bibinfo{year}{2012}\natexlab{}.
\newblock \showarticletitle{Multiparameter sleep monitoring using a depth
  camera}. In \bibinfo{booktitle}{\emph{International Joint Conference on
  Biomedical Engineering Systems and Technologies}}. Springer,
  \bibinfo{pages}{311--325}.
\newblock


\bibitem[\protect\citeauthoryear{Y{\"u}celba{\c{s}}, Y{\"u}celba{\c{s}},
  {\"O}z{\c{s}}en, Tezel, K{\"u}{\c{c}}{\c{c}}{\"u}kt{\"u}rk, and
  Yosunkaya}{Y{\"u}celba{\c{s}} et~al\mbox{.}}{2018}]%
        {yucelbacs2018automatic}
\bibfield{author}{\bibinfo{person}{C{\"u}neyt Y{\"u}celba{\c{s}}},
  \bibinfo{person}{{\c{S}}ule Y{\"u}celba{\c{s}}}, \bibinfo{person}{Seral
  {\"O}z{\c{s}}en}, \bibinfo{person}{G{\"u}lay Tezel}, \bibinfo{person}{Serkan
  K{\"u}{\c{c}}{\c{c}}{\"u}kt{\"u}rk}, {and} \bibinfo{person}{{\c{S}}ebnem
  Yosunkaya}.} \bibinfo{year}{2018}\natexlab{}.
\newblock \showarticletitle{Automatic detection of sleep spindles with the use
  of STFT, EMD and DWT methods}.
\newblock \bibinfo{journal}{\emph{Neural Computing and Applications}}
  \bibinfo{volume}{29}, \bibinfo{number}{8} (\bibinfo{year}{2018}),
  \bibinfo{pages}{17--33}.
\newblock


\end{thebibliography}

\end{document}